\documentclass[12pt]{article}
\pdfoutput=1
\usepackage{putex}
\usepackage{comment}
\usepackage{graphicx}
\usepackage{caption}
\usepackage{amsmath}
\usepackage[usenames,dvipsnames,table]{xcolor}
\usepackage{array}
\usepackage{subcaption}
\usepackage{epstopdf}
\usepackage{enumerate}
\usepackage{cite}
\usepackage{tensor}
\usepackage{slashed}
\usepackage[export]{adjustbox}
\usepackage[aligntableaux=center]{ytableau}
\usepackage{dsfont}
\usepackage[utf8]{inputenc}
\usepackage[
      colorlinks=true,
      linkcolor=blue,
      urlcolor=blue,
      filecolor=black,
      citecolor=red,
      ]{hyperref}
\newcommand{\RNum}[1]{\uppercase\expandafter{\romannumeral #1\relax}}

\newcommand {\be} {\begin {equation}}
\newcommand {\ee} {\end {equation}}

\newcommand {\bes} {\begin {equation*}}
\newcommand {\ees} {\end {equation*}}

\newcommand{\beq}{\begin{equation}}
\newcommand{\eeq}{\end{equation}}

\def\ie{\begin{equation}\begin{aligned}}
\def\fe{\end{aligned}\end{equation}}

\numberwithin{equation}{section}

\def\<{\langle}
\def\>{\rangle}

\begin{document}

%\preprint{PUPT-}

\institution{PU}{Department of Physics, Princeton University, Princeton, NJ 08544, USA}

\title{Monodromy Defects from Hyperbolic Space} 

\authors{Simone Giombi, Elizabeth Helfenberger, Ziming Ji, and Himanshu Khanchandani}

\abstract{We study monodromy defects in $O(N)$ symmetric scalar field theories in $d$ dimensions. After a Weyl transformation, a monodromy defect may be described by placing the theory on $S^1\times H^{d-1}$, where $H^{d-1}$ is the hyperbolic space, and imposing on the fundamental fields a twisted periodicity condition along $S^1$. In this description, the codimension two defect lies at the boundary of $H^{d-1}$. We first study the general monodromy defect in the free field theory, and then develop the large $N$ expansion of the defect in the interacting theory, focusing for simplicity on the case of $N$ complex fields with a one-parameter monodromy condition. We also use the $\epsilon$-expansion in $d=4-\epsilon$, providing a check on the large $N$ approach. When the defect has spherical geometry, its expectation value is a meaningful quantity, and it may be obtained by computing the free energy of the twisted theory on $S^1\times H^{d-1}$. It was conjectured that the logarithm of the defect expectation value, suitably multiplied by a dimension dependent sine factor, should decrease under a defect RG flow. We check this conjecture in our examples, both in the free and interacting case, by considering a defect RG flow that corresponds to imposing alternate boundary conditions on one of the low-lying Kaluza-Klein modes on $H^{d-1}$. We also show that, adapting standard techniques from the AdS/CFT literature, the $S^1\times H^{d-1}$ setup is well suited to the calculation of the defect CFT data, and we discuss various examples, including one-point functions of bulk operators, scaling dimensions of defect operators, and four-point functions of operator insertions on the defect.}

\date{}
\maketitle
\tableofcontents

\section{Introduction and summary}

Conformal defects are extended objects in a conformal field theory, which preserve a subgroup of the original conformal symmetry. A codimension $q$ flat or spherical conformal defect in a $d$-dimensional conformal field theory preserves a $SO(q) \times SO(d - q + 1, 1)$ subgroup of the original $SO(d + 1, 1)$ symmetry group (see \cite{Billo:2016cpy} for an introduction to the subject). $SO(d -q + 1, 1)$ represents the conformal group on the defect, and $SO(q)$ corresponds to rotations around the defect. For the correlation functions of operators on the defect, $SO(q)$ acts as a global symmetry. Even though only a subgroup of the full conformal symmetry is preserved, the correlation functions are highly constrained. There has been a lot of recent activity in analyzing the consequences of these constraints \cite{Billo:2016cpy, Gadde:2016fbj, Lauria:2017wav, Lemos:2017vnx, Guha:2018snh, Isachenkov:2018pef, Lauria:2018klo, Liendo:2019jpu, Herzog:2020bqw}.  

In this paper we study monodromy defects, working with scalar field theories as our main example.\footnote{Conformal defects in free scalar theories were recently discussed also in \cite{Lauria:2020emq, Nishioka:2021uef}.} A monodromy defect is a codimension two defect, so $q = 2$. To describe this codimension two defect in flat space, we can parametrize coordinates as $ x = (r , \theta, \vec{y})$ where $\vec{y}$ represents $d - 2$ dimensions along the defect while $r,\theta$ are polar coordinates perpendicular to the defect. The defect is located at $r = 0$ in these coordinates. A monodromy defect in a CFT with a global symmetry group $G$ may be defined by requiring that the fields satisfy 
\begin{equation}
\phi(r, \theta + 2 \pi, \vec{y}) = \phi^g(r, \theta, \vec{y})
\label{twist-per}
\end{equation} 
where $g$ is a non-identity element of $G$. For a free scalar field and $3d$ Ising model, $G = Z_2$. Such a monodromy defect in the 3d Ising model (where it is a line defect) was introduced in \cite{Billo:2013jda, Gaiotto:2013nva}, and  further studied in \cite{Yamaguchi:2016pbj, Liendo:2019jpu}. In this paper, we consider more generally conformal field theories consisting of $N$ scalars that preserve an $O(N)$ symmetry, so $G = O(N)$. We discuss both free and interacting $O(N)$ model, using the large $N$ and $\epsilon$ expansions . Already in the free theory, the structure is richer compared to the case of a single free scalar. Some results for the monodromy defect in critical $O(N)$ model in $d = 4 - \epsilon$ were also obtained in \cite{Soderberg:2017oaa}.

A codimension two defect may also be studied by mapping the problem to a hyperbolic cylinder, $S^1 \times H^{d-1}$ (for previous examples of conformal mapping to $H^m \times S^n$ spaces to describe conformal defects, see e.g. \cite{Kapustin:2005py, Hung:2014npa, Gadde:2016fbj, Nishioka:2021uef}). A flat defect in flat space can be related by a Weyl transformation to $S^1 \times H^{d -1}$ as follows 
\begin{equation} \label{WeylHyperbolic}
d s_{\mathbb{R}^d}^2 = r^2 \left( \frac{(d \vec{y})^2 + dr^2}{r^2} + d \theta^2 \right) = r^2 d s_{S^1 \times H^{d -1}}^2
\end{equation}
with $\vec{y}$ and $r$ being the Poincar\'e coordinates on $H^{d - 1}$ and $\theta$ being the coordinate on $S^1$. The monodromy defect is then simply described by imposing twisted periodicity conditions (\ref{twist-per}) along $S^1$ in the path integral for the theory on $S^1\times H^{d-1}$ (for instance, in the $Z_2$ case, this just means taking the scalar to be antiperiodic on $S^1$). The defect is located at $r = 0$, which is the boundary of the hyperbolic space. The hyperbolic cylinder setup that we use here is similar to that used in \cite{Klebanov:2011uf, Casini:2010kt, Casini:2011kv, Hung:2011nu, Balakrishnan:2016ttg, Bianchi:2016xvf, Belin:2013dva} to study Renyi entropies and their charged generalizations \cite{Belin:2013uta, Belin:2014mva} for a spherical entangling region.\footnote{In that case, rather than a twisted periodicity condition, one lets the inverse temperature, i.e. the length of $S^1$, be $2\pi q$ to describe the $q$th Renyi entropy. This setup can also be thought in terms of defect CFT, as discussed in \cite{Hung:2014npa, Bianchi:2015liz}.} Starting with the theory on $S^1\times H^{d-1}$, one can then perform a Kaluza-Klein reduction on $S^1$ to obtain a tower of massive fields on $H^{d-1}$ with the defect theory on its boundary. Standard techniques from the AdS/CFT literature may then be used to obtain results for the defect CFT data. For example, the scaling dimensions of the defect operators can be related to the masses on $H^{d-1}$. Other examples of defect CFT data may also be extracted conveniently from the hyperbolic space setup, and we will discuss a few explicit such calculations below. Similar ideas have been used in the literature for boundaries in conformal field theory in \cite{Paulos:2016fap, Carmi:2018qzm, Herzog:2019bom, Herzog:2020lel, Giombi:2020rmc} and for higher codimension defects in \cite{Nishioka:2021uef}.  Field theory on a hyperbolic cylinder was also studied in \cite{Aharony:2015hix, Rodriguez-Gomez:2017kxf, Pittelli:2018rpl}. 

A monodromy defect with spherical geometry may be described in the same way by simply using, instead of the Poincar\'e metric in (\ref{WeylHyperbolic}), the hyperbolic ball metric for $H^{d-1}$, so that the boundary is the sphere $S^{d-2}$. For a spherical defect, it is natural to define its expectation value $\langle \mathcal{D}\rangle$. In the hyperbolic space setup, this expectation value can be obtained in terms of the free energy of the twisted theory on $S^1\times H^{d-1}$ as
%We also use the hyperbolic space to define the expectation value of the defect 
\begin{equation} \label{DefectExpectationDefIntro}
-\log \langle \mathcal{D} \rangle = F_{\textrm{twisted}} - F_{\textrm{untwisted}}
\end{equation}
where $F_{\textrm{twisted}}$ and $F_{\textrm{untwisted}}$ are the free energies on $S^1 \times H^{d -1}$ in the presence and absence of a monodromy defect respectively. The subtraction of the untwisted theory free energy corresponds to normalizing $\langle \mathcal{D}\rangle$ by the partition function of the theory without defect. 

For a range of masses, massive fields on $H^{d-1}$ have two allowed boundary conditions \cite{Klebanov:1999tb}, which in our setup are interpreted as different defect CFTs, which sit at the endpoints of a RG flow localized on the defect. It was conjectured in \cite{Kobayashi:2018lil} (generalizing a similar proposal \cite{Giombi:2014xxa} in the absence of defects) that for a codimension $q$ defect, the quantity defined by 
\begin{equation} \label{DtildeDefinition}
\tilde{\mathcal{D}} \equiv \sin \left( \frac{\pi (d - q) }{2} \right) \log \langle \mathcal{D} \rangle
\end{equation} 
decreases under RG flows localized on the defect. Note that, when the dimension of the defect $d-q$ is even, the expectation value has a logarithmic divergence related to one of the conformal anomaly coefficients of the defect. This divergence appears as a pole when working in dimensional regularization. The sine factor in (\ref{DtildeDefinition}) cancels the pole and $\tilde{\mathcal{D}}$ is a finite quantity proportional to the anomaly coefficient. On the other hand, for odd $d-q$, the expectation value is finite and the sine factor is just an alternating sign. We explicitly check the proposal that $\tilde{\mathcal{D}}$ decreases under defect RG flows in the examples we study, by calculating the defect expectation value on $S^1 \times H^{d-1}$. We verify that whenever the defect RG flow is such that scaling dimensions obey unitarity bounds, $\tilde{\mathcal{D}}$ decreases under the flow, but this does not hold true for non-unitary flows.

We now provide a summary of the rest of this paper. We start in section \ref{SectionFreeTheory} by studying monodromy defects in the free $O(N)$ model. We discuss the most general monodromy defect, and show that it is essentially sufficient to study the monodromy defect in a single free complex scalar field theory, and the results for the most general monodromy defect in free $O(N)$ model follow from the results for this case. We start by describing the monodromy defect in flat space, and then explain how to map the problem to the hyperbolic cylinder $S^1 \times H^{d-1}$ by a Weyl transformation. We also discuss how an alternative Weyl transformation maps the problem to the $d$-dimensional sphere (with twisted periodicity along one angle). We calculate various defect CFT observables, including the dimensions of the defect operators and one-point functions of bulk operators with spin. %The defect is compact on $S^1 \times H^{d-1}$ and on $S^d$, so the defect expectation value is a meaningful quantity. 
We define and calculate the expectation value of a spherical defect using both the $S^1 \times H^{d-1}$ and $S^d$ setups, and show that they agree. In section \ref{SectionDefectRGFlow}, we then study a defect RG flow in the free theory. The two defect CFTs connected by the RG flow correspond to the two possible boundary conditions for one of the low-lying KK modes on $H^{d-1}$ in the $S^1 \times H^{d-1}$ setup. A similar defect RG flow in the case of non-monodromy defects in a free scalar theory was discussed in \cite{Nishioka:2021uef}.   

In section \ref{SectionLargeN}, we study the monodromy defect in the interacting theory of $N$ scalars with a $\phi^4$ $O(N)$ invariant interaction. In the interacting case, we study the simple situation of $2 N$ scalars with $N$ pairs mixing into each other as they go around the defect. This is equivalent to studying a monodromy defect for $N$ interacting complex scalars with the same monodromy for all of them, and it preserves a $U(N)$ symmetry. The case of $Z_2$ monodromy on all scalars, generalizing the Ising case of \cite{Gaiotto:2013nva}, can be obtained as a special case. We develop the large $N$ description of the monodromy defect, and calculate the scaling dimensions of defect operators to leading order at large $N$. We also study the defect expectation value and the expectation value of the bulk stress-tensor, which is proportional to the ``conformal weight" \cite{Kapustin:2005py, Hung:2014npa} of the defect, to leading order at large $N$. We then study the same defect in the Wilson-Fisher $\epsilon$ expansion in $d = 4 - \epsilon$, and provide some checks of the large $N$ analysis. We also calculate correlation functions of the defect operators by computing Witten diagrams in $H^{d-1}$ to leading order in $\epsilon$. We then conclude in section \ref{SectionConclusion} and mention some future directions. Appendices contain some technical details and some useful integrals.

\section{Monodromy defect in free field theory} \label{SectionFreeTheory}
Consider an $O(N)$ symmetric theory of $N$ free scalars in flat space. The most general monodromy defect that we can define imposes that the scalars satisfy \cite{Soderberg:2017oaa}
\begin{equation} \label{MonodromyDefinition}
\phi^I(r, \theta + 2 \pi, \vec{y}) = G^{IJ} \phi^J(r, \theta, \vec{y}), \hspace{0.5cm} G^{IJ} \in O(N).
\end{equation} 
The most general $O(N)$ matrix $G^{IJ}$, can always, by a change of basis, be brought to the following form 
\begin{equation} \label{MonodromyMatrix}
G^{IJ}(\vartheta) = \begin{bmatrix}
						R(\vartheta_1)& & & & &  \\
						&\ddots & & & 0& \\
						& & R(\vartheta_k)& & &\\		
						& & & \pm 1 & &\\
						&0& & & \ddots &\\
						& & & & & \pm1	
					\end{bmatrix}, \hspace{1 cm}
					R(\vartheta) = \begin{bmatrix}
					 \cos \vartheta & - \sin \vartheta \\
					\sin \vartheta & \cos \vartheta
					\end{bmatrix}.
\end{equation}
So there are $k$ pairs of scalars that  mix into each other and the rest either remain unchanged or pick up a minus sign as they go around the defect. We can then combine each pair into a complex combination $\Phi = \phi^1 + i \phi^2$ and the monodromy can be represented as 
\begin{equation} \label{ComplexScalarMonodromy}
\Phi (r, \theta + 2 \pi, \vec{y} ) = e^{i \vartheta} \Phi(r, \theta, \vec{y}) , \hspace{1cm} \vartheta \sim \vartheta + 2 \pi.
\end{equation}
Hence $\vartheta = 0$ describes the trivial defect while  $\vartheta = \pi$ describes the special case when the two fields change a sign as they go around the defect. So in the rest of this section, we will consider a single complex scalar with the monodromy defined in \eqref{ComplexScalarMonodromy}. It has a $U(1) \sim SO(2)$ internal symmetry \footnote{Not to be confused with the group of rotations around the defect, which is a spacetime symmetry in the bulk and is also $SO(2)$.} which is enhanced to $O(2)$ for $\vartheta = 0$ and $\pi$ ($\Phi \rightarrow \bar{\Phi}$, which is a part of $O(2)$ but not $SO(2)$, is also a symmetry for these values of $\vartheta$). One may combine these complex scalars with different $\vartheta$'s to obtain results for free $O(N)$ model with a general monodromy defect (for each minus sign in the monodromy matrix (\ref{MonodromyMatrix}), one can simply set $\vartheta=\pi$ in the result for a complex scalar below, and include an extra factor of $1/2$ to describe a real component instead of a complex one). To make the expressions less cluttered, we define $v = \vartheta/ 2 \pi$ and use either $v$ or $\vartheta$, whichever is convenient. There is a periodicity in $v$ which implies that everything should be invariant under $v \rightarrow v + 1$, but for many calculations, we will fix the range of $v$ to be $0 \le v < 1$. We will write expressions specializing to this range of $v$, so they may not look periodic in $v$.

In a conformal field theory with a defect, in addition to the usual short distance OPE in the bulk, a bulk operator can also be expanded in terms of operators living on the defect. For the complex scalar with monodromy given by \eqref{ComplexScalarMonodromy}, it takes the following form \cite{Billo:2016cpy, Gaiotto:2013nva} 
\begin{equation} \label{BulkDefectOPEGeneral}
\Phi (r, \theta, \vec{y} ) = \sum_{O} C^{\Phi}_{O} \frac{e^{i s_O \theta}}{r^{\Delta_{\Phi} - \Delta_{O}}} \mathcal{B}_{\Delta_O} (r, \vec{\partial}_y) O({\vec{y}}), \hspace{1cm} s_O \in \mathbb{Z} + v.
\end{equation}
As we mentioned in the introduction, $SO(2)$ symmetry of rotations around the defect acts as a global symmetry on the defect. $s_O$ is the charge of the operator $O$ under this global symmetry and we will call it transverse spin or just spin. There is also a longitudinal spin $l$, which is the charge under rotations along the defect, but we will only consider $l = 0$ defect operators in this paper. The remaining conformal invariance fixes the bulk-defect two point function
\begin{equation} \label{BulkDefectTwoPoint}
\langle \Phi (x_1) \bar{O} (\vec{y}_2) \rangle = \frac{C_O^{\Phi} \mathcal{C}_{\Delta_O} e^{i \theta s_O}}{r^{\Delta_{\Phi} - \Delta_O} (r^2 + (\vec{y}_{12})^2)^{\Delta_O}}
\end{equation}
where $\bar{O}$ is a defect operator that has spin $-s_O$ and dimension $\Delta_O$. Consistency of \eqref{BulkDefectOPEGeneral} and \eqref{BulkDefectTwoPoint} fixes the form of the function $\mathcal{B}_s (r, \vec{\partial}_y)$ 
\begin{equation}
\mathcal{B}_{\Delta_O} (r, \vec{\partial}_y) = \sum_{m = 0}^{\infty} \frac{(-1)^m r^{2 m} (\vec{\partial}_y^2)^m}{m! 2^{2 m} \left( \Delta_{O} + 2 - \frac{d}{2} \right)_m}. 
\end{equation}
This is similar to what was done for BCFT in \cite{McAvity:1995zd}. In general, there could be several defect operators of a given spin. But since $\Phi$ is a free field, it satisfies the bulk equation of motion $\nabla^2 \Phi = 0$, which implies 
\begin{equation} \label{DimensionSpinFlat}
\begin{split}
&\left(\frac{\partial^2}{\partial r^2} + \frac{1}{r} \frac{\partial}{\partial r} + \frac{1}{r^2} \frac{\partial^2}{\partial \theta^2} + \frac{\partial^2}{\partial \vec{y}_1^2} \right) \langle \Phi (x_1) \bar{O} (\vec{y}_2) \rangle = 0 \\
&\implies \Delta_O = \Delta_{\Phi} \pm |s_O| = \frac{d}{2} - 1 \pm |s_O|. 
\end{split}
\end{equation}
The unitarity bound for the CFT on the defect requires the dimensions of the defect operator to satisfy
\begin{equation} \label{DefectUnitarity}
\Delta_O \ \ge \ \textrm{max} \left( \frac{d}{2} - 2, 0 \right). 
\end{equation} 
This is always satisfied for the positive sign above (as long as $d > 2$) and we defer the discussion of negative sign until next section. So for every spin, there is a single operator on the defect with dimension $\Delta_s = d/2 - 1 + |s|$. Hence, the bulk-defect OPE of the fundamental fields may be written as a sum over spins
\begin{equation} \label{BulkDefectOPE}
\begin{split}
\Phi (r, \theta, \vec{y} ) &= \sum_{s \in \mathbb{Z} + v} C^{\Phi}_{s} \frac{e^{i s \theta}}{r^{\Delta_{\Phi} - \Delta_{s}}} \mathcal{B}_s (r, \vec{\partial}_y) \Psi_s({\vec{y}}) \\
\bar{\Phi} (r, \theta, \vec{y} ) &= \sum_{s \in \mathbb{Z} + v} (C^{\Phi}_{s})^* \frac{e^{-i s \theta}}{r^{\Delta_{\Phi} - \Delta_{s}}} \mathcal{B}_s (r, \vec{\partial}_y) \bar{\Psi}_s({\vec{y}}).
\end{split}
\end{equation}
In terms of original real scalar fields, $\Psi_s = \psi^1_s + i \psi^2_s$ while $\bar{\Psi}_s =\bar{\psi}^1_{s} - i  \bar{\psi}^2_{s}= \psi^1_{-s} - i \psi^2_{-s}$ where $\psi_s^{i}$ appear in the bulk-defect OPE of the real scalars $\phi^{i}$.  

In the presence of a defect, the two-point function of bulk scalars is fixed up to a function of cross-ratios \cite{Billo:2016cpy, Liendo:2019jpu} 
\begin{equation}
\langle \Phi(x_1) \bar{\Phi} (x_2) \rangle = \frac{\mathcal{F} (\theta_{12}, \xi)}{(r_1 r_2)^{\frac{d}{2} - 1}}, \hspace{0.5cm} \theta_{12} = \theta_{1} - \theta_{2}, \hspace{0.5cm} \xi = \frac{(\vec{y}_1 - \vec{y}_2)^2 + (r_1 - r_2)^2}{4 r_1 r_2}.
\end{equation}
Corresponding to the two OPE limits (i.e. the bulk OPE and the bulk-defect OPE), the function $\mathcal{F}$ can be expanded into bulk and defect channel conformal blocks 
\begin{equation}
\mathcal{F} (\theta_{12}, \xi) = \sum_{\mathcal{O}} {C_{\bar{\Phi}; \Phi;}}^{\mathcal{O}} C^{\mathcal{O}}_{1} g_{\Delta_{\mathcal{O}}, J_{\mathcal{O}}}(\theta_{12}, \xi) = \sum_{O} |C^{\Phi}_O|^2 f_{\Delta_O, s_O} (\theta_{12}, \xi)
\end{equation}
where $g$ and $f$ are the bulk channel and defect channel conformal blocks respectively. The sum on the left runs over the bulk operators that get a non-zero one-point function, and the coefficient is the product of the usual bulk OPE coefficient times the one-point function coefficient of the bulk operator. The sum on the right runs over the defect operators that appear in the bulk-defect OPE of $\Phi$.

As we determined above, the operators appearing in the defect channel have spin $s$ and dimension $\Delta_s = d/2 - 1 + |s|$. The defect channel blocks are known in general \cite{Billo:2016cpy, Liendo:2019jpu}. For the case of a codimension two defect, they simplify and the resulting expression for the two-point function can be written as a sum over defect operators
\begin{equation} \label{TwoPointFunctionFlat}
\langle \Phi(x_1) \bar{\Phi} (x_2) \rangle = G _{\bar{\Phi} \Phi} (x_1, x_2) = \sum_{s \in \mathbb{Z} + v} \frac{ \Gamma \left(\Delta_s \right) 
e^{i s \theta_{12}} \, _2F_1\left(\Delta_s, \Delta_s  +\frac{3 - d}{2};2  \Delta_s + 3 - d ;-\frac{1}{\xi }\right) }{2 (r_1 r_2)^{\frac{d}{2} - 1} \pi^{d/2}  \Gamma ( \Delta_s + 2 -\frac{d}{2})  (4 \xi)^{\Delta_s}}.
\end{equation} 
The sum can be explicitly performed in $d = 4$ to get 
\begin{equation} \label{TwoPointFunction4d}
\begin{split}
&\langle \Phi(x_1) \bar{\Phi} (x_2) \rangle  =  \sum_{s \in \mathbb{Z} + v} \frac{ e^{i s \theta_{12}} }{8 \pi^{2} r_1 r_2 \sqrt{\xi (1 + \xi)} (\sqrt{1 + \xi} + \sqrt{\xi})^{2 \Delta_s - 2}} \\
& =  \frac{(\xi (1 + \xi))^{-1/2}}{8 \pi^{2} r_1 r_2  } \left(\frac{e^{i \theta_{12} v} \left(\sqrt{\xi }+\sqrt{\xi +1}\right)^{2 v}}{-1+e^{i \theta_{12}} \left(2 \xi +2 \sqrt{\xi (1 + \xi)}+1\right)}+\frac{e^{i \theta_{12}   v} \left(\sqrt{\xi }+\sqrt{\xi +1}\right)^{2-2 v}}{-e^{i \theta_{12}}+2 \xi +2 \sqrt{\xi (1 + \xi)}+1} \right).  
\end{split}
\end{equation}
Note that we are using a normalization, such that in the bulk OPE limit, when $x_1 \rightarrow x_2$, the correlator goes like 
\begin{equation}
\langle \Phi(x_1) \bar{\Phi} (x_2) \rangle  \sim \frac{\Gamma\left(\frac{d}{2} - 1 \right)}{2 \pi^{d/2}} \frac{1}{|x_1 - x_2|^{d - 2}}.
\end{equation}   
We normalize defect operators such that $C^{\Phi}_{s} = 1$ in the free theory. The two-point function of the defect operators is then given by 
\begin{equation} \label{DefectTwoPointFree}
\langle \Psi_{s_1}(\vec{y}_1)\bar{\Psi}_{s_2}(\vec{y}_2) \rangle =  \frac{ \delta_{s_1, s_2} \mathcal{C}_{\Delta_{s_1}}}{(\vec{y}^2_{12})^{\Delta_{s_1}} }, \hspace{1cm} \mathcal{C}_{\Delta_{s_1}} = \frac{\Gamma \left(\Delta_{s_1} \right)}{2 \pi^{d/2} \Gamma \left(\Delta_{s_1} + 2 - \frac{d}{2} \right)}. 
\end{equation}

In the bulk channel conformal block decomposition, the operators that appear are the bulk scalar $\bar{\Phi} \Phi$ and the conserved currents of all spins, which can be schematically written as $\bar{\Phi} (\partial_\mu)^J \Phi$. To extract the bulk expansion coefficients, one may use the inversion formula of \cite{Liendo:2019jpu}. Here, we restrict to calculating the one-point function of the first few operators of low spin. The one-point function of the operator $\bar{\Phi} \Phi$ can be extracted from the short distance limit of the correlator \eqref{TwoPointFunctionFlat}
\begin{equation} \label{OnePointPhibarPhi}
\langle \bar{\Phi} \Phi (x) \rangle = \frac{C_1^{\bar{\Phi} \Phi}}{r^{d - 2}}, \hspace{0.5cm} C_1^{\bar{\Phi} \Phi} = \frac{ (d - 1) \Gamma \left( \frac{d}{2}  - v \right) \Gamma \left( \frac{d}{2} - 1 + v \right) \sin \left( v \pi \right) \Gamma \left( \frac{d}{2} \right)}{\pi^{\frac{d}{2} + 1} \Gamma(d) (2 - d)}. 
\end{equation} 
The conserved currents, which are spinning operators, also get one-point functions. The spin one current, which corresponds to the global $U(1)$ symmetry of the theory is given by 
\begin{equation}
J_{\mu} = i  \left( \Phi \nabla_{\mu} \bar{\Phi} - \bar{\Phi} \nabla_{\mu} \Phi \right).
\end{equation}
The one-point function of a parity odd spin one operator in the presence of a defect is fixed by conformal symmetry  \cite{Billo:2016cpy}
\begin{equation} \label{CurrentGeneral}
\langle J_i \rangle = \frac{C^J_1 \epsilon_{ij} n^j}{r^{d}}, \hspace{1cm} \langle J_a \rangle = 0, \hspace{1cm} n_i = x_i/r.
\end{equation}
We parametrize the coordinates as $ x = (x^i, x^a)=$ with $i,j$ now being Cartesian transverse coordinates, $a,b$ being directions along the defect and $\epsilon_{ij}$ is the antisymmetric tensor in transverse directions. We can calculate this one point function by calculating derivatives of \eqref{TwoPointFunctionFlat}, and then taking the short distance limit. Since it is fixed up to a constant, it is enough to do the calculation just for one component. We do it for the $\theta$ component, 
\begin{equation} \label{CurrentOnePointTheta}
\begin{split}
\langle J_{\theta} \rangle &= - \frac{C_1^J}{r^{d - 2}} = i \langle  \Phi \partial_{\theta} \bar{\Phi} - \bar{\Phi} \partial_{\theta} \Phi \rangle = \frac{2^{2 - d} \pi^{-\frac{(d+1)}{2}} \Gamma \left(\frac{3-d}{2}\right)}{r^{d - 2}} \sum_{k = - \infty}^{\infty} \frac{ (k + v) \Gamma \left(\frac{d}{2} - 1 + | k + v| \right)}{ \Gamma \left(2-\frac{d}{2} + | k + v|\right)} \\
& = \frac{(d-2) (2 v-1)}{d-1} \frac{C_1^{\bar{\Phi} \Phi}}{r^{d - 2}}.
\end{split}
\end{equation} 
At $v = 1/2$, we expect the internal $U(1)$ symmetry to be enhanced to full $O(2)$ symmetry which includes $\bar{\Phi} \rightarrow \Phi$, under which $J_{\mu} \rightarrow - J_{\mu}$. So we expect the correlators containing odd powers of $J_{\mu}$ to vanish at $v = 1/2$, and indeed the one-point function vanishes at $v = 1/2$. 

Next let's discuss the stress tensor, which is a spin two conserved current. Conformal invariance, tracelessness and conservation fix the form of its one-point function up to a constant \cite{Billo:2016cpy, Hung:2011nu, Hung:2014npa} 
\begin{equation} \label{StressTensorGeneral}
\langle T_{ij} \rangle_{\mathbb{R}^d} =  \frac{h}{2 \pi} \frac{(d - 1)\delta_{ij} - d n_i n_j}{r^d}, \hspace{0.5cm}\langle T_{ab} \rangle_{\mathbb{R}^d} = - \frac{h}{2 \pi} \frac{\delta_{ab}}{r^d}, \hspace{0.5cm} \langle T_{ai} \rangle_{\mathbb{R}^d} =0. 
\end{equation} 
In analogy with the scaling dimensions of local operators, $h$ is referred to as the conformal weight of the defect \cite{Kapustin:2005py, Hung:2011nu, Hung:2014npa}. It can be determined by doing explicit calculation of any component of the stress tensor and we choose $T_{\theta \theta}$. The canonical stress energy tensor for a free complex scalar in flat space is
\begin{equation}
T_{\mu \nu} = \frac{d}{4 (d-1)} \nabla_{\mu} \bar{\Phi} \nabla_{\nu}\Phi - \frac{1}{4(d-1)} g_{\mu \nu} \nabla \bar{\Phi} \cdot \partial \Phi -  \frac{(d - 2)}{4 (d -1)} \bar{\Phi} \nabla_{\mu} \nabla_{\nu}\Phi + \text{c.c.}.
\end{equation}
This gives
\begin{equation} \label{StressTensorThetaDirect}
\begin{split}
\langle T_{\theta \theta} \rangle &= \frac{\langle \partial_{\theta} \bar{\Phi} \partial_{\theta} \Phi \rangle }{4} - \frac{r^2 \left(\langle \partial_{r} \bar{\Phi} \partial_{r} \Phi \rangle + \langle \vec{\partial}_{y} \bar{\Phi} \vec{\partial}_{y} \Phi \rangle \right) }{4 (d - 1)}  -\frac{d - 2}{4 (d - 1)} \langle \bar{\Phi} \partial^2_{\theta} \Phi \rangle - \frac{r (d - 2)}{4 (d - 1)}\langle \bar{\Phi} \partial_{r} \Phi \rangle + \text{c.c.}.
\end{split}
\end{equation}
Using the two-point function in \eqref{TwoPointFunctionFlat} and taking appropriate derivatives, we get 
\begin{equation} \label{StressTensorThetaDirectResult}
\langle T_{\theta \theta} \rangle = \frac{  \Gamma \left(\frac{3 - d}{2}\right)(1 -v ) v  \left(\csc \pi \left(\frac{d}{2} - v \right)-\csc \pi \left(\frac{d}{2} + v \right) \right) }{ d   (4 \pi)^{\frac{d - 1}{2}}  \Gamma \left( 2 - \frac{d}{2}- v \right) \Gamma \left(1 - \frac{d}{2}+ v \right)} \frac{1}{r^{d - 2}} = -\frac{(d-2)v(1 - v)}{d} \frac{C_1^{\bar{\Phi} \Phi}}{r^{d - 2}}.
\end{equation}
Comparing with \eqref{StressTensorGeneral}, it is easy to see that
\begin{equation} \label{ConformalWeightFree}
\langle T_{\theta \theta} \rangle  = \frac{(d- 1) h}{2 \pi \ r^{d - 2}} \implies h = - \frac{\pi \Gamma \left(\frac{1 - d}{2}\right)(1 -v ) v  \left(\csc \pi \left(\frac{d}{2} - v \right)-\csc \pi \left(\frac{d}{2} + v \right) \right) }{ d   (4 \pi)^{\frac{d - 1}{2}}  \Gamma \left( 2 - \frac{d}{2}- v \right) \Gamma \left(1 - \frac{d}{2}+ v \right)}.
\end{equation}
We checked numerically that this conformal weight $h$ is always positive for $d > 2$. This is consistent with the conjecture proposed in \cite{Lemos:2017vnx} which says that $h \ge 0$ in unitary defect CFTs \footnote{In \cite{Lemos:2017vnx}, stress tensor one-point function was written in terms of $a_T$ which is related to $h$ by $h = - 2 \pi a_T/d$, so they conjectured that $a_T \le 0$.}. We can follow this logic and calculate the one-point function of any higher spin current. We just do it for one more case here, namely the spin 3 symmetric current. The current is given by (explicit expression in $d=4$ can be found in, for example, \cite{Beccaria:2017nco})
\begin{equation}
J_{\mu\nu\rho}=6i \left( \bar{\Phi}\nabla_\mu \nabla_\nu\nabla_\rho \Phi-\frac{3(d+2)}{d-2}\nabla_{(\mu}\bar{\Phi}\nabla_\nu \nabla_{\rho)}\Phi+\frac{6}{d-2} g_{(\mu\nu}\nabla^\gamma\bar{\Phi}\nabla_\gamma\nabla_{\rho)}\Phi \right)+ \text{c.c.}
\end{equation}
where $()$ in the subscript means that the indices are symmetrized.  Its one-point function is also fixed by conformal symmetry up to a number, so we only look at one of its components with all indices equal to $\theta$
\begin{equation}
\begin{split}
J_{\theta \theta \theta} = 6 i \bigg( & \bar{\Phi}\partial^3_{\theta} \Phi -\frac{3 d}{d-2} \partial_{\theta} \bar{\Phi}\partial^2_{\theta} \Phi - 2 \bar{\Phi} \partial_{\theta} \Phi + 3 r \left( \bar{\Phi}\partial _{\theta} \partial _{r} \Phi + \partial _{r} \bar{\Phi}\partial _{\theta} \Phi  \right)  \\
& + \frac{6 r^2}{d - 2} \left(\partial_{r} \bar{\Phi} \partial_{r} \partial_{\theta} \Phi +  \vec{\partial}_{y} \bar{\Phi} \vec{\partial}_{y} \partial_{\theta} \Phi  \right) \bigg) +\text{c.c.}.
\end{split}
\end{equation}
We act with these derivatives on \eqref{TwoPointFunctionFlat} and expand them in the bulk limit $\xi\to 0$ to get
\begin{equation}
\langle J_{\theta\theta\theta} \rangle = 24 (1 - v) v (1 - 2 v)  \frac{C_1^{\bar{\Phi} \Phi}}{r^{d - 2}}.
\end{equation} 
\subsubsection*{Displacement operator} \label{SectionDisplacement}
The presence of a defect breaks the translational symmetry perpendicular to the defect. This leads to the presence of a displacement operator in the spectrum of the defect theory which may be defined as the divergence of stress tensor 
\begin{equation} \label{DisplacementDefinition}
\partial_{\mu}T^{\mu i} = D^{i} (\vec{y}) \delta^{2} (x^i)
\end{equation}
where $i$ represents directions perpendicular to the defect and the delta function is localized on the defect, at $x^i = 0$. This equation fixes the scaling dimension of the displacement equal to $d - 1$, and its $SO(2)$ spin equal to $1$. In the free theory, this requires the displacement to be proportional to the operator $\Psi_v \bar{\Psi}_{-1 + v}$ because this is the only operator with the required spin and conformal dimension. The proportionality constant is also fixed by \eqref{DisplacementDefinition}. In this subsection, we find this constant and calculate the normalization of the displacement operator which is a piece of the defect CFT data.

It is convenient to work with complex coordinates in directions transverse to the defect $z = r e^{- i \theta} $ and call the two components of displacement $D$ and $\bar{D}$ defined by 
\begin{equation}
\partial_{\mu}T^{\mu z} = 2 D (\vec{y}) \delta^{2} (z , \bar{z}), \hspace{1cm} \partial_{\mu}T^{\mu \bar{z}} = 2 \bar{D} (\vec{y}) \delta^{2} (z , \bar{z})
\end{equation}
since $2\delta^2(z , \bar{z}) = \delta^2 (x)$. In terms of these coordinates, the bulk defect OPE takes the form 
\begin{equation}
\Phi (z, \bar{z}, \vec{y} ) = \sum_{s \in \mathbb{Z} + v}  z^{\frac{|s| - s}{2}} \bar{z}^\frac{|s| + s}{2} \mathcal{B}_s (|z|, \vec{\partial}_y) \Psi_s({\vec{y}}).
\end{equation}
Using these definitions, it is easy to see that only the $\partial_z T^{z z}$ term in the divergence of the stress tensor contains a delta function
\begin{equation} \label{DisplacementDefinitionFree}
\begin{split}
&T^{z z} = 4 T_{\bar{z} \bar{z}} \ni \frac{4 v (1 - v)}{\bar{z}} \Psi_{v} \bar{\Psi}_{-1 + v} (\vec{y}) \implies \partial_z T^{z z} \ni 8 \pi v (1-v) \Psi_{v} \bar{\Psi}_{-1 + v} (\vec{y}) \delta^2 (z, \bar{z}) \\
& \implies D(\vec{y}) = 4 \pi v (1-v) \Psi_{v} \bar{\Psi}_{-1 + v} (\vec{y}).
\end{split}
\end{equation}
We can use this to directly calculate coefficient of the displacement operator 
\begin{equation}\label{DisplacementCoefficientFree}
\begin{split}
&\langle D (\vec{y}_1) \bar{D} (\vec{y}_2) \rangle = \frac{C_D}{(\vec{y}_{12}^2)^{d - 1}} = \frac{16 \pi^2 v^2 (1 - v)^2 \mathcal{C}_{\Delta_{v}} \mathcal{C}_{\Delta_{v - 1}}}{(\vec{y}_{12}^2)^{d - 1}}\\
& \implies C_D = \frac{4 \Gamma \left(\frac{d}{2} -1 + v \right) \Gamma \left(\frac{d}{2} - v \right)}{\pi^{d - 2} \Gamma(-v) \Gamma(v - 1)}.
\end{split}
\end{equation}

The displacement operator satisfies a Ward identity involving correlators of the displacement with the bulk operators \cite{Billo:2016cpy}. It implies that if the displacement appears in the bulk-defect OPE of a bulk operator $\mathcal{O}$, then the corresponding bulk-defect OPE coefficient $C^{\mathcal{O}}_D$ must satisfy  \cite{Billo:2016cpy}
\begin{equation} \label{WardIdentity}
\Delta_{\mathcal{O}} C^{\mathcal{O}}_1 = -  \left(\frac{\pi}{4} \right)^{\frac{d}{2} - 1} \frac{\sqrt{\pi}}{\Gamma\left( \frac{d - 1}{2} \right)} C^{\mathcal{O}}_D C_D
\end{equation} 
where $C^{\mathcal{O}}_1$ is the coefficient of the one-point function of $\mathcal{O}$. The displacement operator does appear in the bulk defect OPE of the operator $\bar{\Phi} \Phi$. It can be seen from the defect channel decomposition of the two-point function of $\bar{\Phi} \Phi$, which can be obtained by Wick contraction and contains the following term 
\begin{equation} \label{DisplacementPhiPhi}
\begin{split}
\langle \bar{\Phi} \Phi (x_1) \bar{\Phi} \Phi (x_2) \rangle &\ni \frac{e^{ i (\theta_1 - \theta_2)} \Gamma\left(\Delta_v \right)\Gamma\left(\Delta_{v - 1} \right)}{4^{1 + \Delta_v + \Delta_{v - 1}  } \pi^d \Gamma\left(\Delta_v + 2 -\frac{d}{2} \right) \Gamma\left(\Delta_{v - 1} + 2 - \frac{d}{2} \right) \xi^{\Delta_v + \Delta_{v - 1}} } \\
&\ni \frac{e^{ i (\theta_1 - \theta_2)} \Gamma \left(\frac{d}{2} -1 + v \right) \Gamma \left(\frac{d}{2} - v \right)}{4^{d} \pi^d \Gamma(1 + v) \Gamma(2 - v) \xi^{d-1} } . 
\end{split}
\end{equation}
Comparing it to the form we expect,\begin{equation} \label{DisplacementPhibarPhi}
\langle \bar{\Phi} \Phi (x_1) \bar{\Phi} \Phi (x_2) \rangle \ni (C^{\bar{\Phi} \Phi}_{D})^2 C_D \frac{e^{ i (\theta_1 - \theta_2)}}{(4 \xi)^{d - 1}} \implies (C^{\bar{\Phi} \Phi}_{D})^2 C_D = \frac{\Gamma \left(\frac{d}{2} -1 + v \right) \Gamma \left(\frac{d}{2} - v \right)}{4 \pi^d \Gamma(1 + v) \Gamma(2 - v)}.
\end{equation}
Using results in \eqref{OnePointPhibarPhi}, \eqref{DisplacementCoefficientFree} and \eqref{DisplacementPhibarPhi}, it is is easy to check that the Ward identity \eqref{WardIdentity} is satisfied.  
\subsection{Mapping to $S^1 \times H^{d - 1}$} \label{SectionHyperbolicFree}
As explained in the introduction, the monodromy defect may also be studied on a hyperbolic cylinder by a Weyl transformation as in \eqref{WeylHyperbolic}. The operators also get rescaled under this Weyl transformation. The scalars, for instance, transform as $O_{S^1 \times \mathbb{H}^{d -1}} =  r^{\Delta_O} O_{\mathbb{R}^d}$. 

In order to describe a spherical defect, one may use the hyperbolic ball coordinates on $H^{d - 1}$, obtained from the Poincar\'e coordinates by the following coordinate transformation 
\begin{equation}
r = \frac{1}{\cosh \eta - \Omega_1 \sinh \eta}, \hspace{1cm} y_a = \frac{ \Omega_{a + 1} \sinh \eta}{\cosh \eta - \Omega_1 \sinh \eta} 
\end{equation}
where $(\Omega_1, \hdots, \Omega_{d - 1})$ are the coordinates on a $d - 2$ dimensional sphere with $|\Omega_a|^2 = 1$ and $0 \leq \eta < \infty$. The metric in these coordinates takes the following simple form
\begin{equation} \label{HyperbolicBallMetric}
d s_{S^1 \times \mathbb{H}^{d -1}}^2 = d \theta^2 + d \eta^2 + \sinh^2 \eta \ ds^2 _{S^{d-2}}.
\end{equation}
Note that the defect is compact and is located at the boundary of hyperbolic ball, $\eta \rightarrow \infty$, which is a $d - 2$ dimensional sphere, $S^{d - 2}$.

The complex scalar on $S^1 \times H^{d-1}$ is described by the action
\begin{equation}
\begin{split}
S &= \frac{1}{2} \int d^d x \sqrt{g(x)} \left( g^{\mu \nu} \partial_{\mu} \Phi \partial_{\nu} \bar{\Phi} + \left(\frac{(d - 2)}{4 (d - 1)} \mathcal{R} + m^2 \right) |\Phi|^2 \right)\\
&=\frac{1}{2} \int d^d x \sqrt{g(x)} \left( g^{\mu \nu} \partial_{\mu} \Phi \partial_{\nu} \bar{\Phi} -  \left( \frac{(d - 2)^2}{4} - m^2 \right) |\Phi|^2 \right)
\end{split}
\end{equation} 
with the field $\Phi$ obeying twisted boundary conditions along $S^1$, $\Phi(r, \vec{y}, \theta + 2 \pi) = e^{i \vartheta} \Phi (r, \vec{y}, \theta)$. We will be interested in the conformally coupled case with $m^2 = 0$. An equivalent description of the system can be written in terms of untwisted field $\Psi$ defined by  $\Phi (x) = e^{i v \theta} \Psi(x)$. $\Psi$ has the usual periodic boundary conditions $\Psi(r, \vec{y}, \theta + 2 \pi) = \Psi (r, \vec{y}, \theta)$. The action in terms of $\Psi$ can be written as 
\begin{equation}
S = \frac{1}{2} \int d^d x \sqrt{g(x)} \left( \bigg| \left(\partial_{\theta} + i v \right) \Psi \bigg|^2 + g^{\alpha \beta} \partial_{\alpha} \Psi \partial_{\beta} \bar{\Psi} - \frac{(d - 2)^2}{4} |\Psi|^2 \right)
\end{equation} 
where $\alpha, \beta$ are the coordinates on $H^{d - 1}$. This shows that having a monodromy defect is equivalent to having a constant background gauge field in the $\theta$ direction. Taking derivatives with $v$ is equivalent to inserting the $\theta$ component of the $U(1)$ current 
\begin{equation} \label{U1CurrentHyperbolic}
-\frac{\delta \log Z}{\delta v} =  \frac{i}{2} \int d^d x \sqrt{g(x)} \langle \left( \Phi \partial_{\theta} \bar{\Phi} - \bar{\Phi} \partial_{\theta} \Phi \right)  \rangle = \frac{1}{2} \int d^d x \sqrt{g(x)} \langle J_{
\theta} \rangle   
\end{equation}
where $Z$ is the partition function in presence of the defect.

We then perform a Kaluza-Klein (KK) reduction on $S^1$ to get a tower of massive scalar fields on $H^{d - 1}$. The bulk field can be expanded into KK modes as $\Phi (r, \vec{y}, \theta) = \sum e^{i s \theta}\Phi_s (r, \vec{y})$ where $s \in \mathbb{Z} + v$ and modes $\Phi_s (r, \vec{y})$ have mass $s^2 - (d - 2)^2/4$. Since the defect is located on the boundary of $H^{d - 1}$, we can use the standard AdS/CFT dictionary to calculate the dimensions of the defect operators of spin $s$ induced by $\Phi$ 
\begin{equation} \label{DimensionSpinHyp}
\Delta_{s} (\Delta_s - d + 2) = s^2 - (d - 2)^2/4 \implies \Delta_s = \frac{d}{2} - 1 \pm |s|.
\end{equation}
As before, we leave the discussion of $-$ sign until the next section. The two-point function on $S^1 \times H^{d - 1}$ can then be written as a sum over KK modes with the two-point function of each KK mode being just the usual bulk-bulk propagator on $H^{d - 1}$. This gives 

\begin{equation} \label{TwoPointHyperbolicFree}
\begin{split}
\langle \Phi(\vec{y}_1, r_1, \theta_1) \bar{\Phi}(\vec{y}_2, r_2, \theta_2) \rangle &= G_{\bar{\Phi} \Phi} (x_1, x_2)= \sum_{s \in \mathbb{Z} + v} \frac{2 e^{i s \theta_{12}}}{2 \pi} G^{bb}_{\Delta_s} \\
&  =  \sum_{s \in \mathbb{Z} + v} \frac{ 2 \Gamma \left(\Delta_s \right) 
e^{i s \theta_{12}} \, _2F_1\left(\Delta_s, \Delta_s  +\frac{3 - d}{2};2  \Delta_s + 3 - d ;-\frac{1}{\xi }\right) }{4 \pi^{d/2}  \Gamma ( \Delta_s + 2 -\frac{d}{2})  (4 \xi)^{\Delta_s}}.
\end{split}
\end{equation}
This is related by a Weyl transformation to the two-point function in flat space \eqref{TwoPointFunctionFlat}.

\begin{figure}
\centering
\includegraphics[scale=0.6]{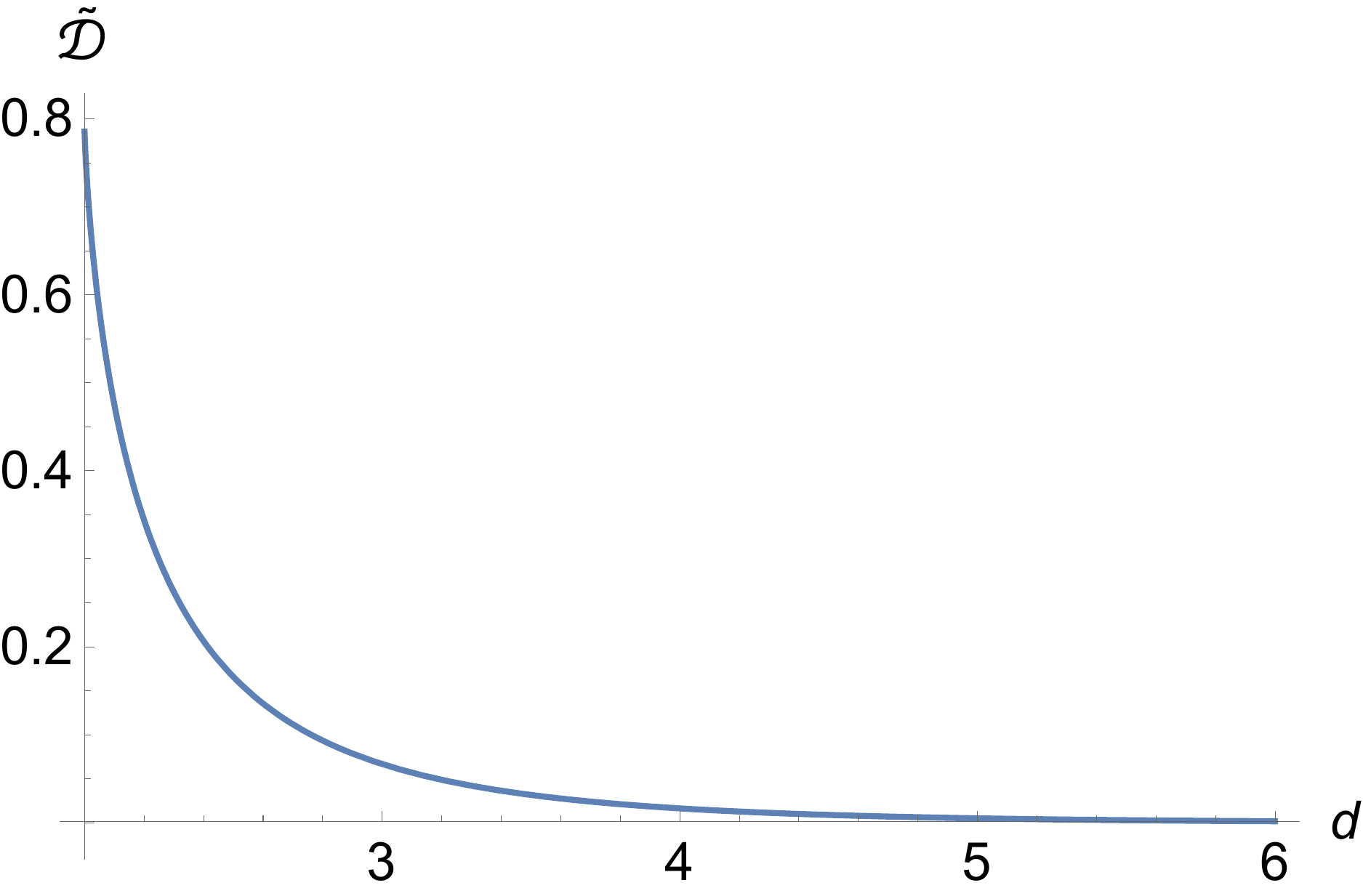}
\caption{A plot of $\tilde{\mathcal{D}}$ for a single free complex scalar between dimensions $2 < d < 6$ when $v = 1/2$.}
\label{FigureDTildeFree}
\end{figure}

A quantity of interest is the free energy on the hyperbolic space, since this is related to the expectation value of the spherical monodromy defect. In the free theory, it is given by the following determinant
\begin{equation}
F_{\textrm{twisted}} (\vartheta) = \textrm{tr} \log \left( - \nabla^2 + m^2 - \frac{(d-2)^2}{4}  \right). 
\end{equation}
The eigenfunctions of the Laplacian on $S^1 \times H^{d -1}$ may be written as $\Phi_{H^{d - 1}}(y^i, r) e^{i s \theta}$ with $\Phi_{H^{d - 1}}$ being the eigenfunction on the $d - 1$ dimensional hyperbolic space. The corresponding eigenvalues are $\lambda + (d - 2)^2/4 + s^2$ with a degeneracy given by \cite{doi:10.1063/1.530850, Bytsenko:1994bc}
\begin{equation}
    D(\lambda) d \lambda = \frac{\textrm( \textrm{Vol}(H^{d-1}))}{(4 \pi)^{\frac{d - 1}{2}} \Gamma(\frac{d - 1}{2})} \frac{|\Gamma(i \sqrt{\lambda} + \frac{d-2}{2})|^2}{|\Gamma(i \sqrt{\lambda})|^2} \frac{d \lambda}{\sqrt{\lambda}}. 
\end{equation}
Using this, we can compute the twisted free energy on the hyperbolic space as 
\begin{equation} \label{FTwistedHyperbolic}
\begin{split}
F_{\textrm{twisted}} (\vartheta) &=  \int_0^{\infty} d \lambda D(\lambda) \sum_{s \in \mathbb{Z} + v}  \log \left( \lambda + m^2 + s^2 \right)\\
& = \frac{ \textrm{Vol}(H^{d-1})}{  (4 \pi)^{\frac{d - 1}{2}} \Gamma(\frac{d - 1}{2})}  \int_{-\infty}^{\infty} d \nu \frac{|\Gamma(i \nu + \frac{d-2}{2})|^2}{|\Gamma(i \nu)|^2}  \sum_{s \in \mathbb{Z} + v} \log \left( \nu^2 + s^2 + m^2 \right).
\end{split}
\end{equation}
This can be used to calculate the expectation value of the defect, and it is natural to normalize it by the partition function of the untwisted theory. In the conformally coupled case, it gives
\begin{equation} \label{DefectExpectationHyperbolic}
\begin{split}
&-\log \langle \mathcal{D} \rangle = F_{\textrm{twisted}} - F_{\textrm{untwisted}}\\
& = \frac{ \textrm{Vol}(\mathbb{H}^{d-1})}{  (4 \pi)^{\frac{d - 1}{2}} \Gamma(\frac{d - 1}{2})}  \int_{-\infty}^{\infty} d \nu \frac{|\Gamma(i \nu + \frac{d-2}{2})|^2}{|\Gamma(i \nu)|^2} \left( \sum_{n \in \mathbb{Z} + v} - \sum_{n \in \mathbb{Z}} \right) \log \left( \nu^2 + n^2  \right)\\
& = \frac{  \textrm{Vol}(\mathbb{H}^{d-1})}{ (4 \pi)^{\frac{d - 1}{2}} \Gamma(\frac{d - 1}{2})} \int_{-\infty}^{\infty} d \nu \frac{|\Gamma(i \nu + \frac{d-2}{2})|^2}{|\Gamma(i \nu)|^2}  \log \left(\frac{1}{2} \text{csch}^2(\pi  \nu ) \left(\cosh (2 \pi  \nu )-\cos (2 \pi v )\right)\right) .
\end{split}
\end{equation}
To derive the above formula, we had to use the sum \cite{Klebanov:2011uf}
\begin{equation}
\sum_{k \in \mathbb{Z}} \log \left( (k + \alpha)^2  + a^2 \right) = \log \left( 2 \cosh (2 \pi a) - 2 \cos (2 \pi \alpha) \right). 
\end{equation}
When $d$ is even, the analytic form is easy to obtain by doing the integral over $\nu$ first in the second line of \eqref{DefectExpectationHyperbolic} and regularizing the sum by a Zeta function regularization
\begin{equation}
\log \langle \mathcal{D} \rangle = \left( \sum_{n \in \mathbb{Z} + v} - \sum_{n \in \mathbb{Z}} \right) \frac{\partial}{\partial \alpha} \left[ \frac{ \textrm{Vol}(\mathbb{H}^{d-1})}{  (4 \pi)^{\frac{d - 1}{2}} \Gamma(\frac{d - 1}{2})}  \int_{-\infty}^{\infty} d \nu \frac{|\Gamma(i \nu + \frac{d-2}{2})|^2}{|\Gamma(i \nu)|^2} \frac{1}{( \nu^2 + n^2 )^{\alpha}} \right]_{\alpha \rightarrow 0}.
\end{equation}
It gives
\begin{equation}
\begin{aligned}
-\log \langle \mathcal{D} \rangle_{d=2}&=\textrm{Vol}(\mathbb{H}^{1})\left(\zeta \left(-1,v\right)+\zeta \left(-1,1-v\right)+\frac{1}{6}\right) = v (1 - v) \textrm{Vol}(\mathbb{H}^{1})  \\
-\log \langle \mathcal{D} \rangle_{d=4}&=\textrm{Vol}(\mathbb{H}^{3})\left(\frac{-60 \zeta \left(-3,v\right)-60 \zeta \left(-3,1-v\right)+1}{360 \pi }\right) =  \frac{v^2 (1 - v)^2}{12 \pi} \textrm{Vol}(\mathbb{H}^{3})  \\
-\log \langle \mathcal{D} \rangle_{d=6}&=\textrm{Vol}(\mathbb{H}^{5})\left(\frac{\zeta \left(-5,v \right)+\zeta \left(-5,1-v\right)}{60 \pi^2} - \frac{\zeta \left(-3,v\right) + \zeta \left(-3,1-v\right)}{36 \pi^2} + \frac{1}{1680 \pi ^2}\right) \\
& = - \frac{v^2 (1 - v)^2 (-3 - v (1 - v))}{180 \pi^2} \textrm{Vol}(\mathbb{H}^{5}).
\end{aligned}
\end{equation}
Note that the factors of the hyperbolic space volume here are logarithmically divergent \cite{Diaz:2007an, Casini:2011kv}, see eq.~(\ref{HyperbolicVolume}). The quantity $\tilde{\mathcal D}$ defined in (\ref{DtildeDefinition}) is however finite and it is proportional to the quantities multiplying the volume factors above. Indeed using the above result \eqref{DefectExpectationHyperbolic}, $\tilde{\mathcal{D}}$ can be seen to be a smooth and finite function of $d$. We plot it for $2<d<6$ and in the special case of $Z_2$ monodromy, $v=1/2$, in figure \ref{FigureDTildeFree}.
For future reference, let us also list some explicit values of $\tilde{\mathcal{D}}$ in various $d$ for $v=1/2$ which can be directly obtained using \eqref{DefectExpectationHyperbolic}
\begin{equation}
\begin{aligned}
&\tilde{\mathcal{D}}|_{d=3}= \frac{\log 2}{4} - \frac{7 \zeta(3)}{8 \pi^2} \,,\qquad 
\tilde{\mathcal{D}}|_{d=4}= \frac{\pi}{192} \,,\qquad \\
&\tilde{\mathcal{D}}|_{d=5}= \frac{\log 2}{64} - \frac{5 \zeta(3)}{192 \pi^2} - \frac{31 \zeta(5)}{128 \pi^4}  \,,\qquad 
\tilde{\mathcal{D}}|_{d=6}= \frac{13 \pi}{23040}. 
\end{aligned}
\end{equation}
To do the above calculation and obtain the plot, we had to use the regularized volume of the hyperbolic space \cite{Diaz:2007an, Casini:2011kv}
\begin{equation} \label{HyperbolicVolume}
\textrm{Vol}(H^{d-1}) = \pi^{\frac{d}{2} - 1} \Gamma \left(1 - \frac{d}{2} \right).
\end{equation}
For odd values of $d$ and generic $v$, it is more convenient to use the sphere geometry to obtain analytic results, as we show in \eqref{DefectExpectation3d} for $d = 3$. Using the round sphere setup below, we will also obtain an expression for $\log\langle \mathcal{D}\rangle$ valid in continuous $d$. 
 
\subsubsection{One-point functions}
One-point functions of bulk operators can also be readily obtained in the hyperbolic space. For the scalar $\bar{\Phi} \Phi$, it is a constant given by 
\begin{equation} \label{OnePointHyperboliFree}
C_1^{\bar{\Phi} \Phi} = \frac{1}{\pi \textrm{Vol} (\mathbb{H}^{d - 1})} \frac{ \partial F_{\textrm{twisted}} (\vartheta)}{\partial m^2} \bigg|_{m^2 = 0}
\end{equation} 
The mass derivative of the free energy can be calculated as follows
\begin{equation} \label{SpectralIntegralOnePoint}
\begin{split}
\frac{ \partial F_{\textrm{twisted}} (\vartheta)}{\partial m^2} &= \sum_{k = - \infty}^{\infty} \frac{ \textrm{Vol}(H^{d-1})}{  (4 \pi)^{\frac{d - 1}{2}} \Gamma \left(\frac{d - 1}{2}\right)}  \int_{-\infty}^{\infty} d \nu \frac{|\Gamma(i \nu + \frac{d-2}{2})|^2}{|\Gamma(i \nu)|^2} \left( \frac{1}{\nu^2 +  \left( k + v \right)^2 + m^2} \right)\\
& = \frac{ \textrm{Vol}(H^{d-1}) \Gamma \left(\frac{3 - d}{2}\right) }{  (4 \pi)^{\frac{d - 1}{2}} }  \sum_{k = - \infty}^{\infty}  \frac{ \Gamma \left(\frac{d}{2} - 1 + \sqrt{m^2 + \left( k + v\right)^2 }\right)}{ \Gamma \left(2-\frac{d}{2} + \sqrt{m^2 + \left( k + v \right)^2 }\right)} \\
&= \frac{ \textrm{Vol}(H^{d-1}) \Gamma \left(\frac{3 - d}{2}\right) }{  (4 \pi)^{\frac{d - 1}{2}} }  \sum_{s} \frac{ \Gamma \left(\Delta_s \right)}{ \Gamma \left(3-d + \Delta_s \right)}.
\end{split}
\end{equation}
To perform the integral, we had to close the contour in the $\nu$ plane and sum over residues \cite{Carmi:2018qzm}. The arc at infinity can only be dropped for $d < 3$, but the final result can be analytically continued in dimensional regularization. One of the Gamma function introduces poles at $\nu = i (d/2 - 1 + 2 \kappa)$ for integer $\kappa$, which all lie in the upper half plane for $d > 2$ and need to be summed to get the final result. For $m \rightarrow 0$, we get 
\begin{equation} \label{OnePointHyperboliFreeRes}
\begin{split}
\frac{ \partial F_{\textrm{twisted}} (\vartheta)}{\partial m^2} \bigg|_{m^2 = 0} &= \frac{ \textrm{Vol}(H^{d-1}) \Gamma \left(\frac{3 - d}{2}\right) }{ (4 \pi)^{\frac{d - 1}{2}} }  \sum_{k = - \infty}^{\infty}  \frac{ \Gamma \left(\frac{d}{2} - 1 + | k + v| \right)}{ \Gamma \left(2-\frac{d}{2} + | k + v|\right)}  \\
& = \frac{ (d - 1) \textrm{Vol}(H^{d-1})  \Gamma \left( \frac{d}{2}  - v \right) \Gamma \left( \frac{d}{2} - 1 + v \right) \sin \left( v \pi \right) }{  (4 \pi)^{\frac{d - 1}{2}} (2 - d) \Gamma \left(\frac{d + 1}{2} \right)} \\
\implies C_1^{\bar{\Phi} \Phi} &= \frac{ (d - 1) \Gamma \left( \frac{d}{2}  - v \right) \Gamma \left( \frac{d}{2} - 1 + v \right) \sin \left( v \pi \right) \Gamma \left( \frac{d}{2} \right)}{\pi^{\frac{d}{2} + 1} \Gamma(d) (2 - d)}. 
\end{split}
\end{equation}
This of course agrees with the flat space result \eqref{OnePointPhibarPhi}. For the spin one $U(1)$ current, using its general form in \eqref{CurrentGeneral}, it is easy to see that the one-point function of its $\theta$ component is a constant on hyperbolic cylinder. It may be calculated as in \eqref{U1CurrentHyperbolic}, by taking the derivative of free energy with $v$
\begin{equation}
\begin{split}
\langle J_{\theta} \rangle &= \frac{1}{\pi \textrm{Vol}(\mathbb{H}^{d-1})} \frac{\partial F_{\textrm{twisted}} (\vartheta)}{\partial v} = \sum_{k = - \infty}^{\infty} \frac{2^{2 - d} \pi^{-\frac{(d+1)}{2}}}{ \Gamma \left(\frac{d - 1}{2}\right)}  \int_{-\infty}^{\infty} d \nu \frac{|\Gamma(i \nu + \frac{d-2}{2})|^2}{|\Gamma(i \nu)|^2} \left( \frac{k + v}{\nu^2 +  \left( k + v \right)^2 } \right) \\
&= 2^{2 - d} \pi^{-\frac{(d+1)}{2}} \Gamma \left(\frac{3-d}{2}\right) \sum_{k = - \infty}^{\infty} \frac{ (k + v) \Gamma \left(\frac{d}{2} - 1 + | k + v| \right)}{ \Gamma \left(2-\frac{d}{2} + | k + v|\right)} \\
&= \frac{(d-2) (2 v-1)}{d-1} C_1^{\bar{\Phi} \Phi},
\end{split}
\end{equation}
which again agrees with the flat space result in \eqref{CurrentOnePointTheta}. 

Similarly, for the stress tensor, the general form \eqref{StressTensorGeneral} tells us that $T_{\theta \theta}$ should have a constant one-point function on the hyperbolic cylinder. There is a simpler way to calculate it on the hyperbolic cylinder \cite{Hung:2014npa, Lewkowycz:2013laa}. We start by keeping the length of $S^1$ to be a variable $\beta$ instead of fixing it to $2 \pi$. This is equivalent to rescaling the metric component  $g_{\theta \theta}$ by $( \beta / 2 \pi)^2$. So if we compute the free energy for arbitrary $\beta$ and then take a derivative with respect to $\beta$, this is the same as inserting $T_{\theta \theta}$ in the path integral 
\begin{equation} \label{ConformalWeightFreeEnergy}
T_{\mu \nu} = \frac{2}{\sqrt{g}} \frac{\delta S}{\delta g^{\mu \nu}} \implies \langle T_{\theta \theta} \rangle_{S^1 \times H^{d - 1}} = - \frac{1}{\textrm{Vol}(H^{d-1})} \frac{\partial F_{\textrm{twisted}} (\vartheta , \beta)}{\partial \beta} \bigg|_{\beta = 2 \pi}.
\end{equation}
In practice, we can calculate the free energy for a general $\beta$ by rescaling $n$ by $2 \pi / \beta$
\begin{equation}
F_{\textrm{twisted}} (\vartheta , \beta) = \frac{ \textrm{Vol}(H^{d-1})}{  (4 \pi)^{\frac{d - 1}{2}} \Gamma(\frac{d - 1}{2})}  \int_{-\infty}^{\infty} d \nu \frac{|\Gamma(i \nu + \frac{d-2}{2})|^2}{|\Gamma(i \nu)|^2}  \sum_{n \in \mathbb{Z} + v}  \log \left( \nu^2 + \frac{4 \pi^2 n^2}{\beta^2} \right).
\end{equation}
We can use this to calculate the stress-tensor one-point function 
\begin{equation}
\begin{split}
\langle T_{\theta \theta} \rangle_{S^1 \times H^{d - 1}}  &= \sum_{k = - \infty}^{\infty} \frac{1}{ \pi (4 \pi)^{\frac{d - 1}{2}} \Gamma \left(\frac{d - 1}{2}\right)}  \int_{-\infty}^{\infty} d \nu \frac{|\Gamma(i \nu + \frac{d-2}{2})|^2}{|\Gamma(i \nu)|^2}  \frac{\left( k + v \right)^2}{\nu^2 +  \left( k + v \right)^2 } \\
& = \frac{  \Gamma \left(\frac{3 - d}{2}\right) }{ \pi (4 \pi)^{\frac{d - 1}{2}} }  \sum_{k = - \infty}^{\infty}  \frac{ \Gamma \left(\frac{d}{2} - 1 + | k + v| \right) \left( k + v \right)^2}{ \Gamma \left(2-\frac{d}{2} + | k + v|\right)}  \\
& =  \frac{  \Gamma \left(\frac{3 - d}{2}\right)(1 -v ) v  \left(\csc \pi \left(\frac{d}{2} - v \right)-\csc \pi \left(\frac{d}{2} + v \right) \right) }{ d   (4 \pi)^{\frac{d - 1}{2}}  \Gamma \left( 2 - \frac{d}{2}- v \right) \Gamma \left(1 - \frac{d}{2}+ v \right)}
\end{split}
\end{equation}
which is consistent with what we got above by a direct calculation in flat space in \eqref{StressTensorThetaDirectResult}. This hyperbolic space technique will be useful below, when we try to calculate the conformal weight in the interacting theory. 

\subsection{Twisted free energy on $S^d$}
Another useful way to study a codimension two defect is to map the problem to a $d$-dimensional sphere $S^d$ which is related by a Weyl transformation to flat space. Indeed starting from the hyperbolic cylinder described by \eqref{HyperbolicBallMetric}, we perform a coordinate transformation $\sinh \eta = \tan \tau$ to get 
\begin{equation} \label{MetricSphereTwist}
ds^2_{S^1 \times H^{d - 1}} =\frac{1}{\cos^2 \tau} ( \cos^2 \tau d \theta^2 + d \tau^2 + \sin^2 \tau ds^2_{S^{d - 2}}) = \frac{1}{\cos^2 \tau} ds^2_{S^d}. 
\end{equation}
The sphere is spanned by $0 \leq \theta < 2 \pi$, $0 \leq \tau < \pi/2$ and the usual $S^{d - 2}$ coordinates. The defect is located at $\tau = \pi/2$, and the monodromy action along $\theta$ is the same as in the hyperbolic cylinder, $\Phi(\tau, \theta + 2 \pi) = e^{i \vartheta} \Phi(\tau, \theta)$, where we suppressed the $S^{d - 2}$ coordinates. These coordinates are related to the usual $S^d$ coordinates by the transformation $\sin \tau = \sin \theta_1 \sin \theta_2, \ \cos \tau \sin \theta = \cos \theta_1$ which gives us the usual $S^d$ metric 
\begin{equation}
ds^2_{S^d} = d \theta_1^2 + \sin^2 \theta_1 (d \theta_2^2 + \sin^2 \theta_2 ds^2_{S^{d - 2}})
\end{equation}
where $0 \leq \theta_{1,2} < \pi$. 

The twisted free energy of a complex scalar on $d-$ dimensional unit sphere is 
\begin{equation}
F_{\textrm{twisted}} (\vartheta) =\textrm{tr} \log \left( - \nabla^2  + \frac{d(d-2)}{4}  \right)
\end{equation}
where the trace is taken over the eigenfunctions obeying twisted boundary condition. We calculate their eigenvalues and degeneracy in appendix \ref{AppendixLaplacianSpectrum}.  Using the result in \eqref{DegenraciesSphere}, the twisted free energy in $d = 3$ is given by
\begin{equation}
\begin{split}
F_{\textrm{twisted}} (\vartheta) &= \left( \sum_{n = k + v} + \sum_{n = k + 1 - v} \right) d_n \log \left( n (n + d - 1) + \frac{d (d - 2)}{4} \right) \\
&= \sum_{k = 0}^{\infty} \frac{(k + 1)(k + 2)}{2}  \log \left( \left(k + v\right) \left(k + v + 2\right) + \frac{3}{4} \right) + v \rightarrow 1 - v \\
& =  \sum_{k = 1}^{\infty} k^2 \left[  \log \left( k + v - \frac{1}{2} \right) \left( k - v + \frac{1}{2} \right)  \right].
\end{split}
\end{equation}
The sum can be regulated using zeta-function regularization 
\begin{equation}
\begin{split}
&F_{\textrm{twisted}} (\vartheta) =
- \frac{d}{d s} \left[  \sum_{k = 1}^{\infty} \frac{k^2}{ \left( k + v - \frac{1}{2} \right)^s}  + \sum_{k = 1}^{\infty} \frac{k^2}{ \left( k - v + \frac{1}{2} \right)^s} \right] \\
& = -\zeta' \left(- 2, v + \frac{1}{2} \right) + 2 \left(v - \frac{1}{2} \right) \zeta' \left(- 1, v + \frac{1}{2} \right) - 
\left(v - \frac{1}{2} \right)^2 \zeta' \left(0, v + \frac{1}{2} 
\right) + v \rightarrow 1 - v.
\end{split}
\end{equation}
Using identities for derivatives of Hurwitz zeta function (see for e.g. \cite{Klebanov:2011uf}), we may write the result as 
\begin{equation}
\begin{split}
&F_{\textrm{twisted}} (\vartheta) = \sum_{n = 1}^{\infty} (-1)^n \left( \frac{\cos n \vartheta}{2 \pi^2 n^3} + \frac{(\vartheta - \pi)}{2 \pi^2 } \frac{\sin n \vartheta}{n^2} - \frac{(\vartheta - \pi)^2}{4 \pi^2} \frac{\cos n \vartheta}{n}  \right) \\
& = \frac{(\pi -\vartheta )^2 \log \left(2 \cos \vartheta/2 \right)- i (\pi -\vartheta) \left(\text{Li}_2\left(-e^{-i \vartheta }\right)-\text{Li}_2\left(-e^{i \vartheta }\right)\right)+ \text{Li}_3\left(-e^{-i \vartheta }\right)+ \text{Li}_3\left(-e^{i \vartheta }\right)}{4 \pi ^2}
\end{split}
\end{equation}
where $Li_s(z)$ is the Polylog. This gives us the expectation value of the defect in $3$ dimensions 
\begin{equation} \label{DefectExpectation3d}
\begin{split}
- \log \langle \mathcal{D} \rangle_{d = 3} &= \frac{(\pi -\vartheta )^2 \log \left(2 \cos \vartheta/2 \right)- i (\pi -\vartheta) \left(\text{Li}_2\left(-e^{-i \vartheta }\right)-\text{Li}_2\left(-e^{i \vartheta }\right)\right) }{4 \pi ^2} \\
& + \frac{\text{Li}_3\left(-e^{-i \vartheta }\right)+ \text{Li}_3\left(-e^{i \vartheta }\right)}{4 \pi ^2} - \frac{\log 2}{4} + \frac{3 \zeta(3)}{8 \pi^2}  
\end{split}
\end{equation}
where we used the result for the untwisted free energy of a complex scalar in three dimensions from \cite{Giombi:2014xxa}. It can be numerically checked to agree with the result obtained from hyperbolic space calculation in \eqref{DefectExpectationHyperbolic}. For general $d$, we get
\begin{equation}
\begin{split}
F_{\textrm{twisted}} (\vartheta) &= \sum_{k = 0}^{\infty} \frac{ \Gamma(k + d)}{\Gamma(k + 1) \Gamma(d)} \log \left( \left(k + v \right)  \left(k + v + d - 1 \right) + \frac{d(d - 2)}{4} \right) + (v \rightarrow 1 -v) \\
& = \sum_{k = 0}^{\infty} \frac{ \Gamma(k + d)}{\Gamma(k + 1) \Gamma(d)} \log \left(k + v + \frac{d}{2} \right) \left(k + v + \frac{d}{2} - 1 \right) + (v \rightarrow 1 - v) \\
&= \sum_{k = 0}^{\infty} \frac{ \Gamma(k + d)}{\Gamma(k + 1) \Gamma(d)} \log \frac{\Gamma \left(k + v + d - \delta \right)}{ \Gamma \left(k + v + \delta \right)} + (v \rightarrow 1 - v), \hspace{0.5cm} \delta = \frac{d}{2} - 1
\end{split}
\end{equation}
where we introduced $\delta$. To do the sum, it is more convenient to treat $\delta$ as a new variable, and perform the sum after taking a derivative with respect to $\delta$ \cite{Diaz:2007an, Giombi:2014xxa}
\begin{equation}
\begin{split}
\frac{\partial F_{\textrm{twisted}} (\vartheta, \delta)}{\partial \delta} & = - \sum_{k = 0}^{\infty} \frac{ \Gamma(k + d)}{\Gamma(k + 1) \Gamma(d)} \left(  \psi \left(k + v + d - \delta \right) + \psi \left(k + v + \delta \right) \right) + (v \rightarrow 1 - v)  \\ 
&=  \Gamma(-d) \left(\frac{\Gamma \left( v + d - \delta \right)}{\Gamma \left( v  - \delta \right) } + \frac{\Gamma \left( v + \delta \right)}{\Gamma \left( v + \delta - d \right) } \right) + (v \rightarrow 1 - v)
\end{split}
\end{equation}
which implies that 

\begin{equation}
\begin{split}
F_{\textrm{twisted}} (\vartheta) &=  \Gamma(-d) \int_{d/2}^{d/2 - 1} d \delta \left(\frac{\Gamma \left( v + d - \delta \right)}{\Gamma \left( v - \delta \right) } + \frac{\Gamma \left( v + \delta \right)}{\Gamma \left( v  + \delta - d \right) } + (v \rightarrow 1 - v) \right) \\
& = -  \Gamma(-d) \int_0^1 d w \left(\frac{\Gamma \left( v + \frac{d}{2} + w \right)}{\Gamma \left( v - \frac{d}{2} + w \right)} + \frac{\Gamma \left( v + \frac{d}{2} - w \right)}{\Gamma \left( v - \frac{d}{2} - w \right)} + (v \rightarrow 1 - v) \right). 
\end{split}
\end{equation}
We can use this to calculate the defect expectation value 
\begin{equation} \label{DefectExpectationSphere}
\begin{split}
-\log \langle \mathcal{D} \rangle &= F_{\textrm{twisted}} - F_{\textrm{untwisted}} \\
& =  -  \int_0^1 d w \bigg[ \Gamma(-d)\left(\frac{\Gamma \left( v + \frac{d}{2} + w \right)}{\Gamma \left( v - \frac{d}{2} + w \right)} + \frac{\Gamma \left( v + \frac{d}{2} - w \right)}{\Gamma \left( v - \frac{d}{2} - w \right)} + (v \rightarrow 1 - v) \right)  \\
& - \frac{ 2 \Gamma \left(\frac{d}{2} \right) \Gamma \left(1 - \frac{d}{2} \right)}{\pi \Gamma(1 + d)} w \sin (\pi w) \Gamma \left( \frac{d}{2} + w \right) \Gamma \left( \frac{d}{2} - w \right) \bigg]
\end{split}
\end{equation}
In $d = 3$, it agrees with the result in \eqref{DefectExpectation3d}. It can also be checked numerically, to agree with the result obtained on hyperbolic cylinder in \eqref{DefectExpectationHyperbolic}. 

Similar to what we did on the hyperbolic cylinder, we can also calculate the one-point function of $\bar{\Phi}\Phi$ on the sphere by introducing a mass term and calculating the derivative of the free energy with mass
\begin{equation}
\begin{split}
&\frac{ \partial F_{\textrm{twisted}} (\vartheta)}{\partial m^2} = \sum_{k = 0}^{\infty} \frac{ \Gamma(k + d)}{\Gamma(k + 1) \Gamma(d)} \left[\frac{1}{ \left(k + v + \frac{d - 1}{2} \right)^2 + m^2 - \frac{1}{4} } + (v \rightarrow 1 - v) \right] \\
& = \frac{\Gamma\left( \frac{d + 1 - 2 v}{2} + \sqrt{\frac{1}{4} - m^2} \right)\Gamma\left( \frac{d - 1 + 2 v}{2} - \sqrt{\frac{1}{4} - m^2} \right) \sin \pi \left(  \sqrt{\frac{1}{4} - m^2} + \frac{1}{2} -v\right) }{2 \Gamma(d) \sin \left( \frac{\pi d}{2} \right) \sqrt{\frac{1}{4} - m^2}} + (v \rightarrow 1 - v). 
\end{split}
\end{equation}
The Weyl factor involved in going from hyperbolic cylinder to $S^d$ is $\cos \tau$, so the one-point function on $S^d$ is $C_1^{\bar{\Phi} \Phi}/(\cos \tau)^{d - 2}$. Hence, 

\begin{equation}
\begin{split}
    \frac{\partial F}{\partial m^2} \bigg|_{m^2 = 0} &=  C_1^{\bar{\Phi} \Phi}  \pi \textrm{Vol} (S^{d-2}) \int_0^{\pi/2} d \tau (\sin \tau)^{d-2} (\cos \tau)^{-d+3} \\
  \implies    C_1^{\bar{\Phi} \Phi} &= \frac{ \Gamma \left( \frac{d}{2} + 1 - v \right) \Gamma \left( \frac{d}{2} - 1 + v\right)  \sin \left( \pi v \right) \Gamma \left( \frac{d}{2} \right)}{\pi^{\frac{d}{2} + 1} \Gamma(d) (2 - d)} + (v \rightarrow 1 - v)  \\
    & = \frac{ (d - 1) \Gamma \left( \frac{d}{2}  - v \right) \Gamma \left( \frac{d}{2} - 1 + v \right) \sin \left( \pi v \right) \Gamma \left( \frac{d}{2} \right)}{\pi^{\frac{d}{2} + 1} \Gamma(d) (2 - d)}
\end{split}
\end{equation}
in agreement with the flat space \eqref{OnePointPhibarPhi} and hyperbolic space results \eqref{OnePointHyperboliFreeRes}.

\section{Alternate boundary condition on $H^{d-1}$ and a defect RG flow} \label{SectionDefectRGFlow}
As we discussed in the previous section in \eqref{DimensionSpinFlat} and \eqref{DimensionSpinHyp}, dimensions of the defect operators are related to their spin by the relation
\begin{equation}
\Delta_{s} (\Delta_s - d + 2) = s^2 - (d - 2)^2/4 \implies \Delta^{\pm}_s = \frac{d}{2} - 1 \pm |s|.
\end{equation}
So far, we considered the $+$ sign above for all values of $s \in \mathbb{Z} + v$. But the $-$ sign may also be consistent with the defect unitarity bound \eqref{DefectUnitarity} as long as $|s| < 1$ which happens when $s = v$ or $-1 + v$ \footnote{In most of this paper, we use $+$ boundary condition for all the modes, so we avoid using the superscript $+$. We use superscript $-$ whenever we impose a $-$ boundary condition.}\textsuperscript{,}\footnote{The idea that imposing alternate boundary condition for low lying Kaluza-Klein modes leads to non-trivial defects was also discussed in \cite{Nishioka:2021uef, Lauria:2020emq}.}. In the free theory, the two cases are allowed by unitarity in the following range of $v$ 
\begin{equation} \label{UnitarityFlow}
\begin{split}
\Delta_v^{-} \ \ \textrm{allowed}: v < \frac{d}{2} - 1, \hspace{1cm } \Delta^{-}_{-1 + v}  \ \ \textrm{allowed}: v > 2 - \frac{d}{2}.
\end{split}
\end{equation}
Note that there is some range of dimensions between $3$ and $4$ where both $s = v$ and $-1 + v$ modes are allowed to have a $\Delta^-$ boundary condition. In $d = 3$ , $s = v$ mode is allowed to have a $\Delta^-$ boundary condition only for $v<1/2$, while $s = -1 + v$ mode is allowed to have a $\Delta^-$ boundary condition only for  $v>1/2$. In the hyperbolic space setup, we can think of the two possible values of $\Delta$ as the two possible boundary conditions for the bulk massive scalar after we have performed KK reduction on $S^1$. In the defect theory, it defines a defect RG flow where we start with a $\Delta^-_{v}$ operator in the UV and flow with a relevant deformation $\bar{\Psi}_{v} \Psi_{v}$ to the IR where the original operator is replaced by an operator with dimension $\Delta^+_{v}$. This is similar to the study of double-trace flows in AdS/CFT \cite{Witten:2001ua}. We can define a similar flow with $\bar{\Psi}_{-1 + v} \Psi_{-1 + v}$ operator under which the dimension of $s = -1 + v$ operator changes from $\Delta^-_{-1 + v}$ to $\Delta^+_{-1 + v}$. Both of these flows preserve the reality of the action and the $U(1) \sim SO(2)$ internal symmetry of the original action. As we mentioned in the previous section, at $v = 1/2$, the theory has an enhanced $O(2)$ symmetry, and to preserve the full $O(2)$ symmetry, we have to add a sum of these two deformations. We now study how some of the defect CFT observables change under such a flow. We only do the calculation for the $\bar{\Psi}_{v} \Psi_{v}$ flow so only $s = v$ operator has a $\Delta^-$ boundary condition. The generalization to the case when either just $s = -1 + v$ or both $s = v$ and $s = -1 + v$ operators have a $\Delta^-$ boundary condition is straightforward. Another interesting scenario is when the defect spectrum contains both $\Delta^+$ and $\Delta^-$ operators for the same $s$. This defines a more non-trivial defect CFT with extra interactions localized on the defect. Some remarks on this possibility were made in \cite{Lauria:2020emq}. We will not study this scenario here. 

First, let us calculate how the defect expectation value \eqref{DefectExpectationHyperbolic} changes under this flow. The change is equal to the change in twisted free energy. As we said above, this defect RG flow is analogous to a double trace flow in the $d-2$ dimensional defect CFT driven by the square of a primary operator. The general result for the change in free energy under such a flow is known \cite{Giombi:2014xxa, Diaz:2007an} 
\begin{equation}
F_{d-2 - \Delta} - F_{\Delta}  = - \frac{2}{\sin \left( \frac{\pi (d-2)}{2} \right) \Gamma(d-1)} \int_0^{\Delta + 1 - \frac{d}{2}} du \ u \sin \pi u \Gamma \left(\frac{d}{2} - 1 + u \right) \Gamma \left(\frac{d}{2} - 1 - u \right)  
\end{equation}
where the operator driving the flow has dimension $\Delta$ in the UV and $d-2-\Delta$ in the IR. In our case, we consider the flow by operator $\bar{\Psi}_{v} \Psi_{v}$, so the change in the defect expectation value is given by
\begin{equation}
\log \langle \mathcal{D} \rangle^{-} - \log \langle \mathcal{D} \rangle^{+} = \frac{2}{\sin \left( \frac{\pi (d-2)}{2} \right) \Gamma(d-1)} \int_0^{v} du \ u \sin \pi u \Gamma \left(\frac{d}{2} - 1 + u \right) \Gamma \left(\frac{d}{2} - 1 - u \right).
\end{equation}
Here and elsewhere the superscript $-$ corresponds to the case when $s = v$ operator has a $\Delta^-$ boundary condition while all the other operators have a $\Delta^+$ boundary condition. As we mentioned in the introduction, the quantity $\tilde{\mathcal{D}}$ defined in \eqref{DtildeDefinition} should decrease under a defect RG flow. So we expect that 
\begin{equation} \label{DifferenceDTilde}
\tilde{\mathcal{D}}^- - \tilde{\mathcal{D}}^+ = \frac{2}{ \Gamma(d-1)} \int_0^{v} du \ u \sin \pi u \Gamma \left(\frac{d}{2} - 1 + u \right) \Gamma \left(\frac{d}{2} - 1 - u \right)
\end{equation}
should always be positive. It can be checked numerically for various values of $v$ and $d$ that this is always positive whenever the flow is allowed by the unitarity \eqref{UnitarityFlow}. But this is not true when the flow is non-unitary, so for instance, in $d = 3$, \eqref{DifferenceDTilde} is not positive for $v > 1/2$.    

The bulk two-point function also changes under the defect RG flow. We have the defect channel conformal block decomposition in \eqref{TwoPointHyperbolicFree}. The change only comes from $s = v$ piece and is given by
\begin{equation}
\begin{split}
&G^{-}_{\bar{\Phi} \Phi} - G^{+}_{\bar{\Phi} \Phi}  = \frac{e^{i v (\theta_1-\theta_2)}}{4 \pi^{d/2}} \left( \frac{ 2 \Gamma \left(\Delta^{-}_v \right) 
 \, _2F_1\left(\Delta^-_v, \Delta^-_v  +\frac{3 - d}{2};2  \Delta^-_v + 3 - d ;-\frac{1}{\xi }\right) }{ 4^{\Delta^-_v} \Gamma ( \Delta^-_v + 2 -\frac{d}{2})  \xi^{\Delta^-_v}} - \Delta^-_v \rightarrow \Delta^+_v \right) \\
 &= \frac{ e^{i v (\theta_1-\theta_2)} 2^{1-d} \Gamma \left(\frac{3}{2}-\frac{d}{2}\right) \left(\csc \left(\frac{1}{2} \pi  (d+2 v)\right)-\csc \left(\frac{1}{2} \pi  (d-2 v)\right)\right)}{ \pi^{\frac{d - 1}{2}} \Gamma \left(-\frac{d}{2}-v+2\right) \Gamma \left(-\frac{d}{2}+v+2\right)} \left(1 + O(\xi) \right)
\end{split}
\end{equation}
where we expanded the result in small $\xi$. The constant $\xi$ independent piece tells us the change in the one-point function coefficient of $\bar{\Phi} \Phi$
\begin{equation} \label{OnePointPhibarPhiMinus}
\begin{split}
&(C_{1}^{\bar{\Phi} \Phi})^{-} -(C_{1}^{\bar{\Phi} \Phi})^{+} =  \frac{  \Gamma \left(\frac{3}{2}-\frac{d}{2}\right) \left(\csc \left(\frac{1}{2} \pi  (d+2 v)\right)-\csc \left(\frac{1}{2} \pi  (d-2 v)\right)\right)}{ 2^{d-1} \pi^{\frac{d - 1}{2}} \Gamma \left(-\frac{d}{2}-v+2\right) \Gamma \left(-\frac{d}{2}+v+2\right)} \\
&\implies (C_{1}^{\bar{\Phi} \Phi})^{-} = \frac{\Gamma \left(\frac{d}{2}-v-1\right) \Gamma \left(\frac{d}{2}+v\right)}{2^{d-2} \pi ^{\frac{d-1}{2}} (d - 2) \Gamma \left(\frac{d-1}{2}\right) \Gamma (v) \Gamma (1-v)}
\end{split}
\end{equation}
where we used the result from \eqref{OnePointPhibarPhi}.

Using the change in the bulk two-point function, it is possible to also calculate the bulk one-point function of other operators. Let's just mention the result for the stress tensor. Its one-point function is specified by the conformal weight \eqref{StressTensorGeneral} and the change in conformal weight under the flow is as follows: 
\begin{equation}
\begin{split}
h^{-} - h^{+} &= -\frac{ v^2 \Gamma \left(\frac{1}{2}-\frac{d}{2}\right) \left(\csc \left(\frac{1}{2} \pi  (d+2 v)\right)-\csc \left(\frac{1}{2} \pi  (d-2 v)\right)\right)}{ 2^{d-1} \pi^{\frac{d - 3}{2}} \Gamma \left(-\frac{d}{2}-v+2\right) \Gamma \left(-\frac{d}{2}+v+2\right)} \\
h^{-} &= -\frac{ v (v+1) \Gamma \left(\frac{1}{2}-\frac{d}{2}\right) \left(\csc \left(\frac{1}{2} \pi  (d-2 v)\right)-\csc \left(\pi  \left(\frac{d}{2}+v\right)\right)\right)}{ 2^{d-1} \pi^{\frac{d - 3}{2}} d \Gamma \left(-\frac{d}{2}-v+1\right) \Gamma \left(-\frac{d}{2}+v+2\right)}.
\end{split}
\end{equation}
As long as the flow is unitary, the conformal weight $h^-$ can be checked to be positive, consistent with the conjecture made in \cite{Lemos:2017vnx}.

Finally, let's discuss the displacement operator in the theory with $\Delta_v^-$ boundary condition. Based on the fact that it must have dimension $d - 2$ and spin $1$, it must be now proportional to $\bar{\Psi}_{v} \Psi_{1 + v}$. To more directly calculate it similar to what we did in for the $\Delta^+$ case in section \ref{SectionFreeTheory},  note that the bulk defect OPE contains 
\begin{equation}
\Phi \ni z^{-v} \Psi_{v} + \bar{z}^{1 + v} \Psi_{1 + v}.
\end{equation}
Then, using the definition of stress -tensor in the free theory, it is easy to see that 
\begin{equation} \label{DisplacementDefinitionFreeMinus}
\begin{split}
&T^{z z} = 4 T_{\bar{z} \bar{z}} \ni -\frac{4 v (1 + v)}{\bar{z}} \bar{\Psi}_{v} \Psi_{1 + v} (\vec{y}) \implies \partial_z T^{z z} = -8 \pi v (1+v)  \bar{\Psi}_{v} \Psi_{1 + v} (\vec{y}) \delta^2 (z, \bar{z}) \\
& \implies D(\vec{y}) = -4 \pi v (1+v)  \bar{\Psi}_{v} \Psi_{1 + v} (\vec{y}).
\end{split}
\end{equation}
This definition gives the following two-point function coefficient

\begin{equation} \label{DisplacementCoeffMinus}
C_D^- = \frac{4 v (1 + v) \Gamma \left(\frac{d}{2} -1 - v \right) \Gamma \left(\frac{d}{2} + v \right)}{\pi^{d - 2} \Gamma(1-v) \Gamma(v)}.
\end{equation}
We can also check that these coefficients satisfy the ward identity \eqref{WardIdentity} by noting that the displacement contribution to the two-point function of $\bar{\Phi} \Phi$ now takes the form 
\begin{equation} 
\langle \bar{\Phi} \Phi (x_1) \bar{\Phi} \Phi (x_2) \rangle \ni \frac{e^{ i (\theta_1 - \theta_2)} \Gamma\left(\Delta^-_v \right)\Gamma\left(\Delta_{1 + v} \right)}{4^{1 + \Delta^-_v + \Delta_{1 + v}  } \pi^d \Gamma\left(\Delta^-_v + 2 -\frac{d}{2} \right) \Gamma\left(\Delta_{1 + } + 2 - \frac{d}{2} \right) \xi^{d - 1} }. 
\end{equation}
This gives a relation between bulk defect OPE coefficient for $\bar{\Phi} \Phi$ with displacement and the displacement two-point function coefficient
\begin{equation} \label{DisplacementPhiPhiMinus}
(C^{\bar{\Phi} \Phi - }_{D})^2 C^-_D = \frac{\Gamma \left(\frac{d}{2} -1 - v \right) \Gamma \left(\frac{d}{2} + v \right)}{4 \pi^d \Gamma(1 - v) \Gamma(2 + v)}.
\end{equation}
The coefficients \eqref{OnePointPhibarPhiMinus}, \eqref{DisplacementCoeffMinus} and \eqref{DisplacementPhiPhiMinus} satisfy the Ward identity \eqref{WardIdentity}.

\begin{comment}
which gives the change under the RG flow 
\begin{equation}
C_D^- - C_D^+ = \frac{4 v }{\pi^{d - 2} \Gamma(1-v) \Gamma(v)} \left( (1 + v)\Gamma \left(\frac{d}{2} -1 - v \right) \Gamma \left(\frac{d}{2} + v \right) - v \rightarrow - v  \right).
\end{equation}
\end{comment}

\section{Monodromy defect at large $N$} \label{SectionLargeN}
In this section, we study the monodromy defect in the interacting $O(2 N)$ model (it will soon be clear why we choose $2N$ instead of $N$ here) at large $N$. The $S^1 \times H^{d - 1}$ setup provides a convenient way to study the problem. The action may be written as \footnote{We assume that the mass terms have been tuned away so that the bulk is always critical.}
\begin{equation}
S =  \int d^d x \sqrt{g(x)} \left( \frac{g^{\mu \nu} \partial_{\mu} \phi^A \partial_{\nu} \phi^A}{2} - \frac{(d - 2)^2}{8} \phi^A \phi^A + \frac{\lambda}{4} (\phi^A \phi^A)^2 \right)
\end{equation}
where $A$ now goes from $1$ to $2N$. We again consider monodromy defect defined as in \eqref{MonodromyDefinition}. We want to do a large $N$ analysis, and to accomplish that, we want to preserve a large symmetry group. The simplest such case is when we fix the matrix $G^{AB}$ in \eqref{MonodromyMatrix} to consist of $N$ identical $2 \times 2$ blocks, so that all $\vartheta_{i} = \vartheta$. This is the only case we consider in this paper. Then, as before, it is convenient to package these $2 N$ real scalars into $N$ complex scalars as $\Phi^I = \phi^{2 I -1} + i \phi^{2 I}$, where $I$ goes from $1$ to $N$ and all $N$ complex scalars have the same monodromy as in \eqref{ComplexScalarMonodromy}. The original theory has $O(2 N)$ symmetry, and the defect breaks it down to $U(N)$. However for $\vartheta = 0$ and $\pi$ which correspond to a trivial defect and $Z_2$ monodromy defect respectively, the symmetry is enhanced and the defect preserves full $O(2 N)$ symmetry. The action in terms of complex variables is
\begin{equation} \label{ActionO2NEpsilon}
S =  \int d^d x \sqrt{g(x)} \left( \frac{g^{\mu \nu} \partial_{\mu} \bar{\Phi}_I \partial_{\nu} \Phi^I}{2} - \frac{(d - 2)^2}{8} \bar{\Phi}_I \Phi^I + \frac{\lambda}{4} (\bar{\Phi}_I \Phi^I)^2 \right).
\end{equation} 
At large $N$, we can use the well-known Hubbard-Stratonovich transformation to write this in terms of auxiliary field $\sigma (x)$ 
\begin{equation}
S =  \int d^d x \sqrt{g(x)} \left( \frac{g^{\mu \nu} \partial_{\mu} \bar{\Phi}_I \partial_{\nu} \Phi^I}{2} - \frac{(d - 2)^2}{8} \bar{\Phi}_I \Phi^I + \frac{1}{2} \sigma \bar{\Phi}_I \Phi^I \right).
\end{equation} 
We dropped a $\sigma^2/ 4 \lambda $ term above, which can be consistently done in the critical limit (see for example \cite{Giombi:2016ejx}, for a review). We can then integrate out the fields $\Phi^I$ since the action is quadratic in $\Phi^I$ to get 
\begin{equation}
Z = \exp[- F_{\textrm{twisted}}] = \int [d \sigma] \exp \left[-N \textrm{tr} \log \left( - \nabla^2 + \sigma - \frac{(d - 2)^2}{4} \right) \right]
\end{equation}
At large $N$, we can use a saddle point approximation to do the integral over $\sigma$ and look for a saddle with a constant value for the field $\sigma (x)$. This constant is the one-point function of $\sigma(x)$ which is a constant on the hyperbolic cylinder \footnote{In the flat space, this one-point function is $\langle \sigma (x) \rangle = \frac{\sigma^*}{r^2}$ with $\sigma^*$ being the constant one-point function on the hyperbolic cylinder}. So at leading order at large $N$, the field $\sigma(x)$ only contributes through its one-point function, and acts as a mass term for $\Phi^I$. Similar to the case of free theory \eqref{FTwistedHyperbolic}, the free energy in the interacting theory at leading order at large $N$ may then be written as
\begin{equation}
F_{\textrm{twisted}} (\vartheta) = \frac{ N \textrm{Vol}(H^{d-1})}{  (4 \pi)^{\frac{d - 1}{2}} \Gamma(\frac{d - 1}{2})}  \int_{-\infty}^{\infty} d \nu \frac{|\Gamma(i \nu + \frac{d-2}{2})|^2}{|\Gamma(i \nu)|^2}  \sum_{n \in \mathbb{Z} + v} \log \left( \nu^2 + n^2 + \sigma \right).
\end{equation}
The value of $\sigma$ at the large $N$ fixed point, $\sigma^*$ can be obtained by solving the saddle point equation which says that the following derivative should vanish

\begin{equation}\label{FDerivativeSigma}
\begin{split}
\frac{ \partial F_{\textrm{twisted}} (\vartheta)}{\partial \sigma} \bigg|_{\sigma = \sigma^*} &= \frac{ N \textrm{Vol}(H^{d-1})}{  (4 \pi)^{\frac{d - 1}{2}} \Gamma(\frac{d - 1}{2})}  \int_{-\infty}^{\infty} d \nu \frac{|\Gamma(i \nu + \frac{d-2}{2})|^2}{|\Gamma(i \nu)|^2}  \sum_{n \in \mathbb{Z} + v} \frac{1}{\nu^2 + n^2 + \sigma^*} \\
& =  \frac{ N \textrm{Vol}(H^{d-1}) \Gamma \left(\frac{3 - d}{2}\right) }{  (4 \pi)^{\frac{d - 1}{2}} }  \sum_{k = - \infty}^{\infty}  \frac{ \Gamma \left(\frac{d}{2} - 1 + \sqrt{\sigma^* + \left( k + v \right)^2 }\right)}{ \Gamma \left(2-\frac{d}{2} + \sqrt{\sigma^* + \left( k + v \right)^2 }\right)}  \\
&= \frac{ N \textrm{Vol}(H^{d-1}) \Gamma \left(\frac{3 - d}{2}\right) }{  (4 \pi)^{\frac{d - 1}{2}} }  \sum_{s \in \mathbb{Z} + v}  \frac{ \Gamma \left(\Delta_s \right)}{ \Gamma \left(3- d + \Delta_s \right)}
\end{split}
\end{equation} 
where the integral over $\nu$ is similar to the one in \eqref{SpectralIntegralOnePoint} and can be performed with similar techniques. In the last line, we used the usual AdS/CFT dictionary to write the expression in terms of the dimensions of defect operators 
\begin{equation} \label{MassDimensionLargeN}
\Delta_s (\Delta_s - d + 2) = s^2 + \sigma^* - \frac{(d-2)^2}{4} \implies \Delta_s^{\pm} = \frac{d}{2} - 1 \pm \sqrt{\sigma^* + s^2}.
\end{equation}
Note that we used $\Delta_s^+$ solution to write the above expression of $\partial F / \partial \sigma$ in \eqref{FDerivativeSigma}. This is because the spectral representation of the free energy is only valid for $\Delta_s > d/2 - 1$. However, written in terms of $\Delta_s$, the expression in \eqref{FDerivativeSigma} can be analytically continued and also used for the case when we impose $\Delta^-$ boundary condition on one or more of the operators. We also want the dimensions of all the defect operators to be real which requires 
\begin{equation} \label{DefectPositiveSigma}
\sigma^* \ge \ \textrm{max} \left( - v^2, - (1 - v)^2 \right).
\end{equation}

Another equivalent way to derive this large $N$ saddle point equation is to look at the two-point function of $\Phi$ in the bulk OPE limit. As we discussed above \eqref{TwoPointHyperbolicFree}, the two-point function on the hyperbolic cylinder is given by the sum over bulk-bulk propagators 
\begin{equation}
\langle \Phi^I (x_1) \bar{\Phi}_J(x_2) \rangle = \sum_{s \in \mathbb{Z} + v} {\delta^I}_J \frac{   \Gamma \left(\Delta_s \right) 
e^{i s \theta_{12}} \, _2F_1\left(\Delta_s, \Delta_s  +\frac{3 - d}{2};2  \Delta_s + 3 - d ;-\frac{1}{\xi }\right) }{2 \pi^{d/2}  \Gamma ( \Delta_s + 2 -\frac{d}{2})  (4\xi)^{\Delta_s}}
\end{equation}
and here $\Delta_s$ is given by \eqref{MassDimensionLargeN}. In the bulk OPE limit, the two point function behaves as 
\begin{equation}
\langle \Phi^I (x_1) \bar{\Phi}_J(x_2) \rangle =  \frac{{\delta^I}_J }{\pi (4 \pi)^{\frac{d - 1}{2}}} \sum_{s \in \mathbb{Z} + v} \left[ \frac{\Gamma\left( \frac{d-3}{2} \right)}{\xi^{\frac{d -3}{2}}} \left(1 + O(\xi) \right) + \frac{ \Gamma \left(\frac{3 - d}{2}\right) \Gamma \left(\Delta_s \right)}{ \Gamma \left(3- d + \Delta_s \right)} \left(1 + O(\xi) \right) \right].
\end{equation}
The constant $\xi$ independent piece in the second term represents the presence of operator $\Phi^I \bar{\Phi}_I$ of dimension $d - 2$ in the bulk OPE. Recall that in the large $N$ critical $U(N)$ model, this operator is replaced by the operator $\sigma$ of dimension $2$. This should still be true in the presence of the defect, and demanding that this term vanishes is equivalent to the saddle point equation written in \eqref{FDerivativeSigma}.   

When we impose $\Delta^+$ boundary condition on all the operators, $\sigma^*$ can be determined by solving the following equation from \eqref{FDerivativeSigma}
\begin{equation} \label{SaddleEquationPlus}
\frac{ \textrm{Vol}(H^{d-1}) \Gamma \left(\frac{3 - d}{2}\right) }{  (4 \pi)^{\frac{d - 1}{2}} }  \sum_{k = - \infty}^{\infty}  \frac{ \Gamma \left(\frac{d}{2} - 1 + \sqrt{\sigma^* + \left( k + v \right)^2 }\right)}{ \Gamma \left(2-\frac{d}{2} + \sqrt{\sigma^* + \left( k + v \right)^2 }\right)}  = 0.
\end{equation}
It is hard to perform this sum analytically as a function of $d$. To proceed, we separate out the sum into a divergent piece at large $k$ and a finite piece. The divergent piece of the sum can be performed by dimensional regularization and analytically continued in $d$. For the finite piece, it is harder to perform the sum as a function of $d$. However in $d = 4 - \epsilon$, we can first do a series expansion in $\epsilon$ \footnote{since $\sigma^* = 0$ for $\epsilon = 0$, we assume that $\sigma^*$ is of order $\epsilon$ and then the solution to the saddle point equation justifies this assumption.} and then the sum can be performed  up to first two orders in $\epsilon$. This procedure gives the following saddle equation 
\begin{equation}
\frac{\sigma^*}{\epsilon} - v(v-1) - \frac{\sigma^*}{2} \left(2 + \psi(v) + \psi(1-v) \right) + \frac{\epsilon}{2} \left[ -1 + v (v - 1) (-1 + \psi(v) + \psi(1-v) )   \right] = 0
\end{equation}
where $\psi$ is the polygamma function. This equation can be solved to give
\begin{equation} \label{SigmaStarEpsP}
\sigma^* = v \left( v - 1 \right) \epsilon +  \frac{3 v (v - 1) + 1}{2} \epsilon^2 + O(\epsilon^3).
\end{equation}
Notice that the order $\epsilon^2$ term does not vanish at $v = 0$ and $1$. But the defect becomes trivial at $v = 0$ and $1$, so all one-point functions, including the one-point function of $\sigma$ should vanish at these values of $v$. This problem arises because we are doing an expansion in $\epsilon$ so we assumed that $v >> \epsilon$, and we expect this problem to be resolved by higher order terms in $\epsilon$. Indeed when we calculate $\sigma^*$ numerically in $d = 3.9$ below in figure \ref{FigureSigmaStar39dPlus}, we will see that it vanishes at $v = 0$ and $1$. One possibility is that the higher order terms in $\epsilon$ are singular at $v = 0$ and $1$. For example, consider the $\epsilon$ expansion of the following simple function 
\begin{equation}
\frac{\epsilon v^{\frac{3}{2}}}{\sqrt{v + \epsilon}} = \epsilon v - \frac{\epsilon^2}{2} + \frac{3 \epsilon^2}{2 v} + O(\epsilon^4).
\end{equation} 
The function vanishes at $v = 0$ for any fixed $\epsilon$, but when we expand at small $\epsilon$, the limit $v \rightarrow 0$ becomes problematic. This is similar to what we see here. 

Plugging this $\sigma^*$ into \eqref{MassDimensionLargeN}, we get the dimension of defect operators at large $N$ and leading orders in $\epsilon$
\begin{equation} \label{DimensionsLargeNEps}
\Delta_s^+ = 1 - \frac{\epsilon}{2} + |s| + \frac{v (v - 1)}{2 |s|} \epsilon + \frac{3 v (v - 1) + 1}{4 |s|} \epsilon^2 - \frac{v^2 (v - 1)^2 }{8 |s|^3} \epsilon^2 + O(\epsilon^3)
\end{equation}
where we assumed $s^2 > \sigma^*$ to do the expansion in $
\epsilon$. This is consistent with what we find in the $\epsilon$ expansion in \eqref{AnomalousDimensionTwoPointDef}. We can also calculate the twisted free energy at the large $N$ fixed point
\begin{equation} \label{TwistedFreeLargeNEps}
\begin{split}
F_{\textrm{twisted}} &= F_{\textrm{twisted}}\bigg|_{\sigma = 0} + \int_0^{\sigma^*} d \sigma \frac{\partial F_{\textrm{twisted}}}{\partial \sigma} \\
&= F_{\textrm{twisted}}\bigg|_{\sigma = 0} + \frac{ N\textrm{Vol} (H^3)}{4 \pi} \int_0^{\sigma^*} d \sigma \left( - \frac{\sigma}{ \epsilon} + v \left( v - 1 \right)  \right) \\
& = N F_{\textrm{twisted}}^{\textrm{free}} + \frac{N \textrm{Vol} (H^3)  v^2 \left( v - 1 \right)^2}{8 \pi}  \epsilon 
\end{split}
\end{equation}
where we used the fact that $F_{\textrm{twisted}}(\sigma = 0)$ is the same as twisted free energy in the free theory. This is also consistent with the $\epsilon$ expansion result we obtain below in \eqref{TwistedFreeEpsilon}. The comparison of the large $N$ and the epsilon expansion calculations of free energy and defect dimensions only involves the  $O(\epsilon)$ piece in $\sigma^*$. In order to check $\epsilon^2$ piece, we look at the bulk OPE coefficient times the one-point function coefficient ${C_{\bar{\Phi}; \Phi}}^{\sigma} \sigma^*$. Since the operator $\sigma$ replaces $\bar{\Phi} \Phi$ in the large $N$ theory, this should match $ {C_{\bar{\Phi}; \Phi}}^{\bar{\Phi} \Phi} C_1^{\bar{\Phi} \Phi}$ calculated in $\epsilon$ expansion. From the large N results, the bulk OPE coefficient for the $O(2 N)$ model is \cite{Petkou:1994ad}, 
\begin{equation} \label{PetkouLargeNConsistency}
    {C_{\bar{\Phi}; \Phi}}^{\sigma} = \sqrt{\frac{1}{N C_{\sigma}} \frac{\Gamma(d - 2)}{\Gamma(3 - \frac{d}{2}) \Gamma^3(\frac{d}{2}-1)}} \  C_{\Phi}
\end{equation}
where $C_{\Phi}$ and $C_{\sigma}$ are the normalization of the bulk two-point functions, and in our conventions, they are given by (see for e.g \cite{Giombi:2016ejx})
\begin{equation} \label{SigmaTwoPointLargeN}
    C_{\sigma} =  \frac{2^{d + 1} \Gamma(\frac{d - 1}{2}) \sin(\frac{\pi d}{2})}{ N \pi^{\frac{3}{2}} \Gamma(\frac{d}{2} -2)}, \hspace{1cm} C_{\Phi} =  \frac{\Gamma\left(\frac{d}{2} - 1 \right)}{2 \pi^{d/2}}.
\end{equation}
Combining it with the result for $\sigma^*$ in \eqref{SigmaStarEpsP}, this gives the following result
\begin{equation} \label{OnePointLargeNEps}
{C_{\bar{\Phi}; \Phi}}^{\sigma} \sigma^* =  \frac{(v-1) v}{4 \pi ^2}  + \frac{(v-1) v (\gamma +3+\log \pi ) + 1 }{8 \pi ^2} \epsilon + O(\epsilon^2)
\end{equation}
which is consistent with the $\epsilon$ expansion result in \eqref{OPEOnePointPhibarPhi}.

Away from $d = 4$, the finite piece of the sum in \eqref{SaddleEquationPlus} can be performed numerically, for a given $d, \sigma^*$ and $v$. We start with $d = 3.9$, so that we can compare it with the prediction in $d = 4 - \epsilon$. We evaluate the sum for a range of values of $\sigma$ and $v$, and then find the root of the equation on the real $\sigma$ axis for different values of $v$. We then interpolate in $v$ and plot the value of $\sigma^*$ in figure \ref{FigureSigmaStar39dPlus}. We also compare the result with the result in $d = 4 - \epsilon$ in \eqref{SigmaStarEpsP} at $\epsilon = 0.1$.

\begin{figure}
\centering
\includegraphics[scale=0.6]{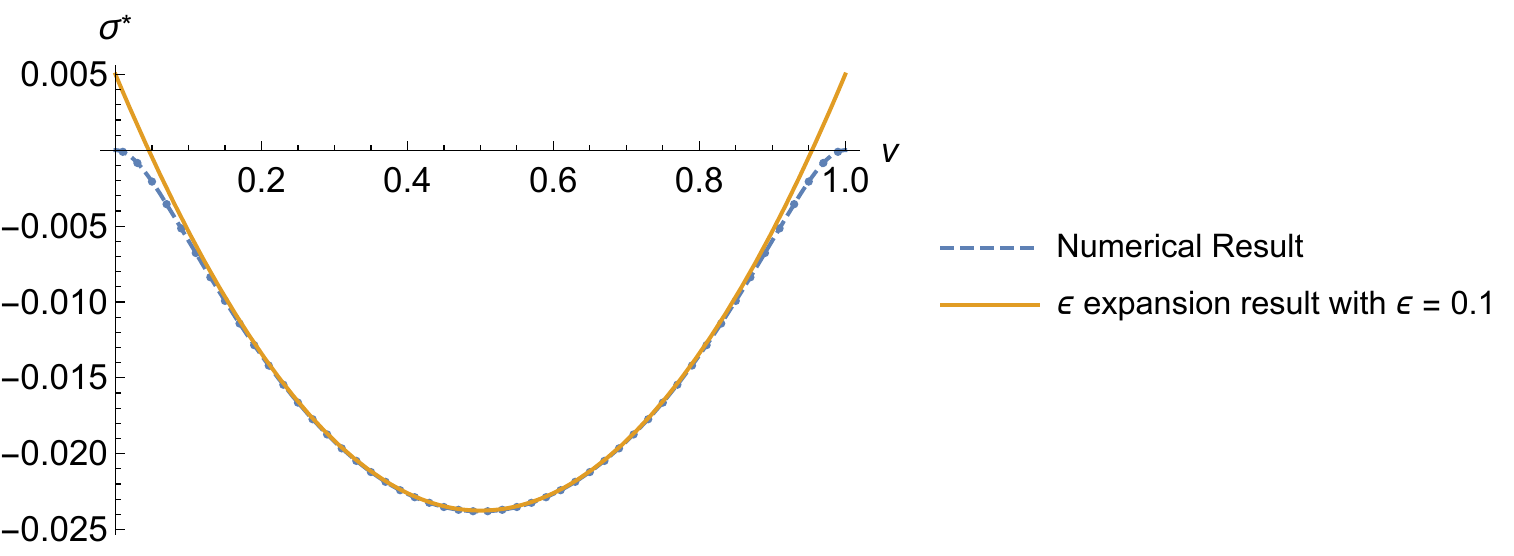}
\caption{Saddle point value of $\sigma^*$  in $d = 3.9$ critical theory with $\Delta^+$ boundary condition at leading order in  large $N$. Dashed line represents smooth interpolation of the numerical result. For comparison, we also plot the analytic result in $d = 4 - \epsilon$ at $\epsilon = 0.1$.}
\label{FigureSigmaStar39dPlus}
\end{figure}

\begin{figure}
\centering
\begin{subfigure}{0.48\textwidth}
\includegraphics[width = \textwidth]{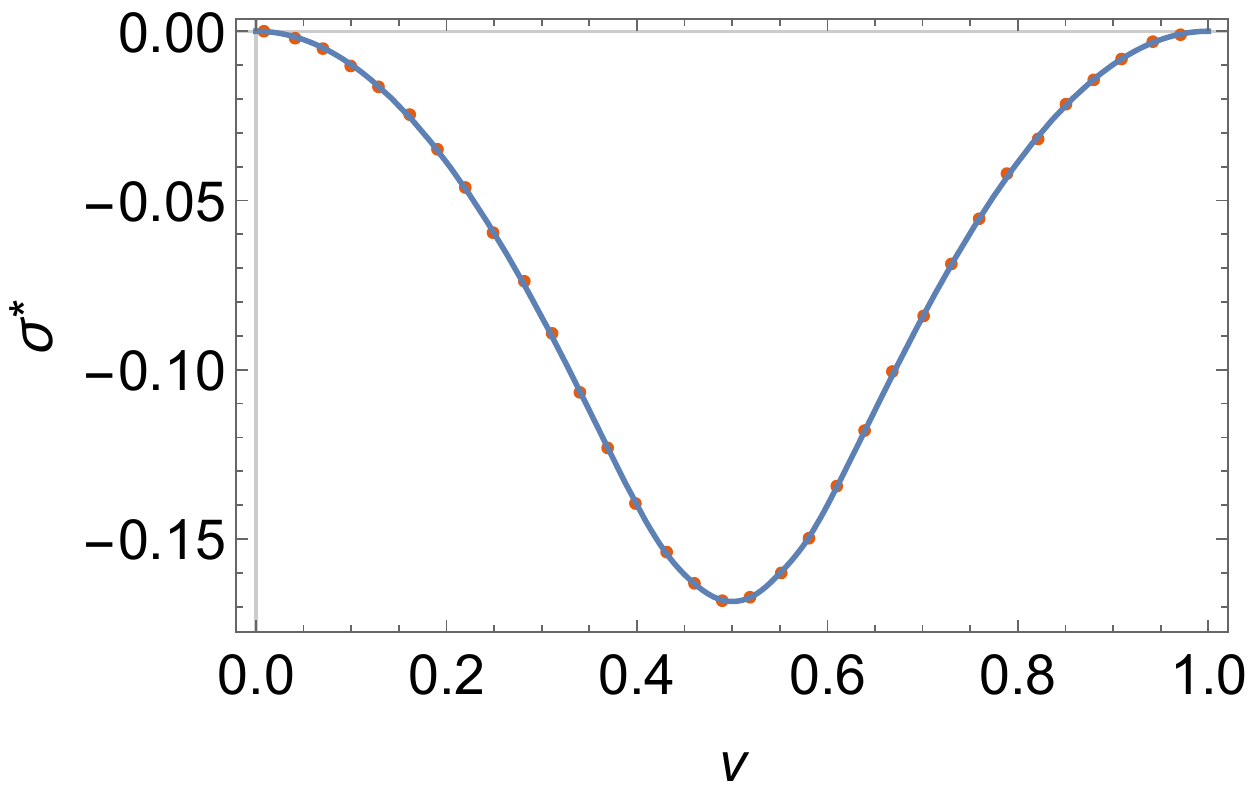}
\end{subfigure}
\begin{subfigure}{0.50\textwidth}
\includegraphics[width = \textwidth]{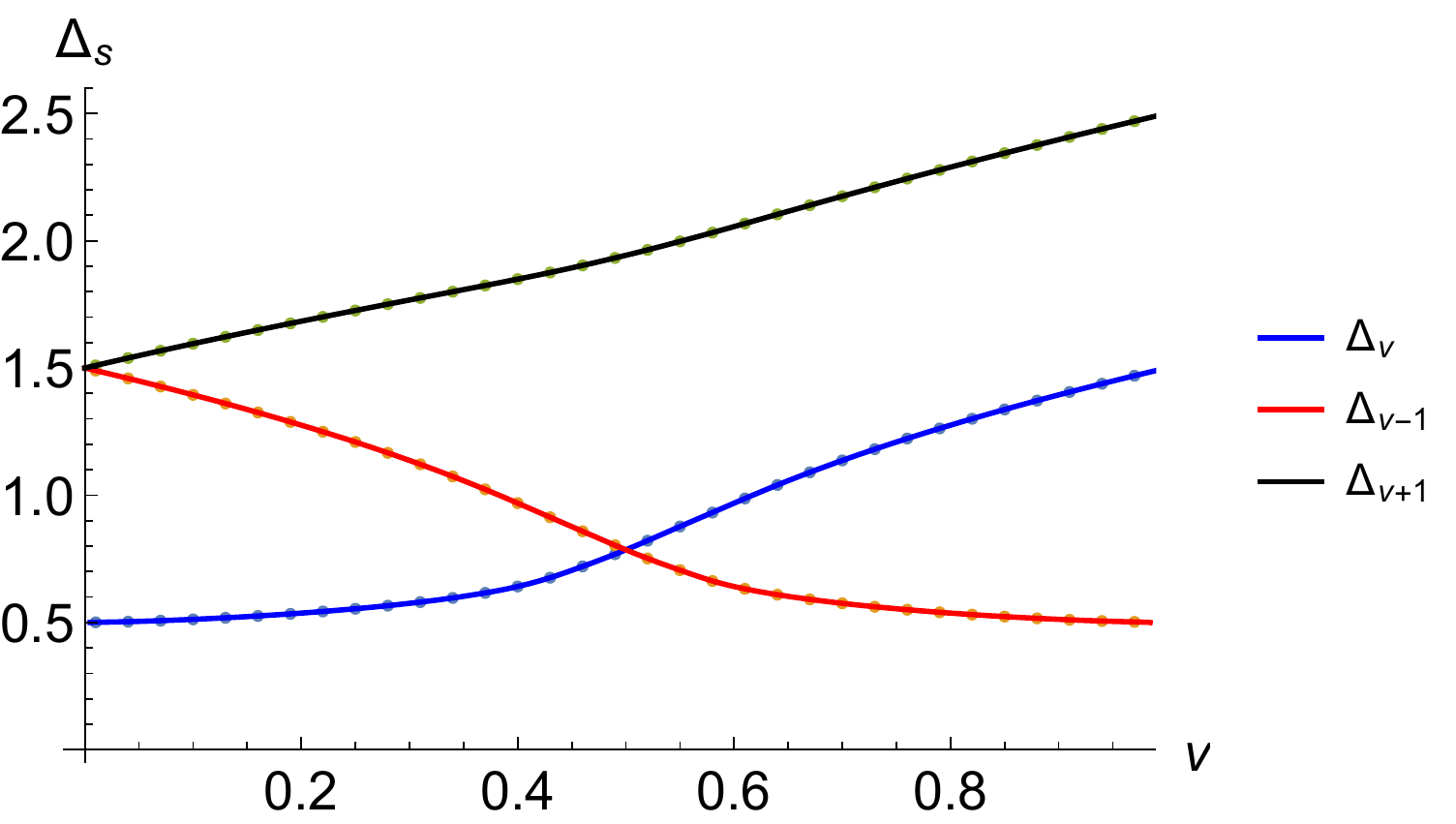}
\end{subfigure}
\caption{Saddle point value of $\sigma^*$ and the dimensions of defect operators in three dimensional critical theory at large $N$. The solid lines are smooth interpolation of the numerical results. Note that the defect is one-dimensional, therefore the unitarity bound just requires the defect dimensions to be positive.}
\label{FigureSigmaStar3d}
\end{figure}

A similar method also works in  $d = 3$ to solve the saddle point equation in \eqref{SaddleEquationPlus} numerically. We plot the solution in figure \ref{FigureSigmaStar3d}. We can get some analytic control if we also take $v$ to be small. Since we know that $\sigma^* = 0$ when $v = 0$, we can expand \eqref{SaddleEquationPlus} for small $v$ and $\sigma^*$ and in $d = 3$, we find the following saddle point equation and the solution to leading order in $v$
\begin{equation}
\begin{split}
&- \frac{(v^2 + \sigma^*)^{\frac{1}{2}}}{2} + v^2 +  \sigma^* \log 2 - \frac{(v^2 + \sigma^*)^{\frac{3}{2}} \pi^2}{6} = 0\\
&\implies \sigma^* = - v^2 + 4 (1 - \log 2)^2 v^4.
\end{split}
\end{equation}

Once we know $\sigma^*$, we can calculate the dimensions of defect operators using \eqref{MassDimensionLargeN}. We also plot the dimensions of three low-lying defect operators in figure \ref{FigureSigmaStar3d}. For $v = 1/2$, corresponding to $Z_2$ monodromy on all scalars, we get 
\begin{equation}
\sigma_*= - 0.168 , \hspace{0.5cm} \Delta_{1/2} = 0.786, \hspace{0.5cm} \Delta_{3/2} = 1.943 
\end{equation}
to leading order at large $N$. We are doing a large $N$ analysis, but it is interesting to compare the result with the Monte Carlo  results for monodromy defect in $ d = 3$ Ising model in \cite{Billo:2013jda}. They found $\Delta_{1/2} = 0.918$ and $\Delta_{3/2} = 1.99$.

\begin{figure}
\centering
\includegraphics[scale=0.6]{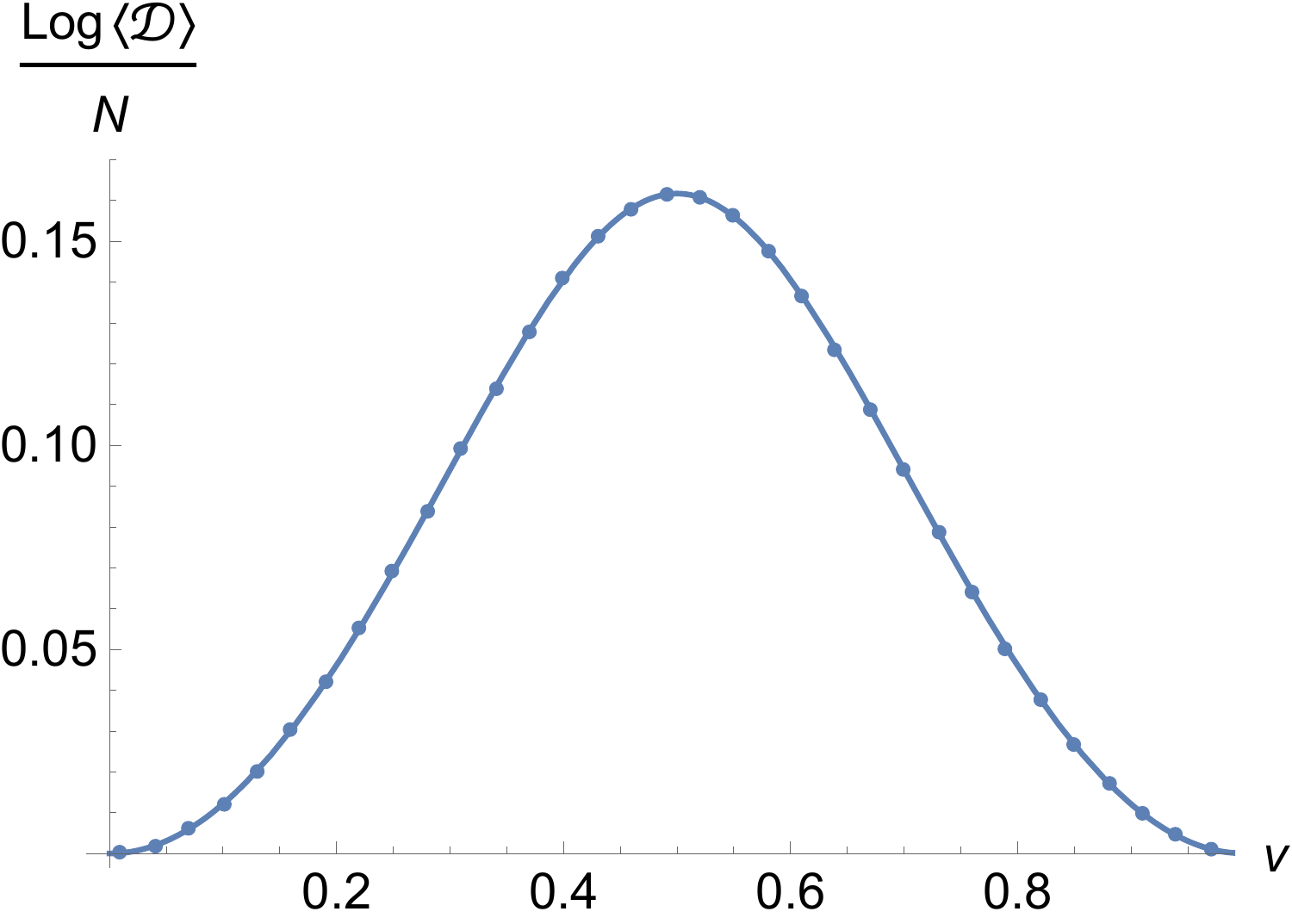}
\caption{Defect expectation value in $d = 3$ critical theory at leading order in  large $N$. Solid line interpolates the numerical results. }
\label{FigureDefectExpectation3d}
\end{figure}

\begin{figure}
\centering
\includegraphics[scale=0.8]{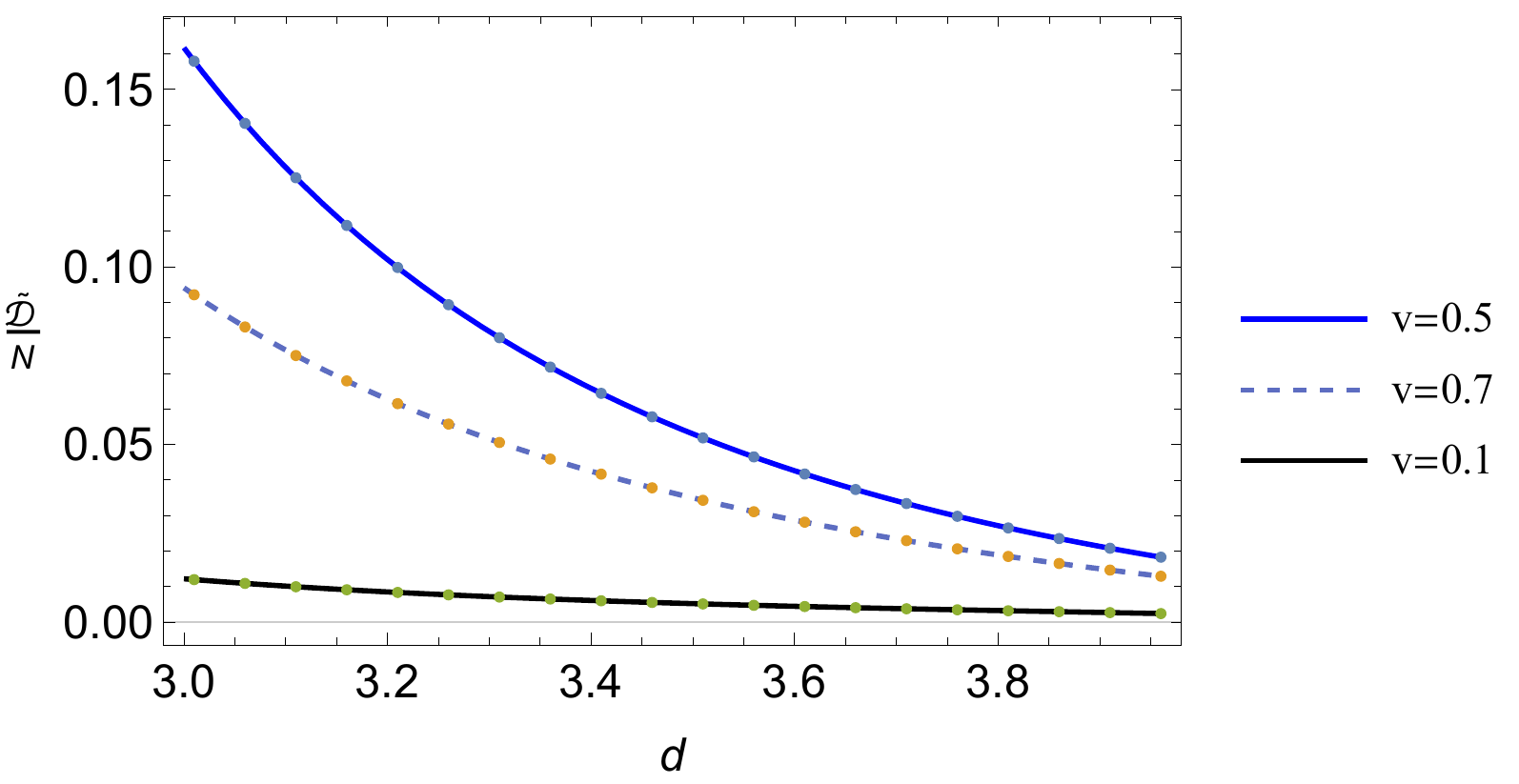}
\caption{ A plot of $\tilde{\mathcal D}$ in the large $N$ critical theory between $3 < d < 4$. Solid and dashed lines interpolate numerical results.}
\label{FigureDefectExpectationDimension}
\end{figure}

Using $\sigma^*$, we can also calculate the expectation value of the defect with spherical geometry. It is defined in the same way as the free theory \eqref{DefectExpectationHyperbolic}, as the negative of the difference between the twisted and the untwisted free energy. However, recall that in the untwisted theory, the one-point functions vanish, so $\sigma^* = 0$. This implies that the corrections due to the interactions for the untwisted free energy start at order $1/N$. So at leading order, we can just use the untwisted free energy of the free theory and the interacting corrections to the defect expectation value are due to the corrections in the twisted free energy
\begin{equation} \label{DefectExpectationLargeN}
\begin{split}
- \log \langle \mathcal{D} \rangle = - N \log \langle \mathcal{D} \rangle^{\textrm{free}} + \int_0^{\sigma^*} d \sigma \frac{\partial F_{\textrm{twisted}}}{\partial \sigma}.
\end{split}
\end{equation}
The first term above is the free theory result we have from \eqref{DefectExpectationHyperbolic}. %In $d = 3$, we can also use the explicit result in \eqref{DefectExpectation3d}. 
We can numerically integrate \eqref{FDerivativeSigma} using the numerical results for $\sigma^*$ to evaluate the second term. We plot the result in figure \ref{FigureDefectExpectation3d} for the case of $d = 3$, corresponding to a circular defect. We used the standard regularized volume of $H^2$, ${\rm Vol}(H^2)=-2\pi$, that can be obtained \eqref{HyperbolicVolume}. We also plot the quantity $\tilde{\mathcal D}$ defined in (\ref{DtildeDefinition}) as a function of $d$ between $3 < d < 4$ for several values of $v$ in figure \ref{FigureDefectExpectationDimension}. It shows that this quantity is a smooth function of $d$ in this range.


\subsection{Alternate boundary condition on $H^{d-1}$}
Recall from section \ref{SectionDefectRGFlow} that there is another defect fixed point in the free theory which arises when we impose $\Delta^-$ boundary condition for $s = v$ mode. One can flow between the two fixed points by turning on the defect operator $\bar{\Psi}_v \Psi_v$ which is relevant at $\Delta_v^-$ fixed point. In this subsection, we study this fixed point in the large $N$ theory. With the $\Delta^-$ boundary condition, we get the following saddle point equation from \eqref{FDerivativeSigma}
\begin{equation} \label{SaddleEquationMinus}
\frac{ \textrm{Vol}(H^{d-1}) \Gamma \left(\frac{3 - d}{2}\right) }{  (4 \pi)^{\frac{d - 1}{2}} } \left[ \frac{ \Gamma \left(\frac{d}{2} - 1 - \sqrt{\sigma^* +  v^2 }\right)}{ \Gamma \left(2-\frac{d}{2} - \sqrt{\sigma^* + v^2 }\right)} +  \sum_{k \neq 0}  \frac{ \Gamma \left(\frac{d}{2} - 1 + \sqrt{\sigma^* + \left( k + v \right)^2 }\right)}{ \Gamma \left(2-\frac{d}{2} + \sqrt{\sigma^* + \left( k + v \right)^2 }\right)} \right]  = 0.
\end{equation}
Note that since we are working in $3 < d < 4$ the defect unitarity bound \eqref{DefectUnitarity} requires 
\begin{equation} \label{DefectUnitaritySigma}
\Delta^-_v > 0 \implies \sigma^* \le \left( \frac{d}{2} - 1\right)^2 - v^2.
\end{equation}
So we have to look for a solution to the saddle point equation \eqref{SaddleEquationMinus} satisfying both \eqref{DefectPositiveSigma} and \eqref{DefectUnitaritySigma}.

\begin{figure}
\centering
\includegraphics[scale=0.77]{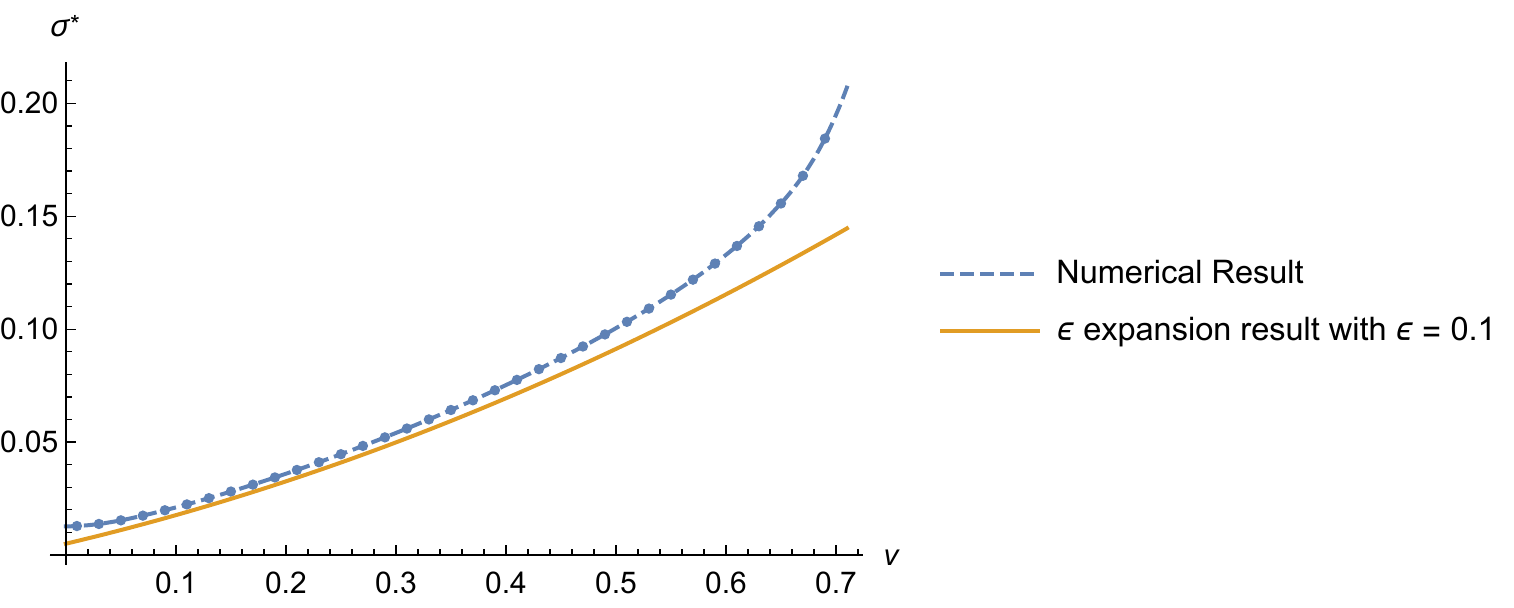}
\caption{Saddle point value of $\sigma^*$  in $d = 3.9$ critical theory with $\Delta^-$ boundary condition and for comparison, the analytic result in $d = 4 - \epsilon$ at $\epsilon = 0.1$.}
\label{FigureSigmaStar39dMinus}
\end{figure}

In $d = 4 - \epsilon $ the saddle point equation in \eqref{SaddleEquationMinus} can be solved perturbatively in $\epsilon$ similar to what we did in the $\Delta^+$ case. The equation and the solution is 
\begin{equation} \label{SigmaStar39dEpsMin}
\begin{split}
&\frac{\sigma^*}{\epsilon} - v(v+1) + \frac{\epsilon}{2} \left[ 1 + v (1- v) + v(v + 1) (\psi(v) + \psi(1-v) )   \right] \\
& - \frac{\sigma^*}{2} \left(2 + \frac{2}{v} + \psi(v) + \psi(1-v) \right)  = 0 \\
& \implies \sigma^* = v \left( v + 1 \right) \epsilon + \frac{3 v (v+1)+1}{2}  \epsilon^2 + O(\epsilon^3).
\end{split}
\end{equation}
As in the $\Delta^+$ case, the order $\epsilon^2$ term does not vanish at $v = 0$ and $1$. However, in this case, as we show numerically in figure \ref{FigureSigmaStar39dMinus}, $\sigma^*$ is nonzero at $v = 0$ also in $d = 3.9$. This shows that already for $v = 0$, we have a defect which is defined by changing the boundary condition of $s = 0$ mode, such that it has dimensions $\Delta_0 = d/2 - 1 - \sqrt{\sigma^*}$. To understand this defect in a perturbation in $\epsilon$ in $d = 4 - \epsilon$ using action \eqref{ActionO2NEpsilon}, notice that the operator $\bar{\Psi}_v \Psi_v$ becomes marginal for $v = 0$. So we have to add a defect localized term $g \bar{\Psi}_0 \Psi_0 $ to the action and look for a fixed point in the $g, \lambda$ plane. We do not do this calculation here, but our numerical results in the large $N$ theory indicate that there should be such a fixed point. A similar calculation was done in \cite{Metlitski:2009iyg} where this model was considered without any monodromy, but with the length of $S^1$ being different from $2 \pi$.

Using $\sigma^*$, we can calculate the dimensions of the defect operators at large $N$
\begin{equation} \label{DimensionsLargeNEpsMinus}
\begin{split}
&\Delta_s^- = 1 - \frac{\epsilon}{2} + |s| + \frac{v (v + 1)}{2 |s|} \epsilon + O(\epsilon^2) ,\hspace{1cm} s \neq v \\
& \Delta_v^- = 1 - \frac{\epsilon}{2} - v - \frac{(v + 1)}{2} \epsilon + O(\epsilon^2).
\end{split}
\end{equation}
These results are consistent with the results in \eqref{AnomalousDimensionTwoPointDefMinOther} and \eqref{AnomalousDimensionTwoPointDefMinv} found using $\epsilon$ expansion. Note that the notation is such that $-$ sign in the superscript above means that we are working in a theory where $s = v$ mode has $\Delta^-$ boundary condition, but all the other modes still have $\Delta^+$ boundary condition. As in $\Delta^+$ case in \eqref{OnePointLargeNEps}, to check the order $\epsilon^2$ term in $\sigma^*$, we look at the product of bulk OPE coefficient and the one-point function
\begin{equation}\label{OnePointLargeNEpsM}
{C_{\bar{\Phi}; \Phi}}^{\sigma} \sigma^* = \frac{v (v+1)}{4 \pi ^2} + \frac{ v (v+1) (\gamma +3+\log \pi )+1}{8 \pi ^2} \epsilon + O(\epsilon^2).
\end{equation} 
which is also consistent with the $\epsilon$ expansion result \eqref{OPEOnePointPhibarPhiM}.

Far from $d = 4$, we have to perform the sum numerically as we did before. In $d = 3.9$, it is possible to numerically find a saddle point subject to constraints \eqref{DefectPositiveSigma} and \eqref{DefectUnitaritySigma} between $0< v < 0.71$. We plot the result in figure \ref{FigureSigmaStar39dMinus}. The match with the analytic exression in $d = 4 - \epsilon$ is not as good as $\Delta^+$ boundary condition, but this may just indicate that $O(\epsilon^3)$ terms are important. 

\begin{figure}
\begin{subfigure} {0.48\textwidth}
\includegraphics[width = \textwidth ]{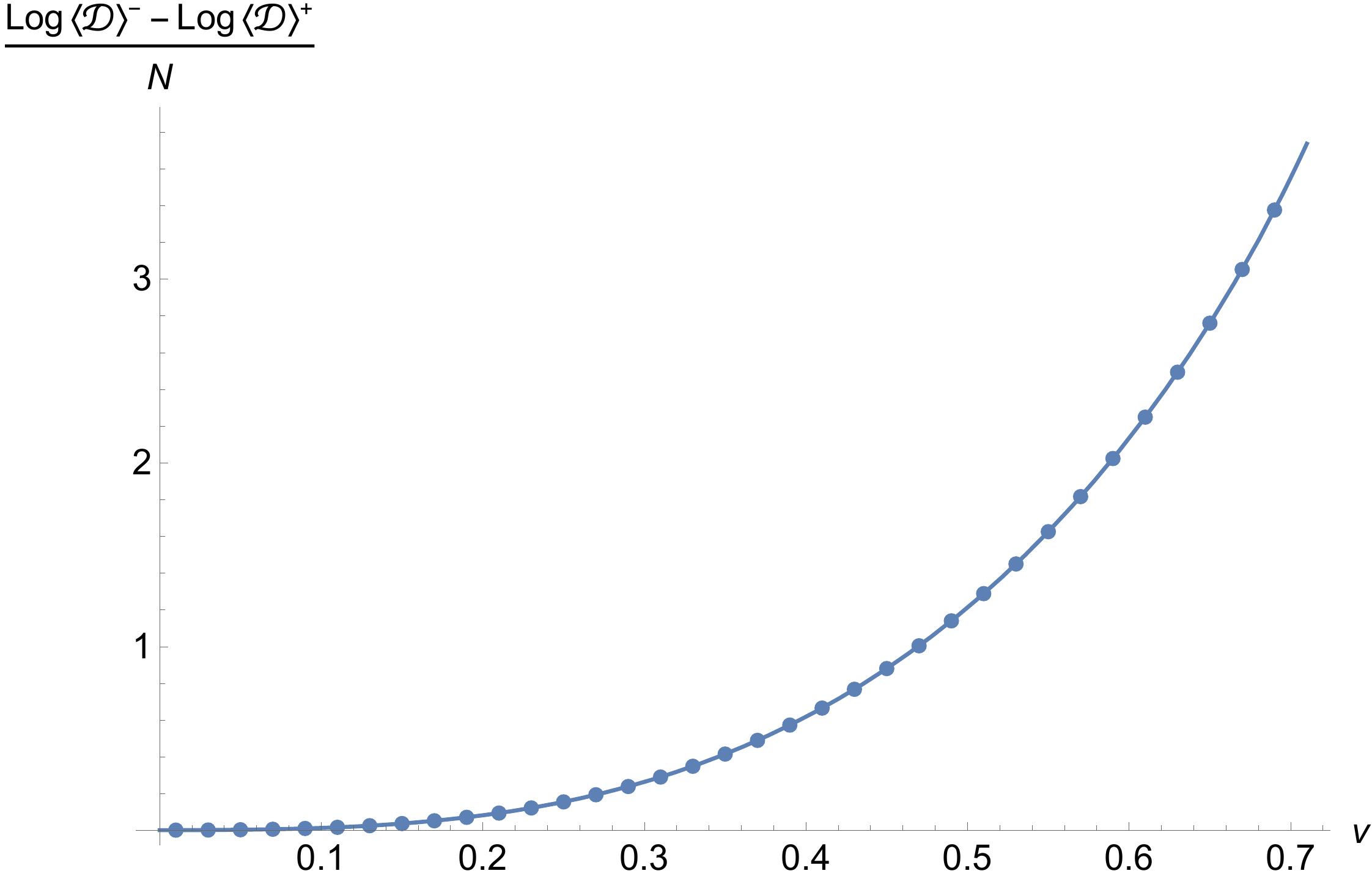}
\end{subfigure}
\begin{subfigure}{0.48\textwidth}
\includegraphics[width = \textwidth, scale = 0.6]{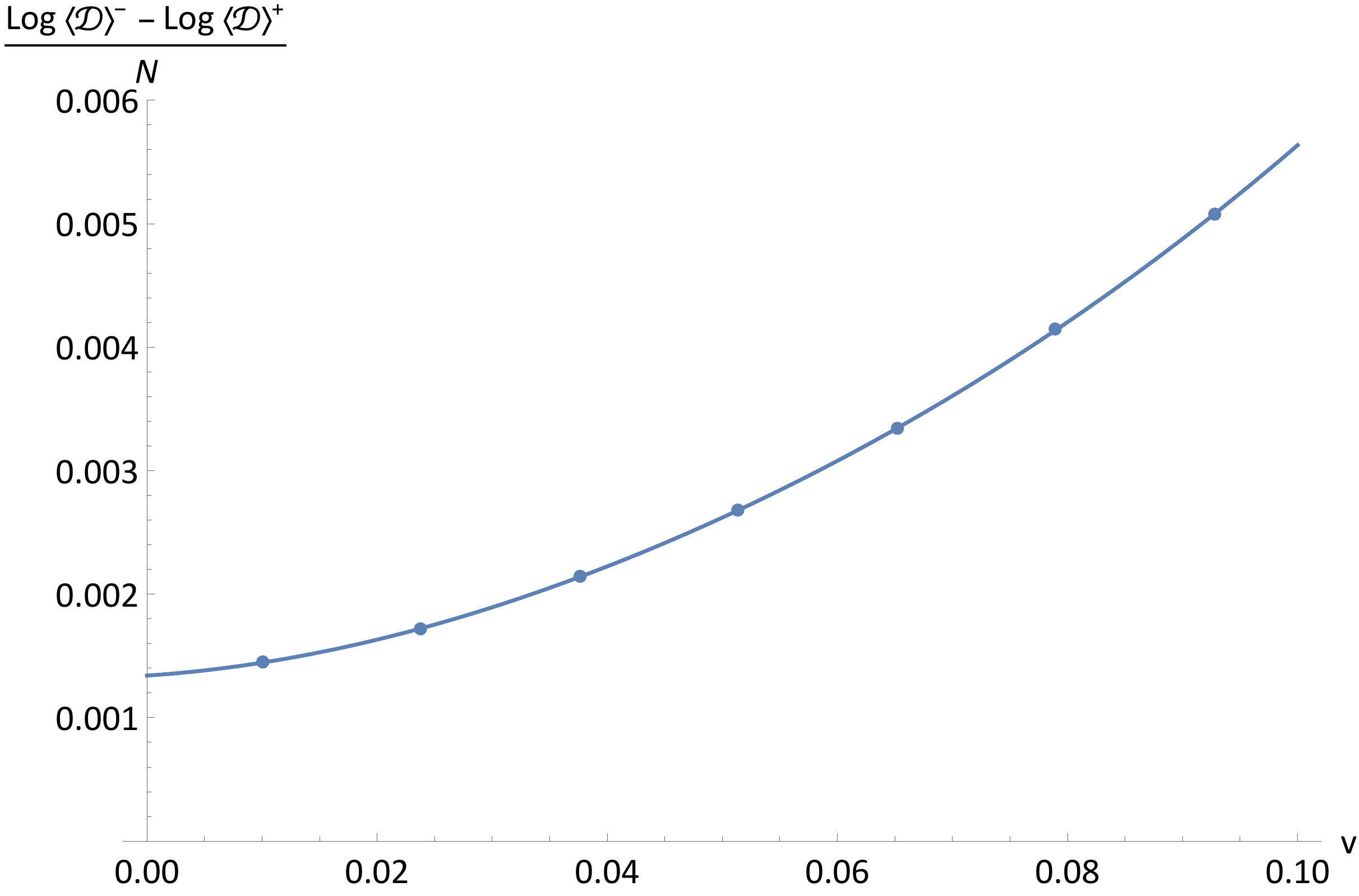}
\end{subfigure}
\caption{The difference in the defect expectation value between the theories with $\Delta^-$ and $\Delta^+$ boundary condition in $d = 3.9$. Note that for $\Delta^-$ boundary condition, as discussed below \eqref{SigmaStar39dEpsMin}, there is a nontrivial defect also at $v = 0$, so the difference in the expectation value should be nonzero at $v = 0$. This becomes clear when we zoom in to the region near $v =0$ which is what we show on the right plot.}
\label{FigureDefectExpectationDiffLargeN}
\end{figure}

Using the value of $\sigma^*$, we can also calculate the expectation value of the defect \eqref{DefectExpectationLargeN}. A quantity of interest is the difference in the logarithm of defect expectation value between the $\Delta^-$ and $\Delta^+$ boundary condition. It should be positive according to the conjectured defect F-theorem \eqref{DtildeDefinition} between $3< d < 4$ \footnote{The sine factor in front in the definition of $\tilde{\mathcal{D}}$ in \eqref{DtildeDefinition} is positive between $3 < d < 4$, so the difference of $\tilde{\mathcal{D}}$ and $\log \langle \mathcal{D} \rangle$ should both be positive.}. We plot it in figure \ref{FigureDefectExpectationDiffLargeN} in $d = 3.9$ dimensions. It is indeed positive for whole range of $v$ where the $\Delta^-$ saddle exists.   

As we go down to $d = 3$, there is no solution to the saddle point equation \eqref{SaddleEquationMinus} consistent with the constraints \eqref{DefectPositiveSigma} and \eqref{DefectUnitaritySigma}. So an interacting unitary fixed point in three dimensions does not exist for $\Delta^-$ boundary condition.

\subsection{Conformal weight }
We can also calculate the conformal weight of the defect in the interacting theory. It is related to $T_{\theta \theta}$ \eqref{ConformalWeightFree}, which in the hyperbolic cylinder approach is given by \eqref{ConformalWeightFreeEnergy}. So we need to know the dependence of the twisted free energy on $\beta$. Similar to the case of free theory, when we change the length of $S^1$ to $\beta$, the expression for the free energy changes to
\begin{equation}
F_{\textrm{twisted}} (\vartheta, \beta) = \frac{ N \textrm{Vol}(H^{d-1})}{  (4 \pi)^{\frac{d - 1}{2}} \Gamma(\frac{d - 1}{2})}  \int_{-\infty}^{\infty} d \nu \frac{|\Gamma(i \nu + \frac{d-2}{2})|^2}{|\Gamma(i \nu)|^2}  \sum_{n \in \mathbb{Z} + v} \log \left( \nu^2 + \frac{4 \pi^2 n^2}{\beta^2} + \sigma \right).
\end{equation}
If we impose $\Delta^+$ boundary condition on all the operators, then the large $N$ saddle point equation is    
\begin{equation} \label{SaddlePointBeta}
\frac{\partial F_{\textrm{twisted}}}{\partial \sigma} \bigg|_{\sigma = \sigma^*} = \frac{ N \textrm{Vol}(H^{d-1}) \Gamma\left( \frac{3 - d}{2} \right)}{  (4 \pi)^{\frac{d - 1}{2}} }\sum_{k=-\infty}^{\infty}\frac{\Gamma \left(\frac{d}{2}-1+\sqrt{\frac{4 \pi^2  }{\beta ^2}(k+v)^2+\sigma}\right)}{\Gamma \left(2-\frac{d}{2}+\sqrt{\frac{4 \pi^2  }{\beta ^2}(k+v)^2+\sigma}\right)}=0
\end{equation} 
Proceeding in the same way as $\beta = 2 \pi$, we first expand the sum in general $d$ in large $k$ and isolate the pieces that diverge as $k \rightarrow \infty$. The divergent piece of the sum can be performed in dimensional regularization and analytically continued in $d$. And for the finite piece, we have to either expand in $\epsilon$ or turn to numerical methods. In $d = 4 - \epsilon$, the solution to the saddle point equation to leading order in $\epsilon$ is 
\begin{equation}
\sigma^* = \left(\frac{2 \pi ^2 (6 (v -1) v +1)}{3 \beta ^2}-\frac{1}{6}\right)\epsilon + \mathcal{O}(\epsilon)^2
\end{equation}
Correction to the twisted free energy, to leading order in $\epsilon$ and $N$ is 
\begin{equation}
\begin{split} \label{FreeEnergyIntBeta}
F_{\textrm{twisted}} (\vartheta, \beta) &= F_{\textrm{twisted}} (\vartheta, \beta) \bigg|_{\sigma = 0} + N\int_0^{\sigma^*} d\sigma\, \frac{\partial F_{\textrm{twisted}}}{\partial \sigma}\\
&= F_{\textrm{twisted}} (\vartheta, \beta) \bigg|_{\sigma = 0} + \textrm{Vol} (H^3)N\frac{ \left(\beta ^2-4 \pi ^2 (6 (v -1) v +1)\right)^2}{576 \pi^2  \beta ^3}\epsilon+\mathcal{O}(\epsilon)^2
\end{split}
\end{equation}
Using \eqref{ConformalWeightFreeEnergy} and \eqref{ConformalWeightFree}, we get the conformal weight in terms of the free energy 
\begin{equation} \label{ConformalWeightFreeEnergyForm}
h = - \frac{1}{\textrm{Vol} (H^{d - 1})} \frac{2 \pi}{(d - 1)} \frac{\partial F_{\textrm{twisted}}}{\partial \beta} \bigg|_{\beta = 2 \pi}.
\end{equation}
This gives the conformal weight in the interacting theory to leading order in $\epsilon$ 
\begin{equation} \label{ConformalWeightLargeNEps}
\begin{split}
h &= - \frac{N \pi \Gamma \left(\frac{1 - d}{2}\right)(1 -v ) v  \left(\csc \pi \left(\frac{d}{2} - v \right)-\csc \pi \left(\frac{d}{2} + v \right) \right) }{ d   (4 \pi)^{\frac{d - 1}{2}}  \Gamma \left( 2 - \frac{d}{2}- v \right) \Gamma \left(1 - \frac{d}{2}+ v \right)} + \frac{N (1 - v) v (9 (1 - v)v - 2) }{72 \pi  }\epsilon \\
&=  \frac{N v^2 (1- v)^2}{12 \pi} + \frac{\epsilon N (1 - v)^2 v^2 }{{144 \pi }}   \left(-6 H^{-v-1}-6 \psi(v-1)+37+6 \log (\pi ) - \frac{4}{v (1 - v)}\right)
\end{split}
\end{equation}
where $\psi$ is the polygamma function and $H^n$ is the $n^{th}$ harmonic number. We used the free theory result for conformal weight \eqref{ConformalWeightFree} in $d = 4- \epsilon$. This agrees with the result from epsilon expansion calculation in the large $N$ limit \eqref{ConformalWeightEpsilon}.

\begin{figure}
\centering
\includegraphics[scale=0.6]{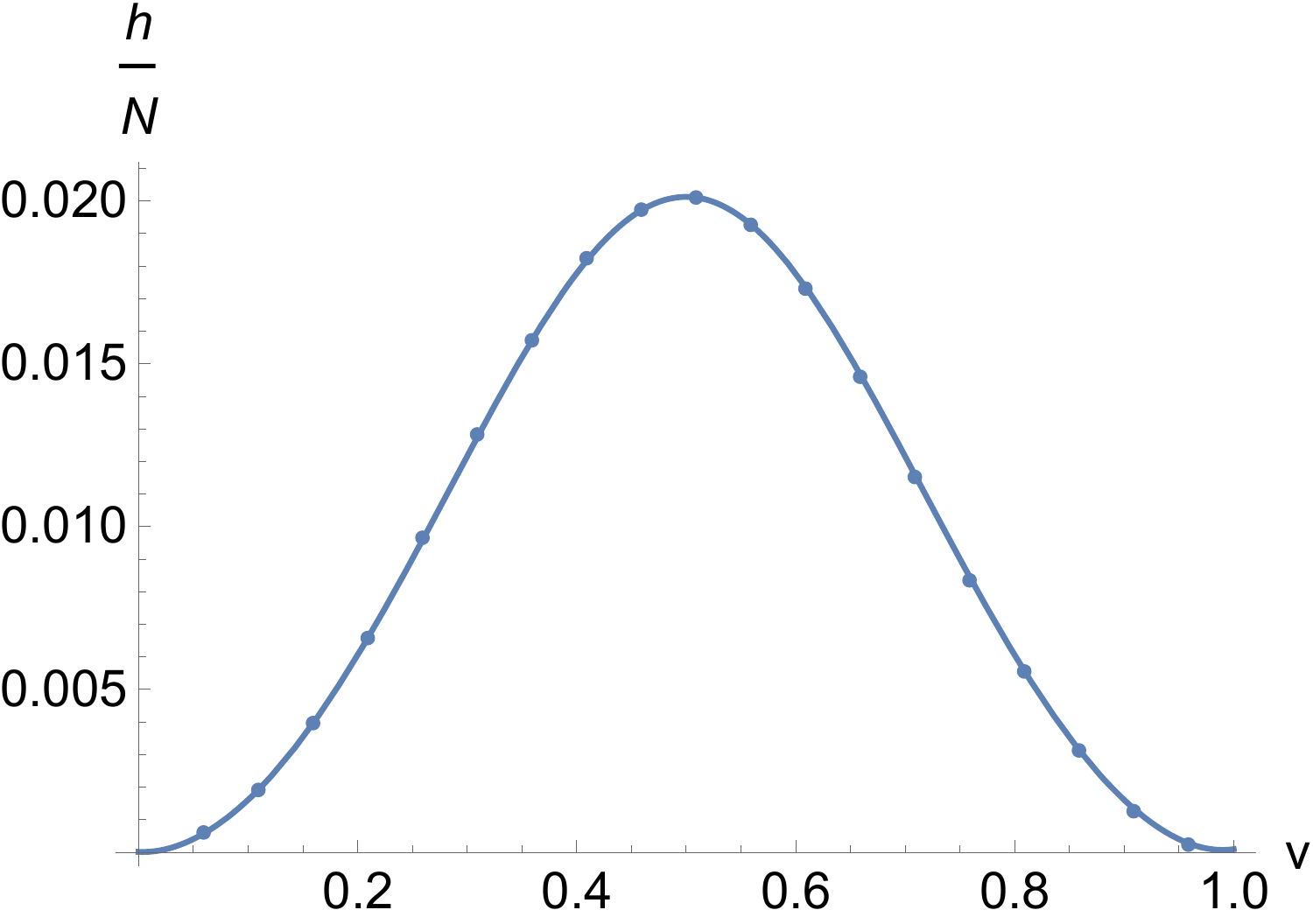}
\caption{Numerical result and a smooth interpolation for conformal weight in $d = 3$.}
\label{ConformalWeight3d}
\end{figure}

Away from $d = 4$, we can still work numerically. For a given $d$, we now have three variables in the sum \eqref{SaddlePointBeta}, namely $\beta, v$ and $\sigma$. We are interested in calculating a derivative with $\beta$ at $\beta = 2 \pi$. So we choose three values of $\beta$ near $2 \pi$ as $\beta = \{ 2 \pi - 0.01, 2 \pi, 2 \pi + 0.01 \}$ and then calculate the sum in \eqref{SaddlePointBeta} over a range of values of $\sigma$ and $v$. We do an interpolation in $\sigma$ and find the root for several values of $v$ and all three values of $\beta$. So we have an analogue of figure \ref{FigureSigmaStar3d} but for three different values of $\beta$. We then use this saddle point solution for $\sigma^*$ to calculate the integral for free energy in \eqref{FreeEnergyIntBeta}. We finally calculate the conformal weight using \eqref{ConformalWeightFreeEnergyForm} where for the derivative, we use the numerical analogue 
\begin{equation}
h = - \frac{1}{\textrm{Vol} (H^{d - 1})} \frac{2 \pi}{(d - 1)} \frac{ F_{\textrm{twisted}} (2 \pi + 0.01) - F_{\textrm{twisted}} (2 \pi - 0.01)}{0.02}.
\end{equation}
We plot the result in $d = 3$ in figure \ref{ConformalWeight3d}. It is positive in accordance with the conjecture made in \cite{Lemos:2017vnx}. 

\section{Monodromy defect in $d = 4 - \epsilon$} \label{SectionEpsilon}
In this section, we study monodromy defect in the critical $O(2N)$ model described in \eqref{ActionO2NEpsilon} in a perturbation theory near $4$ dimensions. In $d = 4 - \epsilon$, there is a fixed point with the fixed point value of the coupling constant given by 
\begin{equation}
\lambda^* = \frac{8 \pi^2}{2 N + 8} \epsilon.
\end{equation}
We will compute defect CFT observables at this fixed point. To start with, let's calculate 
the twisted free energy on the hyperbolic space. To leading order in $\lambda$, it is given by 
\begin{equation}
\begin{split}
F_{\textrm{twisted}} &= N F^{\textrm{free}}_{\textrm{twisted}} + \frac{\lambda}{4} \int d^d x \sqrt{g_x} \langle (\bar{\Phi}_I \Phi^I (x))^2  \rangle \\
& = N F^{\textrm{free}}_{\textrm{twisted}} + \frac{\lambda N (N + 1)}{4} 2 \pi \textrm{Vol} (H^3) (C_1^{\bar\Phi \Phi})^2
\end{split}
\end{equation}
where $F^{\textrm{free}}_{\textrm{twisted}}$ is the free energy of a single free complex scalar in the presence of the twist defect. Working to first order in $\lambda$, we only need $C_1^{\bar\Phi \Phi}$ in the free theory \eqref{OnePointPhibarPhi}. Using that in $d = 4$ and plugging in the fixed point value of the coupling gives
\begin{equation} \label{TwistedFreeEpsilon}
F_{\textrm{twisted}} = N F^{\textrm{free}}_{\textrm{twisted}} + \frac{ \epsilon N (N + 1)\textrm{Vol} (H^3) \left( 1- v \right)^2 v^2 }{8 \pi (N + 4)}.
\end{equation}
This is consistent with the large $N$ result \eqref{TwistedFreeLargeNEps}. Using this, we can compute expectation value of the defect defined in \eqref{DefectExpectationHyperbolic}, by subtracting the free energy of the untwisted energy from the above expression. Note that in the theory without the twist defect, leading correction to the free energy is of order $\lambda^2$, because the one-point functions vanish. So at this order
\begin{equation}
\begin{split}
&-\log \langle \mathcal{D} \rangle = F_{\textrm{twisted}} - F_{\textrm{untwisted}} = - N  \log \langle \mathcal{D} \rangle^{\textrm{free}} + \frac{ \epsilon N (N + 1)\textrm{Vol} (H^3) \left( 1- v \right)^2 v^2 }{8 \pi (N + 4)} \\
&=   \frac{N \textrm{Vol} (H^{3- \epsilon})\left( 1- v \right)^2 v^2 }{12 \pi} + \frac{  \textrm{Vol} (H^3) \epsilon N (N + 1) \left( 1- v \right)^2 v^2 }{8 \pi (N + 4)} - \frac{\textrm{Vol} (H^3) \epsilon N}{4 \pi^2} \\
&\times  \int_{0}^{\infty} d \nu \nu ^2 \left(H^{-i \nu }+\psi(i \nu +1)-2-\log \pi\right) \log \left(\frac{\csch^2(\pi  \nu )}{2} (\cosh (2 \pi  \nu )-\cos (2 \pi v ))\right)  
\end{split}
\end{equation} 
where we expanded the free theory result in \eqref{DefectExpectationHyperbolic} in $d = 4 - \epsilon$. 

We can proceed in the same way to calculate the conformal weight of the defect, defined in \eqref{ConformalWeightFreeEnergy}. It can be calculated from the free energy on $H^{d-1} \times S^1$ as in \eqref{ConformalWeightFreeEnergyForm} by first keeping the length of $S^1$, $\beta$, to be arbitrary and then taking a derivative with respect to $\beta$ at $\beta = 2 \pi$. Generalizing what we did above to arbitrary $\beta$ gives
\begin{equation}
F_{\textrm{twisted}} = N F^{free}_{\textrm{twisted}} + \frac{\lambda N (N + 1)}{4} \beta \textrm{Vol} (H^3) (C_1^{\bar{\Phi} \Phi}(\beta))^2
\end{equation}
Since $\lambda$ is order $\epsilon$, we only need to do the calculation of one-point function of $\bar{\Phi} \Phi$ as a function of $\beta$ in $d = 4$. As in \eqref{OnePointHyperboliFree}, we can calculate this one-point function coefficient by taking a mass derivative of the free energy
\begin{equation}
\begin{aligned}
C_1^{\bar{\Phi} \Phi}(\beta)=&\frac{1}{N}\frac{2}{\beta \textrm{Vol}(H^{d-1})}\frac{\partial F_{\textrm{twisted}}}{\partial m^2}\bigg|_{m^2 = 0}\\
=&\frac{2\Gamma\left(\frac{3-d}{2}\right)}{\beta(4\pi)^{\frac{d-1}{2}}}\sum_{k=-\infty}^{\infty}\frac{\Gamma \left(\frac{d}{2}-1+\frac{2 \pi  }{\beta }| k+v |\right)}{\Gamma \left(2-\frac{d}{2}+\frac{2 \pi  }{\beta }| k+v |\right)}
\end{aligned}
\end{equation}
To do the sum we first expand the sum in general $d$ in large $k$ and isolate the pieces that diverge as $k \rightarrow \infty$. We compute the sum for the divergent pieces keeping $d$ arbitrary, and then analytically continue to $d=4$. The finite piece starts contributing at order $\epsilon$ in $d = 4 - \epsilon$, so we do not need to add it here. The end result is
\begin{equation}
C_1^{\bar{\Phi} \Phi}(\beta) =
\frac{6 (v -1) v +1}{6 \beta ^2}-\frac{1}{24 \pi ^2}.
\end{equation}
We can then use \eqref{ConformalWeightFreeEnergyForm}, to find the the conformal weight in $d = 4 - \epsilon$ 
\begin{equation} \label{ConformalWeightEpsilon}
\begin{split}
h &= - \frac{N \pi \Gamma \left(\frac{1 - d}{2}\right)(1 -v ) v  \left(\csc \pi \left(\frac{d}{2} - v \right)-\csc \pi \left(\frac{d}{2} + v \right) \right) }{ d   (4 \pi)^{\frac{d - 1}{2}}  \Gamma \left( 2 - \frac{d}{2}- v \right) \Gamma \left(1 - \frac{d}{2}+ v \right)} \\
&+ \frac{N (N + 1)(1 - v) v (9 (1 - v)v - 2) }{72 \pi  (N+4)}\epsilon \\
& =  \frac{N v^2 (1- v)^2}{12 \pi} + \frac{\epsilon N (1 - v)^2 v^2 }{{144 \pi }}   \left(-6 H^{-v-1}-6 \psi(v-1)+37+6 \log (\pi ) - \frac{4}{v (1 - v)}\right) \\
&- \frac{3 \epsilon N (1 - v) v (9 (1 - v)v - 2) }{72 \pi  (N+4)}.
\end{split}
\end{equation}
This is also consistent with the large $N$ expansion results \eqref{ConformalWeightLargeNEps}. 

\subsection{Defect two-point function}
In this subsection, we calculate the two-point function of the defect fields $\Psi_s^I$ induced by the bulk field $\Phi^I$ on the defect. In the free theory, it is just given by the $N$ field generalization of \eqref{DefectTwoPointFree}
\begin{equation} 
\langle \bar{\Psi}_{I \ s_1}(\vec{y}_1) \Psi^J_{s_2} (\vec{y}_2) \rangle_0 =  \frac{ {\delta_I}^{J} \delta_{s_1, s_2} \mathcal{C}_{\Delta_{s_1}}}{(\vec{y}^2_{12})^{\Delta_{s_1}} }, \hspace{1cm} \mathcal{C}_{\Delta_{s_1}} = \frac{\Gamma \left(\Delta_{s_1} \right)}{2 \pi^{d/2} \Gamma \left(\Delta_{s_1} + 2 - \frac{d}{2} \right)}. 
\end{equation}
In the interacting theory, it gets corrected by the bulk one-loop Witten diagram
\begin{equation}
\langle \bar{\Psi}_{I \ s_1}(\vec{y}_1) \Psi^J_{s_2} (\vec{y}_2) \rangle_1   \ = \ \  \vcenter{\hbox{\includegraphics[scale=0.4]{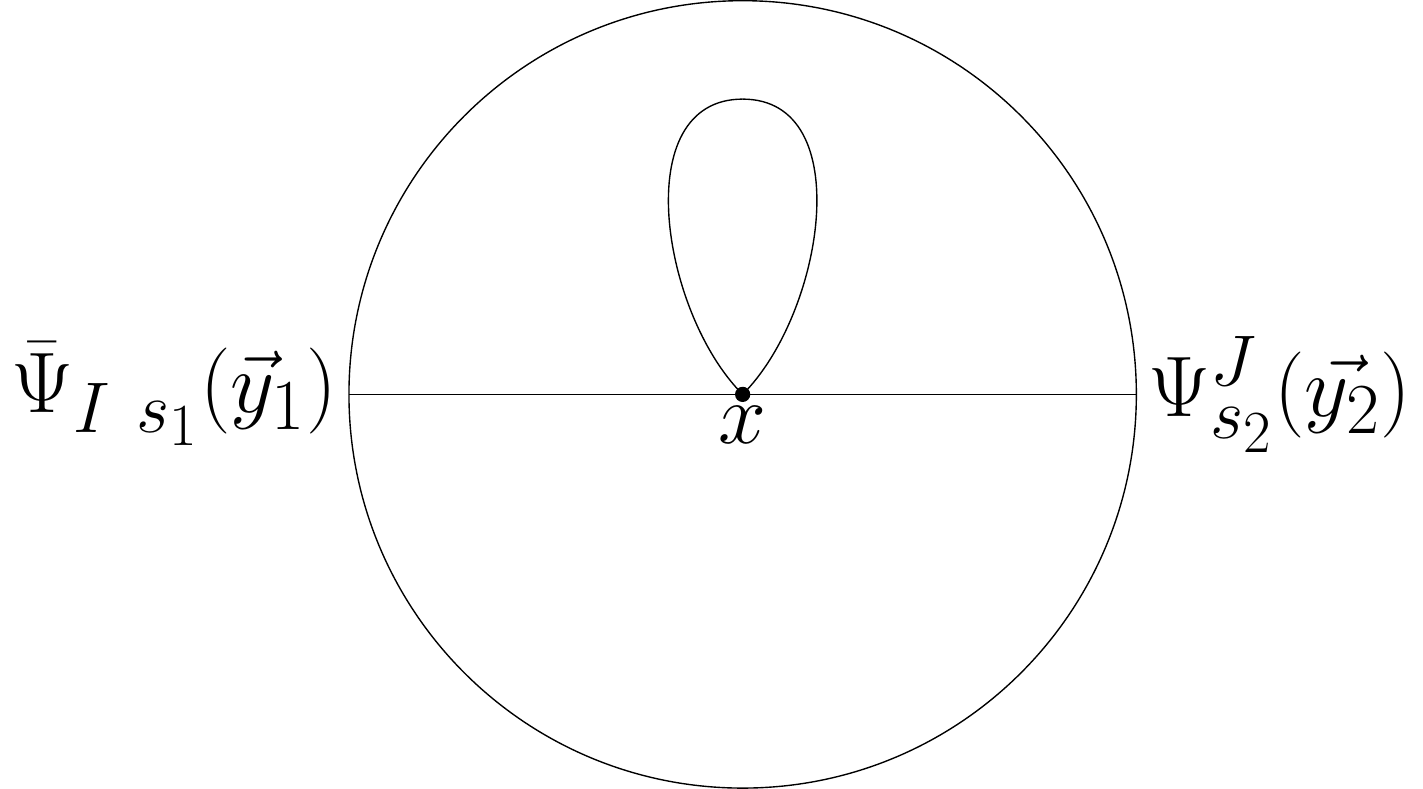}}}.
\end{equation}
The integral involved is
\begin{equation}
\langle \bar{\Psi}_{I \ s_1}(\vec{y}_1) \Psi^J_{s_2} (\vec{y}_2) \rangle_1 = - \pi \lambda (N + 1) {\delta_I}^{J} \delta_{s_1, s_2} C_1^{\bar{\Phi} \Phi} \int_{H^{d - 1}} (\mathcal{C}_{\Delta_{s_1}}  K_{\Delta_{s_1}}  ) (\mathcal{C}_{\Delta_{s_2}}  K_{\Delta_{s_2}}  )   
\end{equation}
where we already performed the bulk integral over $S^1$. The bulk-boundary propagator is normalized as 
\begin{equation}
 K_{\Delta_{s_i}} (\vec{y}_i, \vec{y}, r) = \left( \frac{r}{r^2 + (\vec{y}_i - \vec{y})^2} \right)^{\Delta_{s_i}}.
\end{equation}
Plugging in $d = 4$, the integral may be written as
\begin{equation}
\begin{split}
\langle \bar{\Psi}_{I \ s_1}(\vec{y}_1) \Psi^J_{s_2} (\vec{y}_2) \rangle_1 &= - \frac{\lambda (N + 1) {\delta_I}^{J} \delta_{s_1, s_2}  (C_1^{\bar{\Phi} \Phi}) }{4 \pi^3}  \int   \frac{ r^{2 \Delta_{s_1} - 3 + \eta} d r d^2 \vec{y}}{(r^2 + (\vec{y} - \vec{y}_1)^2)^{\Delta_{s_1}}(r^2 + (\vec{y} - \vec{y}_2)^2)^{\Delta_{s_2}} } \\
&=- \frac{\lambda (N + 1) {\delta_I}^{J} \delta_{s_1, s_2}  (C_1^{\bar{\Phi} \Phi})}{4 \pi^3}  \int   \frac{ r^{2 |s_1| - 1 + \eta} d r d^2 \vec{y}}{(r^2 + (\vec{y} - \vec{y}_1)^2)^{|s_1| + 1}(r^2 + (\vec{y} - \vec{y}_2)^2)^{|s_2| + 1} }
\end{split}
\end{equation} 
where we added a regulator $\eta$ to regulate the $r \rightarrow 0 $ divergence on the boundary of hyperbolic space. The integral over $\vec{y}$ can be performed by introducing Feynman parameter $\alpha$. The remaining integral is
\begin{equation}
\begin{split}
\langle \bar{\Psi}_{I \ s_1}(\vec{y}_1) \Psi^J_{s_2} (\vec{y}_2) \rangle_1 &= - \frac{\lambda (N + 1){\delta_I}^{J} \delta_{s_1, s_2}  (C_1^{\bar{\Phi} \Phi})  \Gamma(1 + 2 |s_1|) }{4 \pi^2 \Gamma(1 + |s_1|)^2}  \\
& \times \int_0^1 d \alpha \int d r \frac{(\alpha(1 - \alpha))^{|s_1|} r^{2 |s_1| - 1 + \eta}}{(r^2 + \alpha (1 - \alpha) \vec{y}_{12}^2)^{1 + 2 |s_1|}} \\
& = - \frac{\lambda (N + 1) {\delta_I}^{J} \delta_{s_1, s_2}  (C_1^{\bar{\Phi} \Phi})  }{4 \pi^2 |s_1| (\vec{y}_{12}^2)^{1 +  |s_1|}} \left( \frac{2}{\eta} + \log \left(\vec{y}_{12}^2 \right) - \frac{1}{|s_1|}  \right) \\
&= \frac{\epsilon (N + 1) v (1 - v) {\delta_I}^{J} \delta_{s_1, s_2} }{4 \pi^2 (N + 4) |s_1| (\vec{y}_{12}^2)^{1 +  |s_1|} } \left( \frac{2}{\eta} + \log \left(\vec{y}_{12}^2 \right) - \frac{1}{|s_1|}  \right)
\end{split}
\end{equation}
up to terms that vanish as $\eta \rightarrow 0$. The anomalous dimension of the defect operators can be read off from the coefficient of the $\log$, which gives the corrected dimension
\begin{equation} \label{AnomalousDimensionTwoPointDef}
\Delta_{s} = 1 + |s| - \frac{\epsilon}{2} + \frac{v (v - 1) (N + 1)}{2 (N + 4) |s|} \epsilon.
\end{equation}
This result is consistent with the results from large $N$ expansion \eqref{DimensionsLargeNEps}. 

Let's also do this calculation when the KK mode with $s = v$ has a $\Delta^-_v$ boundary condition. First let us look at the correction to the two-point function of defect operators $\Psi_{s}$ with $s \neq v$. Then the integrals above remain the same, but the one-point function coefficient of $\bar{\Phi} \Phi$ changes. So the correction to the two-point function is 
\begin{equation}
\begin{split}
\langle \bar{\Psi}_{I \ s_1}(\vec{y}_1) \Psi^J_{s_2} (\vec{y}_2) \rangle^-_1 &= - \frac{\lambda (N + 1) {\delta_I}^{J} \delta_{s_1, s_2}  (C_1^{\bar{\Phi} \Phi})^-}{4 \pi^2 |s_1| (\vec{y}_{12}^2)^{1 +  |s_1|}} \left( \frac{2}{\eta} + \log \left(\vec{y}_{12}^2 \right) - \frac{1}{|s_1|}  \right) \\
&= - \frac{\epsilon (N + 1) v (1 + v) {\delta_I}^{J} \delta_{s_1, s_2} }{4 \pi^2 (N + 4) |s_1| (\vec{y}_{12}^2)^{1 +  |s_1|} } \left( \frac{2}{\eta} + \log \left(\vec{y}_{12}^2 \right) - \frac{1}{|s_1|}  \right)
\end{split}
\end{equation}
where we used the value of the one-point function of $\bar{\Phi} \Phi$ in the free theory with $\Delta^-$ boundary condition \eqref{OnePointPhibarPhiMinus}. This gives us the dimension of the $s \neq v$ operators to leading order in $\epsilon$ 
\begin{equation} \label{AnomalousDimensionTwoPointDefMinOther}
\Delta^-_{s} = 1 + |s| - \frac{\epsilon}{2} + \frac{v (1 + v) (N + 1)}{2 (N + 4) |s|} \epsilon.
\end{equation}
For the $s = v$ operator, the leading correction to the two-point function is given by 
\begin{equation}
\begin{split}
\langle \bar{\Psi}_{I v}(\vec{y}_1) \Psi^J_{v} (\vec{y}_2) \rangle^-_1 &= - \frac{\lambda (N + 1) {\delta_I}^{J} (C_1^{\bar{\Phi} \Phi})^- }{4 \pi^3}  \int   \frac{ r^{-2 v - 1 + \eta} d r d^2 \vec{y}}{(r^2 + (\vec{y} - \vec{y}_1)^2)^{1 - v}(r^2 + (\vec{y} - \vec{y}_2)^2)^{1 - v} } \\
&=  \frac{\lambda (N + 1) {\delta_I}^{J} (C_1^{\bar{\Phi} \Phi})^- }{4 \pi^2 v (\vec{y}_{12}^2)^{1 - v}} \left( \frac{2}{\eta} + \log \left(\vec{y}_{12}^2 \right) + \frac{1}{v}  \right) \\
&=  \frac{\epsilon (N + 1) (1 + v) {\delta_I}^{J} }{4 \pi^2 (N + 4) (\vec{y}_{12}^2)^{1 -  v} } \left( \frac{2}{\eta} + \log \left(\vec{y}_{12}^2 \right) + \frac{1}{v}  \right).
\end{split}
\end{equation}
This gives the corrected dimension of the $s = v $ operator 
\begin{equation} \label{AnomalousDimensionTwoPointDefMinv}
\Delta^-_{v} = 1 - v - \frac{\epsilon}{2} - \frac{(1 + v) (N + 1)}{2 (N + 4) } \epsilon.
\end{equation}
These results are also consistent with the large $N$ expansion  calculation \eqref{DimensionsLargeNEpsMinus}.

\subsection{Bulk two-point function} \label{SectionBulkTwoPoint}
We now calculate the correction to the bulk two-point function of $\Phi$. In the bulk OPE limit, when the two fields are close to each other, we expect the following behavior 
\begin{equation}
\langle \bar{\Phi}_I(x_1) \Phi^J (x_2) \rangle = \frac{\Gamma\left( \frac{d}{2} - 1 \right)}{2 \pi^{\frac{d}{2}}} \frac{1}{|x_{12}|^{d - 2}} + {C_{\bar{\Phi};{\Phi}};}^{\bar{\Phi} \Phi} (C_1^{\bar{\Phi} \Phi}) + ... 
\end{equation}
up to terms subleading in $x_{12} = x_1 - x_2$. We will calculate the bulk one-point function $C_1^{\bar{\Phi} \Phi}$ to leading order in the perturbation theory in $\lambda$. In the free theory, the two-point function is given by \eqref{TwoPointFunctionFlat}. The leading correction in the interacting theory is given by a tadpole diagram in $H^{d-1} \times S^1$
\begin{equation}
\langle \bar{\Phi}_I(x_1) \Phi^J (x_2) \rangle_1 = - \frac{\lambda (N + 1) (C_1^{\bar{\Phi} \Phi}) {\delta_{I}}^{J} }{2} \int d^d x \sqrt{g_x} \ G _{\bar{\Phi} \Phi} (x_1, x) G _{\bar{\Phi} \Phi} (x, x_2) .
\end{equation}
To leading order in $\lambda$, we only need the bulk propagator in  $d = 4$, 
\begin{equation}
G_{\bar{\Phi} \Phi} (x_1, x_2) = e^{i s \theta_{12}} \sum_{s \in \mathbb{Z} + v} \frac{(\xi(1 + \xi))^{- \frac{1}{2}}}{8 \pi^2  (\sqrt{\xi} + \sqrt{1 + \xi})^{2 \Delta_s - 2}}. 
\end{equation}  
Then following \cite{Gaiotto:2013nva}, we place the two operators at the same position on $S^1$, $(\theta_1 = \theta_2)$ and at the same distance from the defect $(r_1 = r_2 = r')$. We place them at a separation of $y'$ along the defect and parameterize their position as $ \vec{y}_1 = (y'/2,0)$ and $\vec{y}_2 = (- y'/2, 0)$. We can then do the integral over $S^1$ and are left with the following integral on $H^{d - 1}$
\begin{equation} \label{BulkTwoPointCorrSum}
\langle \bar{\Phi}_I(x_1) \Phi^J (x_2) \rangle_1 = - \frac{\lambda (N + 1) (C_1^{\bar{\Phi} \Phi}) {\delta_{I}}^{J} }{4 \pi^3} \sum_{s} \int \frac{d r d y d z}{r} \frac{(r')^2 (4 r' r)^{2 \Delta_s - 2}}{d_+ d_- e_+ e_- ((d_+ + d_-)(e_+ + e_-))^{2 \Delta_s - 2}}
\end{equation}
where the integral runs over the $H^{d-1}$ coordinated $0 \le r < \infty$ and $-\infty < y,z < \infty$. As in \cite{Gaiotto:2013nva}, we defined
\begin{equation} \label{dpmDefinition}
d_{\pm} = \sqrt{ \left(y - \frac{y'}{2} \right)^2 + z^2 +  (r \pm r')^2}, \hspace{1 cm} e_{\pm} = \sqrt{\left(y + \frac{y'}{2} \right)^2 + z^2 + (r \pm r')^2}.
\end{equation}
We can then perform the sum over spins to get
\begin{equation} 
\begin{split}
\langle \bar{\Phi}_I(x_1) \Phi^J (x_2) \rangle &= - \frac{\lambda (N + 1) (C_1^{\bar{\Phi} \Phi}) {\delta_{I}}^{J} }{4 \pi^3} \times \\
&  \int \frac{d r d y  d z}{r} \frac{ (r')^2 (4 r' r)^{2 - 2 v} (d_+ + d_-)^{2 v} (e_+ + e_-)^{2 v} }{d_+ d_- e_+ e_- \left( (d_+ + d_-)^{2}(e_+ + e_-)^{2} - (4 r' r)^2 \right)} + (v \rightarrow 1 - v).
\end{split}
\end{equation}
We are interested in extracting the one-point function of the operator $\bar{\Phi} \Phi$ from this two-point function. For that purpose, we define $\mu = y'/r'$ and then look at the integral in the limit $\mu \rightarrow 0$. Such a analysis was done in \cite{Gaiotto:2013nva} for $v = 1/2$ and can be repeated for arbitrary values of $v$. We do this in appendix \ref{AppendixBulkTwoPoint}, and here just report the result \eqref{AppendixresultTwoPoint}
\begin{equation}
\langle\bar{\Phi}_I(x_1) \Phi^J (x_2) \rangle_1
= - \frac{\lambda (N + 1) (C_1^{\bar{\Phi} \Phi}) {\delta_{I}}^{J}}{8 \pi^2}  \left( -2 \log \mu -  \left(H^{1 - v}+H^{v}\right)+ \frac{1}{v (1 - v)} \right)
\end{equation}
where $H^n$ is the $n$-th harmonic number. The constant, $\mu$ independent, piece above contributes to the one-point function coefficient of $\bar{\Phi} \Phi$. Combining this with the free theory result \eqref{OnePointPhibarPhi} in $d = 4- \epsilon$ gives the result  \footnote{Note that we work in a normalization such that ${C_{\bar{\Phi};{\Phi}};}^{\bar{\Phi} \Phi} = 1$ in the free theory of a single complex scalar.} 
\begin{equation}
\begin{split}
{C_{\bar{\Phi};{\Phi}};}^{\bar{\Phi} \Phi} (C_1^{\bar{\Phi} \Phi}) &= \frac{(v-1) v}{4 \pi ^2} +   \frac{(v-1) v (\gamma +3+\log \pi ) + 1 }{8 \pi ^2} \epsilon \\
& +  \frac{3 \ \epsilon \ v (v - 1)}{8 \pi^2 (N + 4)}  \left(-  \left(H^{1 - v}+H^{v}\right)+ \frac{1}{v (1 - v)}\right)
\end{split}
\end{equation}
At large $N$, this gives 
\begin{equation} \label{OPEOnePointPhibarPhi}
{C_{\bar{\Phi};{\Phi}};}^{\bar{\Phi} \Phi} (C_1^{\bar{\Phi} \Phi}) = \frac{(v-1) v}{4 \pi ^2}  + \frac{(v-1) v (\gamma +3+\log \pi ) + 1 }{8 \pi ^2} \epsilon
\end{equation}
consistent with \eqref{OnePointLargeNEps}. For general $N$, but for the $Z_2$ twist defect with $v = 1/2$, this gives 
\begin{equation}
{C_{\bar{\Phi};{\Phi}};}^{\bar{\Phi} \Phi} (C_1^{\bar{\Phi} \Phi}) = -\frac{1}{16 \pi^2} + \frac{\epsilon}{32 \pi^2} (1 - \gamma - \log \pi) - \frac{3 \epsilon}{8 \pi^2 (N + 4)} \log 2.
\end{equation}
This is the result for $O(2 N)$ model, which should give the Ising result for $N = 1/2$, and it matches with the result in \cite{Gaiotto:2013nva} for that case up to a difference in normalization. Using result for ${C_{\bar{\Phi};{\Phi}};}^{\bar{\Phi} \Phi}$ from \eqref{BulkOPECoefficient}, we get 
\begin{equation} \label{PhiPhiOnePointepsilon}
\begin{split}
C_1^{\bar{\Phi} \Phi} &= \frac{ N(v-1) v}{4 \pi ^2} +    \frac{N\left((v-1) v (\gamma +4+\log \pi ) + 1 \right)}{8 \pi ^2} \epsilon   \\
& +  \frac{3 \epsilon N  v (v - 1)}{8 \pi^2 (N + 4)}  \left(-  \left(H^{1 - v}+H^{v}\right)+ \frac{1}{v (1 - v)} - 1\right).
\end{split}
\end{equation}
There is an extra factor of $N$ in the free part compared to \eqref{OnePointPhibarPhi} because we have $N$ complex fields \footnote{We hope that using the same symbol, $C_1^{\bar{\Phi} \Phi}$, for the one-point function of $\bar{\Phi} \Phi$ in both single field and $N$ field case does not lead to confusion.}.

Let's also discuss how the calculation changes when we use $\Delta^-$ boundary condition for the $s = v$ mode. We now have to use the $(C_1^{\bar{\Phi} \Phi})^-$ from \eqref{OnePointPhibarPhiMinus} for the free theory one-point function. And when we perform the sum in \eqref{BulkTwoPointCorrSum}, we need to use $\Delta^-$ for the $s = v$ mode. So the correction due to interactions, to the bulk two-point function is given by 

\begin{equation}
\begin{split}
&\langle \bar{\Phi}_I(x_1) \Phi^J (x_2) \rangle_1^-
= - \frac{\lambda (N + 1) (C_1^{\bar{\Phi} \Phi})^- {\delta_{I}}^{J}}{8 \pi^2}  \left( -2 \log \mu -  \left(H^{1 - v}+H^{v}\right)+ \frac{1}{v (1 - v)} \right) \\
& - \frac{\lambda (N + 1) (C_1^{\bar{\Phi} \Phi})^- {\delta_{I}}^{J} }{4 \pi^3}  \int \frac{d r d y d z}{r} \frac{(r')^2 }{d_+ d_- e_+ e_- } \left( \frac{(4 r' r)^{- 2 v}}{((d_+ + d_-)(e_+ + e_-))^{- 2 v}} -  (v \rightarrow - v) \right).
\end{split}
\end{equation}
So we now have an additional integral to perform which we also perform in appendix \ref{AppendixBulkTwoPoint} with the result in \eqref{AppendixIntegralBulkTwoResM}. Overall, the result for the interacting correction to the bulk two-point is
\begin{equation}
\langle \bar{\Phi}_I(x_1) \Phi^J (x_2) \rangle_1^- = - \frac{\lambda (N + 1) (C_1^{\bar{\Phi} \Phi})^- {\delta_{I}}^{J}}{8 \pi^2}  \left( -2 \log \mu -  \left(H^{1 - v}+H^{v}\right)+ \frac{2 v - 1}{v (1 - v)} \right).
\end{equation}
The $\mu$ independent piece is the contribution of the bulk interaction to the product of the bulk OPE coefficient and the one-point function of $\bar{\Phi} \Phi$. Adding it to the free theory result in $d = 4 - \epsilon$ \eqref{OnePointPhibarPhiMinus} gives
\begin{equation} \label{OPEOnePointPhibarPhiM}
\begin{split}
{C_{\bar{\Phi};{\Phi}};}^{\bar{\Phi} \Phi} (C_1^{\bar{\Phi} \Phi})^{-} &= \frac{(v+1) v}{4 \pi ^2} +   \frac{(v+1) v (\gamma +3+\log \pi ) + 1 }{8 \pi ^2} \epsilon \\
& +  \frac{3 \ \epsilon \ v (v + 1)}{8 \pi^2 (N + 4)}  \left(-  \left(H^{1 - v}+H^{v}\right)+ \frac{2 v - 1}{v (1 - v)}\right).
\end{split}
\end{equation}
In the large $N$ limit, the leading order in $N$ piece agrees with the result from the large $N$ calculation \eqref{OnePointLargeNEpsM}.

\subsection{Coefficient of displacement} \label{SectionDisplacementCoefficient}
In this subsection, we calculate the coefficient of two-point function of the displacement operator to leading order in $\epsilon$. To do that, we  use the fact that the displacement appears in the bulk-defect OPE  of $\bar{\Phi} \Phi$ as we saw in the free theory \eqref{DisplacementPhiPhi}. So we need to now look at corrections to the bulk two-point function of $\bar{\Phi} \Phi$. In the free theory, it is given by 
\begin{equation}
\langle \bar{\Phi}_I \Phi^I(x_1) \bar{\Phi}_J \Phi^J (x_2) \rangle_0 = N (G_{\bar{\Phi} \Phi} (x_1, x_2))^2.
\end{equation}
It gets corrected in the interacting theory, and the leading correction is given by 
\begin{equation}
\begin{split}
\langle \bar{\Phi}_I \Phi^I(x_1) \bar{\Phi}_J \Phi^J (x_2) \rangle_1 &=  - \frac{\lambda N (N + 1) }{2} \int d^d x \sqrt{g_x} \ G^2 _{\bar{\Phi} \Phi} (x_1, x) G^2_{\bar{\Phi} \Phi} (x, x_2) \\
& = - \sum_{s_1, s_2, s_3 s_4} \frac{\lambda N (N + 1) }{2 (8 \pi^2)^4} \int d^d x \sqrt{g_x} e^{i (s_1 - s_2) (\theta_1 - \theta) + i (s_3 - s_4) (\theta_2 - \theta )} \\
& \times \frac{ \xi^{-1}_{x_1 x} (1 + \xi_{x_1 x})^{-1} \xi_{x_2 x}^{-1} (1 + \xi_{x_2 x})^{-1} }{ (\sqrt{\xi_{x_1 x}} + \sqrt{ 1 + \xi_{x_1 x}})^{2(|s_1| + |s_2|) } (\sqrt{\xi_{x_2 x}} + \sqrt{ 1 + \xi_{x_2 x}})^{2(|s_3| + |s_4|) }}.  
\end{split}
\end{equation}
The integral over $\theta$ gives a delta function $\delta(s_1 - s_2 + s_3 - s_4)$. The remaining sum over spins then captures the contribution of various defect operators appearing in the bulk defect OPE of $\bar{\Phi} \Phi$. Let us then isolate the contribution of the displacement operator, which appears for $s_1 =  s_4 = v$ and $s_2 = s_3 = -1 + v$
\begin{equation} \label{DisplacementIntegral}
\begin{split}
\langle \bar{\Phi}_I \Phi^I(x_1) \bar{\Phi}_J \Phi^J (x_2) \rangle_1 & \ni  -  \frac{\lambda N (N + 1) }{2 (8 \pi^2)^4} \int  \frac{ d^{d - 1} x \sqrt{g_x} \ e^{i \theta_{12} } \xi^{-1}_{x_1 x} (1 + \xi_{x_1 x})^{-1} \xi_{x_2 x}^{-1} (1 + \xi_{x_2 x})^{-1} }{ (\sqrt{\xi_{x_1 x}} + \sqrt{ 1 + \xi_{x_1 x}})^{2} (\sqrt{\xi_{x_2 x}} + \sqrt{ 1 + \xi_{x_2 x}})^{2}} \\
&\ni -  \frac{\lambda N (N + 1) }{ \pi^7} \int dr dy dz \frac{  (r')^6 r^3  }{ d_+^2 d_-^2 e_+^2 e_-^2 (d_+ + d_-)^2(e_+ + e_-)^2}.
\end{split}
\end{equation}
To go from the first line to the second line, we fixed the position of operators $x_1,x_2$ as written for the bulk two-point function of $\Phi$ above \eqref{BulkTwoPointCorrSum}. $d_{\pm}, e_{\pm}$ are defined in the same way as before \eqref{dpmDefinition}. But now, we are interested in the contribution of the defect operator, so we define $\kappa = r'/ y' = 1/ \sqrt{4 \xi_{x_1 x_2}}$ and look at the integral in the limit $\kappa \rightarrow 0$. We do this in the appendix \ref{AppendixDisplacement}, with the result being \eqref{AppendixResultDisplacement}  
\begin{equation} \label{DisplacementPhiPhiEps1}
\begin{split}
\langle \bar{\Phi}_I \Phi^I(x_1) \bar{\Phi}_J \Phi^J (x_2) \rangle_1 &\ni \frac{\lambda N (N + 1) \kappa^6 }{ 16 \pi^6} \left(  \log \kappa + \frac{1}{4} \right) \\
&\ni \frac{\epsilon N (N + 1)  }{ 8 \pi^4 (N + 4) (4 \xi)^3} \left( - \log (4\xi) + \frac{1}{2} \right) .
\end{split}
\end{equation}
We should combine this result with the result in \eqref{DisplacementPhiPhi} in $d = 4 -\epsilon$. Note that the result in the first line of \eqref{DisplacementPhiPhi} is written in terms of $\Delta_s$ which also gets corrected by the interactions \eqref{AnomalousDimensionTwoPointDef}, so we need to use the corrected $\Delta_s$. This gives, to leading order in $\epsilon$  
\begin{equation} \label{DisplacementPhiPhiEps0}
\begin{split}
\langle \bar{\Phi}_I \Phi^I(x_1) \bar{\Phi}_J \Phi^J (x_2) \rangle_0 &\ni \frac{N}{4 \pi^4 (4 \xi)^{3 - \epsilon}} +  \frac{\epsilon N (N + 1)  }{ 8 \pi^4 (N + 4) (4 \xi)^3} \log (4 \xi)  \\
&- \frac{\epsilon N }{ 8 \pi^4 (4 \xi)^3} (\psi(1 + v) + \psi (2 - v) - 2 \log \pi)
\end{split}
\end{equation}
where $\psi(z)$ is the polygamma function. Adding \eqref{DisplacementPhiPhiEps0} and \eqref{DisplacementPhiPhiEps1}, the $\log$ piece cancels, which implies that, as expected, the displacement has protected dimension $d - 1$. It gives the following result for the product of bulk-defect OPE coefficient and the displacement two-point function 
\begin{equation}
(C^{\bar{\Phi} \Phi}_{D})^2 C_D = \frac{N}{4 \pi^4} - \frac{\epsilon N }{ 8 \pi^4}  \left(\psi(1 + v) + \psi (2 - v) - \frac{1}{2} - 2 \log \pi \right) -  \frac{3 \epsilon N }{ 16 \pi^4 (N + 4)}.
\end{equation} 
Then, we can use the ward identity  \eqref{WardIdentity}
\begin{equation}
\Delta_{\bar{\Phi} \Phi} C^{\bar{\Phi} \Phi}_1 = -  \left(\frac{\pi}{4} \right)^{\frac{d}{2} - 1} \frac{\sqrt{\pi}}{\Gamma\left( \frac{d - 1}{2} \right)} C^{\bar{\Phi}  \Phi}_D C_D
\end{equation}
and the result for $C^{\bar{\Phi} \Phi}_1$ in \eqref{PhiPhiOnePointepsilon} to get 
\begin{equation}
\begin{split}
C_D &= \frac{4 v^2 (1 - v)^2 N}{\pi^2} \bigg[ 1 + \epsilon \left( \gamma + \frac{7}{4} + \log \pi - \frac{1}{v (1 - v)} + \frac{(H^{1-v}+H^v)}{2} \right)   \\
&+ \frac{ 3 \epsilon}{N + 4} \left(-  \left(H^{1 - v}+H^{v}\right)+ \frac{1}{v (1 - v)} - \frac{7}{4} \right)  \bigg].
\end{split}
\end{equation}
Note that we had to use the bulk result for $O(2 N )$ model \cite{PhysRevD.7.2911} 
\begin{equation}
\Delta_{\bar{\Phi} \Phi} = 2 -  \frac{3 \epsilon}{N + 4}.
\end{equation}

\subsection{Defect four-point function} \label{SectionDefectFourPoint}

In this subsection we calculate the four-point function of the defect operators $\Psi_{s}$ in the $d - 2$ dimensional defect CFT. We use this four-point function to study composite operators on the defect that appear in the corresponding conformal block decomposition. The composite operators can be, schematically, either of the form $\Psi _{s_1} (\vec{\partial}^2)^m (\vec{\partial})^l \bar{\Psi}_{s_2}$ or of form $\Psi _{s_1} (\vec{\partial}^2)^m (\vec{\partial})^l \Psi_{s_2}$. Each of these have free theory dimensions $\Delta_{s_1} + \Delta_{s_2} + 2 m + l$ and longitudinal spin $l$. For the former, the transverse $SO(2)$ spin is $s_1 - s_2$ and hence is an integer, while for the latter, it is $s_1 + s_2$ and hence, is fractional, of the form $\mathbb{Z} + 2 v$. We will study both these cases below. 

\subsubsection*{Integer spin operators on the defect}
Let's first look at the integer spin operators on the defect by considering the following four-point function in the $12 \rightarrow 34$ channel
\begin{equation}
\langle\Psi^I_{s_1}(\vec{y}_1) \bar{\Psi}_{J \ s_2}(\vec{y}_2) \Psi^K_{s_3}(\vec{y}_3) \bar{\Psi}_{L \ s_4}(\vec{y}_4) 
\rangle = {\delta^I}_J {\delta^K}_L \mathcal{G}_{\textrm{sing}}+ \left( {\delta^I}_L {\delta^J}_K - \frac{{\delta^I}_J {\delta^K}_L}{N} \right) \mathcal{G}_{\textrm{adj}}
\end{equation}
where we have decomposed the correlator into singlet and adjoint representations of $U(N)$. The operators that appear in this channel are of form $\Psi^I _{s_1} (\vec{\partial}^2)^m (\vec{\partial})^l \bar{\Psi}_{J \ s_2}$. Let's restrict to the longitudinal spin $l$ being  zero, because to the order we will consider, we will only be able to see anomalous dimensions for $l = 0$ operators. There will be mixing among operators which have the same dimension in the free theory and same $SO(2)$ spin
\begin{equation}
|s_1| + |s_2| + d - 2 + 2 m = |s_3| + |s_4| + d - 2 + 2 n, \hspace{1cm} s_1 - s_2 = s_4 - s_3.
\end{equation} 
To keep the analysis simple, we will also restrict to the case when $\{s_1,s_4\} > 0$ and $\{s_2, s_3\} < 0$ and then the equality of $SO(2)$ spins implies $m = n$, and hence there will only be mixing between operators having equal number of derivatives. So we are only considering composite operators of the form form $O^{s,m}_{\alpha} = \Psi_{\alpha + v}  (\vec{\partial}^2)^m \bar{\Psi}_{- s + \alpha + v}$  for positive integer $s$ and $\alpha = 0, 1,\hdots s - 1 $. These are all degenerate, and we expect this degeneracy to be lifted by interactions. So we expect the following form for their two-point function 
\begin{equation} \label{TwoPointCompositeI}
\langle O^{s,m}_{\alpha} (\vec{y}_1) \bar{O}^{s,m}_{\beta} (\vec{y}_2)  \rangle = \frac{1}{(\vec{y}_{12}^2)^{s + d - 2 + 2 m }} \left( \delta_{\alpha \beta} -  \epsilon \Delta ^{s, m}_{\alpha \beta} \log ( \vec{y}^2_{12}) \right).
\end{equation} 
Note that we have to properly normalize the operators such that their two-point function is unit normalized in the free theory. $\Delta ^{s, m}_{\alpha \beta}$ is the matrix of anomalous dimensions, and its eigenvalues give the anomalous dimensions of all the spin $s$ composite operators with $2 m$ derivatives. To extract this anomalous dimension from the four-point function, we note that in general, such a four-point function can be decomposed into conformal partial waves, or equivalently the conformal blocks \footnote{Note that all the irreducible representations of $U(N)$ will individually have this decomposition.}
\begin{equation}
\begin{split}
&\langle\Psi^I_{s_1}(\vec{y}_1) \bar{\Psi}_{J \ s_2}(\vec{y}_2) \Psi^K_{s_3}(\vec{y}_3) \bar{\Psi}_{L \ s_4}(\vec{y}_4) 
\rangle = \sum_{O} C_{\Psi_{s_1} \Psi_{s_2} O} {C^{O}}_{\Psi_{s_3} \Psi_{s_4}} W_{\Delta_O, l} (\vec{y}_i)\\
&W_{\Delta, l} (\vec{y}_i) =  \left(\frac{\vec{y}_{24}^2}{\vec{y}_{14}^2} \right)^{\frac{\Delta_{12}}{2}} \left(\frac{\vec{y}_{14}^2}{\vec{y}_{13}^2} \right)^{\frac{\Delta_{34}}{2}} \frac{G_{\Delta, l} (u,v)}{(\vec{y}_{12}^2)^{\frac{\Delta_{s_1} + \Delta_{s_2} }{2}} (\vec{y}_{34}^2)^{\frac{\Delta_{s_3} + \Delta_{s_4} }{2}}}  
\end{split}
\end{equation}
where $\Delta_{12} = \Delta_{s_1} - \Delta_{s_2}$ and the conformal block depends on the cross ratios 
\begin{equation}
u = \frac{\vec{y}_{12}^2 \vec{y}_{34}^2}{\vec{y}_{13}^2 \vec{y}_{24}^2}, \hspace{1cm} v = \frac{\vec{y}_{14}^2 \vec{y}_{23}^2}{\vec{y}_{13}^2 \vec{y}_{24}^2}. 
\end{equation}
The particular correlator we are considering then, in perturbation theory in $\epsilon$ will contain
\begin{equation}
\begin{split}
&\langle\Psi^I_{\alpha + v}(\vec{y}_1) \bar{\Psi}_{J \ \alpha - s  + v}(\vec{y}_2) \Psi^K_{- s + \beta + v}(\vec{y}_3) \bar{\Psi}_{L \ \beta + v}(\vec{y}_4) 
\rangle \ni \\
& \sum_{m} C_{\Psi_{s_1} \bar{\Psi}_{s_2} \bar{O}^{s,m}_{\alpha}} {C^{O^{s,m}_{\beta}}}_{\Psi_{s_3} \bar{\Psi}_{s_4}}  \left( \delta_{\alpha \beta} + \frac{\epsilon}{2} \Delta ^{s, m}_{\alpha \beta} \partial_m  \right) W_{d -2 + s + 2 m , 0}(\vec{y}_i).
\end{split}
\end{equation}

With these generalities in mind, we can go on and calculate the four-point function. At tree level, this correlator is just given by the disconnected pieces
\begin{equation}
\langle\Psi^I_{s_1}(\vec{y}_1) \bar{\Psi}_{J \ s_2}(\vec{y}_2) \Psi^K_{s_3}(\vec{y}_3) \bar{\Psi}_{L \ s_4}(\vec{y}_4) 
\rangle_0 = \mathcal{C}_{\Delta_{s_1}} \mathcal{C}_{\Delta_{s_3}}  \bigg[  \frac{{\delta^I}_J {\delta^K}_L \delta_{s_1,s_2} \delta_{s_3, s_4}}{(\vec{y}_{12}^2)^{\Delta_{s_1}} (\vec{y}_{34}^2)^{\Delta_{s_3}}} + \frac{{\delta^I}_L {\delta^J}_K  \delta_{s_1,s_4} \delta_{s_2,s_3}}{(\vec{y}_{14}^2)^{\Delta_{s_1}} (\vec{y}_{23}^2)^{\Delta_{s_3}}} \bigg].
\end{equation}
The first term above just represents the contribution of the identity operator and the second term represents the contribution of composite operators of the type we mentioned. To decompose it into the conformal blocks, we can use the following result from \cite{Fitzpatrick:2011dm}  
\begin{equation}
\frac{\delta_{s_1,s_4} \delta_{s_2,s_3}}{(\vec{y}_{14}^2)^{\Delta_{s_1}} (\vec{y}_{23}^2)^{\Delta_{s_3}}} = \delta_{s_1,s_4} \delta_{s_2,s_3}  \sum_{n,l} c_0^{(12)}(n,l) W_{\Delta = \Delta_{s_1} + \Delta_{s_2} + 2 n + l, l} (\vec{y}_i).
\end{equation} 
The squared OPE coefficients are given by
\begin{equation}
c_0^{(12)}(n,l) = \frac{\left(\Delta_{s_1} + 2 - \frac{d}{2} \right)_n \left(\Delta_{s_2} + 2 - \frac{d}{2} \right)_n \left(\Delta_{s_1} \right)_{l + n} \left(\Delta_{s_2} \right)_{l + n} \left(\Delta_{s_1} + \Delta_{s_2} + 2 n + 2 l - 1\right)_{-l} }{l! n! \left(l + \frac{d}{2}  - 1 \right)_{n} \left(\Delta_{s_1} + \Delta_{s_2} + n - d + 3\right)_n  \left(\Delta_{s_1} + \Delta_{s_2} + n + l - \frac{d}{2} + 1\right)_n }. 
\end{equation}
Note that the individual dimensions appearing in the subscript of partial wave $\Delta_{s_i}$ include corrections from \eqref{AnomalousDimensionTwoPointDef}. Also, the delta function constraint $\delta_{s_1,s_4} \delta_{s_2,s_3}$ is equivalent to $\delta_{\alpha \beta}$ so the disconnected piece only contributes to the diagonal part of the anomalous dimension matrix. This also immediately tells us that, in the $U(N)$ singlet sector for example 
\begin{equation}
 (C_{\Psi_{s_1} \bar{\Psi}_{s_2} \bar{O}^{s,m}_{\alpha}})^2   = \frac{c_0^{(12) } (m,0) \mathcal{C}_{\Delta_{s_1}} \mathcal{C}_{\Delta_{s_2}}}{N} = \frac{\left(\Delta_{s_1} \right)^2_m\left(\Delta_{s_2} \right)^2_m}{4 \pi^4 N (m!)^2 (s + m + 1)_m^2}
\end{equation}

The leading correction to it comes from the contact Witten diagram
\begin{equation}
\begin{split}
\langle\Psi^I_{s_1}(\vec{y}_1) \bar{\Psi}_{J \ s_2}(\vec{y}_2) \Psi^K_{s_3}(\vec{y}_3) \bar{\Psi}_{L \ s_4}(\vec{y}_4) 
\rangle_1 &= \ \vcenter{\hbox{\includegraphics[scale=0.3]{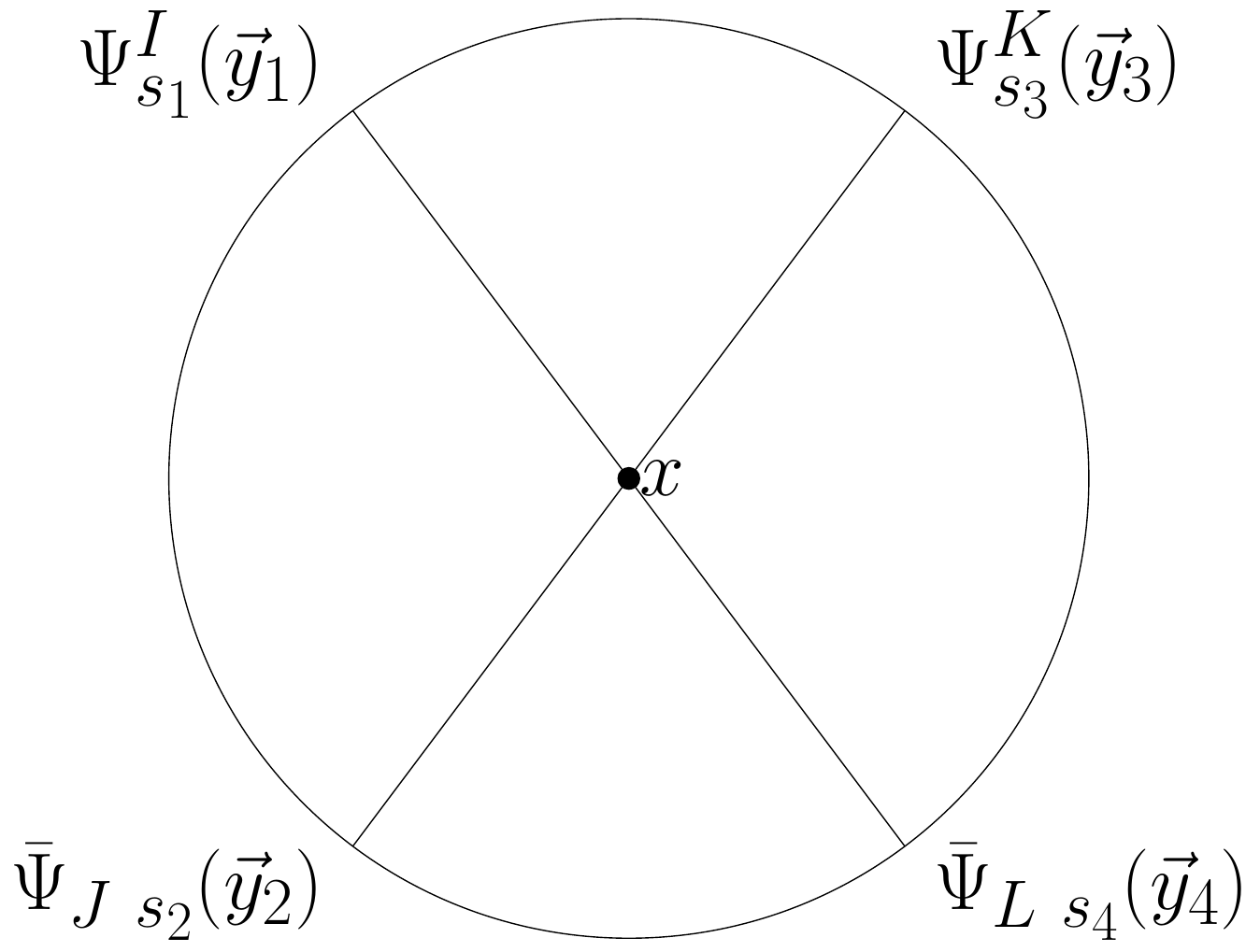}}} \\
&=  -  \pi \lambda ({\delta^I}_J {\delta^K}_L + {\delta^I}_L {\delta^J}_K) \delta(s_1 + s_3 - s_2 - s_4)  \\
&\times \int \prod_{i = 1}^4 \mathcal{C}_{\Delta_{s_i}} K_{\Delta_{s_i}} (\vec{y}_i, \vec{y}, r)
\end{split}
\end{equation}
where the remaining integral is only over $H^{d - 1}$.
The above integral can be evaluated in terms of the well known $D$-functions \cite{Liu:1998ty, DHoker:1999kzh, Dolan:2000ut}
\begin{equation}
\begin{split}
\langle\Psi^I_{s_1}(\vec{y}_1) \bar{\Psi}_{J \ s_2}(\vec{y}_2) \Psi^K_{s_3}(\vec{y}_3) \bar{\Psi}_{L \ s_4}(\vec{y}_4) 
\rangle_1 &= -  \pi \lambda ({\delta^I}_J {\delta^K}_L + {\delta^I}_L {\delta^J}_K) \delta(s_1 + s_3 - s_2 - s_4) \\
& \times \left( \prod_{i = 1}^4 \mathcal{C}_{\Delta_{s_i}} \right) D_{\Delta_{s_1} ,\Delta_{s_2},\Delta_{s_3},\Delta_{s_4}} (\vec{y}_i)
\end{split}
\end{equation}
The $D$-function has the following conformal block decomposition \cite{Hijano:2015zsa,Jepsen:2019svc} 
\begin{equation}
D_{\Delta_{s_1} ,\Delta_{s_2},\Delta_{s_3},\Delta_{s_4}} (\vec{y}_i) = \sum_m P_1^{(12)} (m,0) W_{\Delta_m, 0} (\vec{y}_i) + \sum_{n} P_1^{(34)} (n,0) W_{\Delta_n,0} (\vec{y}_i).
\end{equation}
The dimensions of the operators appearing are $\Delta_m = \Delta_{s_1} + \Delta_{s_2} + 2 m $ and $\Delta_n = \Delta_{s_3} + \Delta_{s_4} + 2 n $ and the squared OPE coefficients are given by 
\begin{equation}
\begin{split}
&P_1^{(12)} (m,0) =\\
& \frac{(-1)^m  \pi^{\frac{d}{2} - 1}(\Delta_{s_1})_m (\Delta_{s_2})_m \Gamma \left(\frac{ \Delta_{s_4} + \Delta_{s_3} + \Delta_m + 2 - d}{2} \right) \Gamma \left(\frac{ \Delta_{s_4} + \Delta_{s_3} - \Delta_m }{2} \right) \Gamma \left(\frac{ \Delta_{43} +  \Delta_m }{2} \right) \Gamma \left(\frac{ \Delta_{34} +  \Delta_m }{2} \right)}{2 m!  (\Delta_{s_1} + \Delta_{s_2} + m + 1 - \frac{d}{2})_m \Gamma \left(\Delta_{s_3} \right) \Gamma \left(\Delta_{s_4} \right) \Gamma \left(\Delta_m \right)}
\end{split} 
\end{equation} 
and similarly for $P_1^{(34)} (n,0)$. The coefficients clearly have a divergence when $\Delta_{s_1} +  \Delta_{s_2} - \Delta_{s_3} - \Delta_{s_4} = 2 k$, an even integer. This is equivalent to the condition that $|s_1| + |s_2| - |s_3| - |s_4| = 2 k$. Note that we also have the $s_1 + s_3 - s_2 - s_4 =0$ constraint coming from the delta function in the contact diagram. For the case we are considering when ${s_1,s_4} > 0$ and ${s_2, s_3} < 0$, the delta function constraint implies that $\Delta_{s_1} +  \Delta_{s_2} = \Delta_{s_3} + \Delta_{s_4}$. In that particular case, the above decomposition for the $D$-function becomes 

\begin{equation} \label{DFunctionDecomposition}
\begin{split}
&D_{\Delta_{s_1} ,\Delta_{s_2},\Delta_{s_3},\Delta_{s_4}} (\vec{y}_i) = \sum_{m} \frac{\pi \left( \prod_{i = 1}^4 (\Delta_{s_i})_m \right) (\Delta_{s_1} + \Delta_{s_2} + 2 m - 1)^2_{-m} }{ 2 ( m!)^2  (\Delta_{s_1} + \Delta_{s_2} + 2 m - 1)} \bigg( \frac{2}{\Delta_{s_1}+\Delta_{s_2}+2 m-1}    \\
& + 2 H^m - 2 H^{m+ \Delta_{s_1}+\Delta_{s_2}-2}+4 H^{2 m+\Delta_{s_1}+\Delta_{s_2}-2}- \sum_{i = 1}^4 H^{m+\Delta_{s_i}-1}  -  \partial_{m} \bigg) W_{ \Delta_{s_1} + \Delta_{s_2} + 2 m, 0} (\vec{y}_i).
\end{split}
\end{equation}
Notice that as we said before, to this order, the contact term only gives anomalous dimensions to $l = 0$ operators.

As we mentioned briefly above, in addition to the contact interaction, there is also a disconnected diagram that contributes to the anomalous dimension of $O^{s,m}_{\alpha}$. This just corresponds to the anomalous dimensions of the two individual $\Psi_{s_i}$ which make up $O^{s,m}_{\alpha}$. This disconnected piece just contributes to the matrix for $\alpha = \beta$. Combining the result from the corrections to the individual dimensions \eqref{AnomalousDimensionTwoPointDef}, and the contact diagram \eqref{DFunctionDecomposition}, we get the following anomalous dimension matrix in the singlet $U(N)$ sector
\begin{equation}
\Delta_{\alpha \beta}^{s, m} =
\begin{cases}
\frac{(N + 1)}{(N + 4) (s + 2 m + 1)} & \text{if $\alpha \neq \beta$} \\
 \frac{(N + 1)}{(N + 4) (s + 2 m + 1)} + \frac{v (v - 1) (N + 1) s}{ 2 (N + 4)(\alpha + v) (s - \alpha - v)} & \text{if $\alpha = \beta$}.
\end{cases}
\end{equation}
To get the dimensions of all the spin $s$ operators of this kind then, we have to find the eigenvalues of the above matrix and add that to the bare dimension $2 - \epsilon + s + 2m$.  
In particular, the displacement operator is non degenerate and corresponds to $|s_1| = v$ and $|s_2| = 1 - v$. So for the displacement, $s = 1, m = 0$ and $\alpha = \beta = 0$ and it is easy to see that it does not get anomalous dimension to this order. In general, we expect the displacement to stay protected to all orders in perturbation theory. Also, notice that the contribution of the contact interaction is the same order in $N$ as the contribution of the disconnected diagram. This is because we are talking about the singlet sector of $U(N)$. In order to calculate this correction in the large $N$ expansion, we have to know the correlator of $\sigma$ operator. We do not do it here. 

\subsubsection*{Fractional spin operators on the defect}
In order to analyze the operators of the form $
\Psi_{s_1} (\vec{\partial}^2)^m \partial^l \Psi_{s_2}$, we have to look at the above correlator in $13 \rightarrow 24$ channel. These operators have $SO(2)$ spin  $s = s_1 + s_2 \in \mathbb{Z} + 2 v$.  Alternatively, we find it more convenient to consider the following four-point function, and look at it in $12 \rightarrow 34$ channel
\begin{equation}
\langle\Psi^I_{s_1}(\vec{y}_1) \Psi^J_{s_2}(\vec{y}_2) \bar{\Psi}_{K \ s_3}(\vec{y}_3) \bar{\Psi}_{L \ s_4}(\vec{y}_4) 
\rangle = \left( \frac{{\delta^I}_K {\delta^J}_L + {\delta^I}_L {\delta^J}_K}{2} \right) \mathcal{G}_{\textrm{sym.}} + \left( \frac{ {\delta^I}_K {\delta^J}_L - {\delta^I}_L {\delta^J}_K}{2} \right) \mathcal{G}_{\textrm{ant}}
\end{equation}
and now we are looking at the operators in the tensor product of two fundamentals of $U(N)$, so the four-point function decomposes into the symmetric and anti-symmetric representations. Let's again restrict to the $l = 0$ case, and also consider the case when $\{s_1, s_2 , s_3, s_4\} > 0$. This is the case when mixing happens only among the operators with the same number of derivatives. So we are going to consider operators of the form $O^{s,m}_{\alpha} = \Psi_{\alpha + v} (\vec{\partial}^2)^m \Psi_{s - \alpha - v}$ where $s = k + 2 v$ for some positive integer $k$ and $\alpha = 0, 1 \hdots \lfloor \frac{k}{2} \rfloor $. We again use the normalization such that the two-point function of the composite operators is given by \eqref{TwoPointCompositeI}.

At tree level, as before, the correlator is given by the disconnected pieces, which can be written as
\begin{equation}
\begin{split}
&\langle\Psi^I_{s_1}(\vec{y}_1) \Psi^J_{s_2}(\vec{y}_2) \bar{\Psi}_{K \ s_3}(\vec{y}_3) \bar{\Psi}_{L \ s_4}(\vec{y}_4) 
\rangle_0 = \mathcal{C}_{\Delta_{s_1}} \mathcal{C}_{\Delta_{s_2}}  \bigg[  \frac{{\delta^I}_K {\delta^J}_L \delta_{s_1,s_3} \delta_{s_2, s_4}}{(\vec{y}_{13}^2)^{\Delta_{s_1}} (\vec{y}_{24}^2)^{\Delta_{s_2}}} + \frac{{\delta^I}_L {\delta^J}_K  \delta_{s_1,s_4} \delta_{s_2,s_3}}{(\vec{y}_{14}^2)^{\Delta_{s_1}} (\vec{y}_{23}^2)^{\Delta_{s_2}}} \bigg] \\
& = \mathcal{C}_{\Delta_{s_1}} \mathcal{C}_{\Delta_{s_2}}  \sum_{n,l} \left( (-1)^l {\delta^I}_K {\delta^J}_L  \delta_{s_1,s_3} \delta_{s_2, s_4}  + {\delta^I}_L {\delta^J}_K \delta_{s_1,s_4} \delta_{s_2,s_3} \right) c_0^{(12)}(n,l) W_{\Delta = \Delta_{s_1} + \Delta_{s_2} + 2 n + l, l} (\vec{y}_i)
\end{split}
\end{equation}
The difference from the previous case is that when we consider operators in the symmetric sector, for $s_1 = s_2$ or equivalently $\alpha = s/2 - v$, both the terms above will contribute and hence the OPE coefficient squared will be twice as much. So, in the symmetric sector, the OPE coefficients are  
\begin{equation}
\begin{split}
( C_{\Psi_{s_1} \Psi_{s_2} \bar{O}^{s,m}_{\alpha}})^2 &=   c_0^{(12)} (m,0) \mathcal{C}_{\Delta_{s_1}} \mathcal{C}_{\Delta_{s_2}}  = \frac{ \left(\Delta_{s_1} \right)^2_m \left(\Delta_{s_2} \right)^2_m }{ 4 \pi^4 (m!)^2 (s + m + 1)_m^2} \\
( C_{\Psi_{s_1} \Psi_{s_1} \bar{O}^{s,m}_{\alpha}})^2 &=  2 c_0^{(12)} (m,0) \mathcal{C}_{\Delta_{s_1}}^2  = \frac{ \left(\Delta_{s_1} \right)^4_m}{ 2 \pi^4 (m!)^2 (s + m + 1)_m^2}.
\end{split}
\end{equation}

The first order correction coming from the interaction is also similar to before 
\begin{equation}
\begin{split}
\langle\Psi^I_{s_1}(\vec{y}_1) \Psi^J_{s_2}(\vec{y}_2) \bar{\Psi}_{K \ s_3}(\vec{y}_3) \bar{\Psi}_{L \ s_4}(\vec{y}_4) 
\rangle_1 = &-  \pi \lambda ({\delta^I}_K {\delta^J}_L + {\delta^I}_L {\delta^J}_K) \delta(s_1 + s_2 - s_3 - s_4)  \\
& \times \left( \prod_{i = 1}^4 \mathcal{C}_{\Delta_{s_i}} \right) D_{\Delta_{s_1} ,\Delta_{s_2},\Delta_{s_3},\Delta_{s_4}} (\vec{y}_i)
\end{split}
\end{equation}
Again, for the case when all  $\{s_1, s_2 , s_3, s_4\} > 0$, above contact interaction is only non-zero when $\Delta_{s_1} + \Delta_{s_2} = \Delta_{s_3} + \Delta_{s_4}$ and the decomposition in \eqref{DFunctionDecomposition} can be used. Using that, and adding the result of the disconnected pieces from \eqref{AnomalousDimensionTwoPointDef}, we can see that in the symmetric sector, we have the following anomalous dimension matrix
\begin{equation}
\Delta_{\alpha \beta}^{s, m} =
\begin{cases}
\frac{2}{(N + 4) (s  + 2 m + 1)} & \text{if $\alpha \neq \beta$ and neither $\alpha$ nor $\beta$ is $\frac{s}{2} - v$ } \\
\frac{\sqrt{2}}{(N + 4) (s  + 2 m + 1)} & \text{if $\alpha \neq \beta$ and either $\alpha$ or $\beta$ is $\frac{s}{2} - v$ } \\
 \frac{2}{(N + 4) (s + 2 m + 1)} + \frac{v (v - 1) (N + 1) s}{ 2 (N + 4)(\alpha + v) (s - \alpha - v)} & \text{if $\alpha = \beta \neq \frac{s}{2} - v$}\\
 \frac{1}{(N + 4) (s + 2 m + 1)} + \frac{2 v (v - 1) (N + 1) }{ s (N + 4)} & \text{if $\alpha = \beta = \frac{s}{2} - v$}.
\end{cases}
\end{equation}

\subsubsection*{$\mathbb{Z}_2$ monodromy defect}
For completeness, we also specialize this four-point function calculation to the case of  
$\mathbb{Z}_2$ twist defect where $v = 1/2$. We consider N real scalars and the monodromy is defined such that they pick up a minus sign as they go around the defect. We have the full unbroken $O(N)$ invariance. We only need to talk about one set of operators on the defect since the reality condition on scalars implies $\bar{\Psi}_s = \Psi_{-s}$. These defect operators have the two-point function given by
\begin{equation} 
\begin{split}
&\langle \Psi^I_{s}(\vec{y}_1) \Psi^J_{-s}(\vec{y}_2) \rangle =  \frac{ \delta^{IJ} \mathcal{C}_{\Delta_{s}}}{(\vec{y}^2_{12})^{\Delta_{s}} }, \hspace{1cm} \mathcal{C}_{\Delta_{s}} = \frac{\Gamma \left(\Delta_{s} \right)}{4 \pi^{d/2} \Gamma \left(\Delta_{s} + 2 - \frac{d}{2} \right)}, \\
&\Delta_{s} = 1 + |s| - \frac{\epsilon}{2} + \frac{v (v - 1) (N + 2)}{2 (N + 8) |s|} \epsilon . 
\end{split}
\end{equation}
where $s$ is an half-integer now. We then consider the following four-point function in the $12 \rightarrow 34$ channel
\begin{equation}
\begin{split}
\langle\Psi^I_{s_1}(\vec{y}_1) \Psi^J_{s_2}(\vec{y}_2) \Psi^K_{s_3}(\vec{y}_3) \Psi^L_{s_4}(\vec{y}_4) 
\rangle &= \delta^{IJ} \delta^{KL} \mathcal{G}_{S}+ \left( \frac{\delta^{IK} \delta^{JL} + \delta^{IL} \delta^{JK}}{2} - \frac{\delta^{IJ} \delta^{KL}}{N} \right) \mathcal{G}_{T} \\
& +  \frac{\delta^{IK} \delta^{JL} - \delta^{IL} \delta^{JK}}{2} \mathcal{G}_{A}
\end{split}
\end{equation}
and we have now decomposed it into singlet, traceless symmetric and anti-symmetric representations of $O(N)$. We restrict to the case with $s_1, s_2 > 0$ and $s_3, s_4 < 0$ and consider operators of the form $O^{s,m}_{\alpha} = \Psi_{\alpha + \frac{1}{2}} (\vec{\partial}^2)^m \partial^l \Psi_{s - \alpha - \frac{1}{2}}$ for $\alpha = 0, 1 \hdots \lfloor \frac{s - 1}{2} \rfloor $. These operators are normalized to have unit two-point function as before. 

As in the previous case, we can calculate the correlator as 
\begin{equation}
\begin{split}
\langle\Psi^I_{s_1}(\vec{y}_1) \Psi^J_{s_2}(\vec{y}_2) \Psi^K_{s_3}(\vec{y}_3) \Psi^L_{s_4}(\vec{y}_4) 
\rangle_0 &= \mathcal{C}_{\Delta_{s_1}} \mathcal{C}_{\Delta_{s_2}}  \bigg[  \frac{\delta^{I K}\delta^{J L}  \delta_{s_1,-s_3} \delta_{s_2, -s_4}}{(\vec{y}_{13}^2)^{\Delta_{s_1}} (\vec{y}_{24}^2)^{\Delta_{s_2}}} + \frac{\delta^{I L}\delta^{J K}  \delta_{s_1,s_4} \delta_{s_2,s_3}}{(\vec{y}_{14}^2)^{\Delta_{s_1}} (\vec{y}_{23}^2)^{\Delta_{s_2}}} \bigg] \\
& = \left( \delta^{I K}\delta^{J L}  \delta_{s_1,-s_3} \delta_{s_2,-s_4} (-1)^l + \delta^{I L}\delta^{J K}  \delta_{s_1,-s_4} \delta_{s_2,-s_3} \right) \\
& \times  \mathcal{C}_{\Delta_{s_1}} \mathcal{C}_{\Delta_{s_2}} \sum_{n,l} c_0^{(12)}(n,l) W_{\Delta = \Delta_{s_1} + \Delta_{s_2} + 2 n + l, l} (\vec{y}_i) 
\end{split}
\end{equation}
and the first order correction 

\begin{equation}
\begin{split}
\langle\Psi^I_{s_1}(\vec{y}_1) \Psi^J_{s_2}(\vec{y}_2) \Psi^K_{s_3}(\vec{y}_3) \Psi^L_{s_4}(\vec{y}_4) 
\rangle_1 &= - 4 \pi \lambda (\delta^{IJ} \delta^{KL} + \delta^{IK} \delta^{JL} + \delta^{IL} \delta^{JK}) \delta(s_1 + s_2 + s_3 + s_4)\\
&\times  \left( \prod_{i = 1}^4 \mathcal{C}_{\Delta_{s_i}} \right) D_{\Delta_{s_1} ,\Delta_{s_2},\Delta_{s_3},\Delta_{s_4}} (\vec{y}_i).
\end{split}
\end{equation}
We can extract the anomalous dimension from the above results similar to the previous cases, and in the $O(N)$ singlet sector, we find
\begin{equation}
\Delta_{\alpha \beta}^{s, m} =
\begin{cases}
\frac{2 (N + 2)}{(N + 8) (s  + 2 m + 1)} & \text{if $\alpha \neq \beta$ and neither $\alpha$ nor $\beta$ is $\frac{s - 1}{2}$ } \\
\frac{\sqrt{2} (N + 2)}{(N + 8) (s  + 2 m + 1)} & \text{if $\alpha \neq \beta$ and either $\alpha$ or $\beta$ is $\frac{s-1}{2} $ } \\
 \frac{2 (N + 2)}{(N + 8) (s + 2 m + 1)} - \frac{(N + 2) s}{ 2 (N + 8)(2 \alpha + 1) (2 s - 2\alpha - 1)} & \text{if $\alpha = \beta \neq \frac{s - 1}{2}$}\\
 \frac{(N + 2)}{(N + 8) (s + 2 m + 1)} - \frac{(N + 2) }{2 s (N + 8)} & \text{if $\alpha = \beta = \frac{s-1}{2}$}.
\end{cases}
\end{equation}
For $N = 1$, they give the results for the twist defect in Ising model, which were also discussed in \cite{Gaiotto:2013nva}. The authors in \cite{Gaiotto:2013nva} did not discuss the case in the second line of the above equation.

\section{Conclusion} \label{SectionConclusion}
In this paper, we studied monodromy defects in free and critical $O(N)$ model. We used conformal symmetry to map the problem to $S^1 \times H^{d-1}$, and saw that the hyperbolic space is well suited to many of the calculations. In particular, we computed the expectation value of a spherical monodromy defect (in 3d CFT, a circular defect) from the free energy on the hyperbolic cylinder, and also from an alternative approach based on mapping to $S^d$. We then studied a defect RG flow related to choosing alternate boundary conditions on $H^{d-1}$, and verified the conjectured ``defect C-theorem" for this RG flow. We studied the monodromy defect in the critical $O(N)$ model both in an large $N$ expansion keeping $d$ arbitrary and in $d = 4 - \epsilon$ dimensions with $\epsilon$ expansion techniques keeping $N$ arbitrary. We performed several consistency checks among these two approaches. 

As a next step, it would be interesting to do a systematic study of the bulk data (the bulk one-point functions for instance). In \cite{Liendo:2019jpu}, a defect CFT inversion formula was used to learn more details about the bulk data for twist defect in the Ising model. It would be interesting to extend the techniques to the more general monodromy defect we considered. It would also be interesting to get more analytic control over the large $N$ results. We relied on numerics to solve the saddle point equation \eqref{SaddleEquationPlus}, but there might be a way to do this analytically.  Understanding the large $N$ solution better may also be a useful first step towards understanding the holographic description of the system \footnote{A holographic description of a similar defect was discussed in \cite{Horowitz:2014gva}.}. Moreover, as we saw in section \ref{SectionLargeN}, an interacting unitary fixed point does not exist in $d = 3$ when we choose $\Delta^-$ boundary condition for $s = v$ mode. However, there does not seem to be any issues in describing it close to $d = 4$. It would be useful to understand why this happens, and an analytic understanding of the large $N$ solution may help us understand it better. It would be also interesting to go beyond the leading order at large $N$, which would entail including the $\sigma$ fluctuations around the saddle point and finding the corresponding propagator. Such analysis in the BCFT case was done in \cite{McAvity:1995zd} (see also \cite{Giombi:2020rmc}), and similar techniques may apply in this case. 

\section*{Acknowledgments}
We thank Xinan Zhou and Nathan Benjamin for discussions on related topics. This research was supported in part by the US NSF under Grants No.~PHY-1914860.

\appendix

\section{Spectrum of Laplacian on $S^d$ with twisted boundary conditions} \label{AppendixLaplacianSpectrum}
In this appendix, we determine the spectrum of Laplacian on $S^d$ with twisted boundary conditions. In the coordinate system specified in \eqref{MetricSphereTwist}, we may write the eigenfunctions of Laplacian  as $ e^{i m_{\theta} \theta} f(\tau) \kappa_{S^{d-2}} $, where $\kappa_{S^{d-2}}$ is the eigenfunction of Laplacian on  $S^{d-2}$ with the usual periodic boundary condition. The required monodromy in $\theta$ direction implies $m_{\theta}  \in \mathbb{Z} + v$. The eigenvalues and their degeneracies for the Laplacian with periodic boundary condition on $S^{d - 2}$ , are well known
\begin{equation} \label{DegeneracyUsual}
  - m_{\phi}(m_{\phi} + d - 3), \ \ \ d_{m_{\phi}} = \frac{(2 m_{\phi} + d -3) \Gamma(m_{\phi} + d -3)}{m_{\phi}! \Gamma(d - 2)}, \ \ \ m_{\phi} = 0,1,2,...
\end{equation}
Hence, the eigenfunction equation for $f(\tau)$ is 
\begin{equation}
    f''(\tau) +((d-2) \cot{\tau} - \tan \tau) f'(\tau) - \bigg( \frac{m_{\theta}^2}{ \cos^2 \tau} + \frac{m_{\phi}(m_{\phi} + d - 3)}{ \sin^2 \tau} \bigg) f(\tau) = \lambda f (\tau).
\end{equation}
The solution regular at $\tau = \pi/2$ looks like 
\begin{equation}
\begin{split}
    f (\tau) = (\cos \tau )^{|m_{\theta}|}  (\sin \tau)^{3 -d - m_{\phi}} \ {}_2F_1\bigg(\frac{5-d+2 |m_{\theta}|-2 m_{\phi} -\sqrt{(d - 1)^2-4 \lambda }}{4}, & \\
   \frac{5-d+2 |m_{\theta}|-2 m_{\phi}+\sqrt{(d-1)^2-4 \lambda}}{4} ,|m_{\theta}|+1, \cos ^2(\tau ) \bigg)&.
\end{split}
\end{equation}
We also want the solution to be regular at $\tau = 0$ in all dimensions. The eigenvalue $\lambda$ can be parametrized as 
\begin{equation}
    \lambda = - n(n + d -1 ), \ \ \ 
\end{equation}
and regularity of the solution at $\tau = 0 $ demands 
\begin{equation}
    n = |m_{\theta}| +m_{\phi} + 2 a, \ \ \ a \in \mathbb{N}
\end{equation}
where $\mathbb{N}$ represents the set of non-negative integers (includes $0$). As mentioned previously, the definition of defect requires $m_{\theta} \in \mathbb{Z} + v $ which implies that $n = k + v$ or $n = k + 1 - v$ with $k$ being a non-negative integer. Using the degeneracies of $m_{\phi}$ in \eqref{DegeneracyUsual}, it is easy to count the degeneracies of $n$ 
\begin{equation} \label{DegenraciesSphere}
\begin{split}    
    n &\in \mathbb{N} , \ \ \ d_{n} = \frac{(2 n + d -1) \Gamma(n + d -1)}{n! \Gamma(d )} \\
    n &= k + v, \ \ \ k \in \mathbb{N}, \ \ \ d_{n} = \frac{ \Gamma (k + d)}{ k! \Gamma (d)}  
\end{split}.    
\end{equation}
When summing over the eigenvalues, we need to sum over both $n = k + v$ and $n = k + 1 - v$. When $v = 1/2$, they both coincide. This effectively means that for $v = 1/2$,  the degeneracy of a given $n$ is twice of what we wrote above. 
\section{ Bulk Integrals } \label{AppendixBulkIntegrals}
\subsection{For bulk two-point function} \label{AppendixBulkTwoPoint}
In order to calculate the one-point function of $\bar{\Phi} \Phi$ to leading order in $\epsilon$ in subsection \ref{SectionBulkTwoPoint}, we need to do the following integral 
\begin{equation}
J = \int \frac{d r d y d z}{r} \frac{ (r')^2 (4 r' r)^{2 - 2 v} (d_+ + d_-)^{2 v} (e_+ + e_-)^{2 v} }{d_+ d_- e_+ e_- \left( (d_+ + d_-)^{2}(e_+ + e_-)^{2} - (4 r' r)^2 \right)} + (v \rightarrow 1 - v)
\end{equation}
where the integral runs over $0 \le r < \infty, -\infty < y < \infty, - \infty < z < \infty$ and 
\begin{equation}
d_{\pm} = \sqrt{\left(y - \frac{y'}{2} \right)^2 + z^2  + (r \pm r')^2}, \hspace{1 cm} e_{\pm} = \sqrt{\left(y + \frac{y'}{2} \right)^2 + z^2  + (r \pm r')^2}.
\end{equation}
This integral was analyzed in the limit $\mu \rightarrow 0$ in \cite{Gaiotto:2013nva} for $v = 1/2$. The same method goes through for arbitrary values of $v$, with some changes. We will do that here. First, let's make the integral over dimensionless variables by making a substitution $y = r' a, z = r' b, r = r' c$, and then extend the integration over all of $\mathbb{R}^3$ 
\begin{equation}
\begin{split}
&J = 2^{3 - 4 v} \int_{\mathbb{R}^3} d a d b d c \frac{ (c^2)^{\frac{1}{2} - v} (\tilde{d}_+ + \tilde{d}_-)^{2 v} (\tilde{e}_+ + \tilde{e}_-)^{2 v} }{\tilde{d}_+ \tilde{d}_- \tilde{e}_+ \tilde{e}_- \left( (\tilde{d}_+ + \tilde{d}_-)^{2}(\tilde{e}_+ + \tilde{e}_-)^{2} - (4 c)^2 \right)} + (v \rightarrow 1 - v) \\
&\tilde{d}_{\pm} = \sqrt{\left(a - \frac{\mu}{2} \right)^2 + b^2  + (c \pm 1)^2}, \hspace{1 cm} \tilde{e}_{\pm} = \sqrt{\left(a + \frac{\mu}{2} \right)^2 + b^2  + (c \pm 1)^2}.
\end{split}
\end{equation}
When $\mu \rightarrow 0$, this integral has logarithm divergences around $(a,b,c) = (0,0, \pm 1)$. So we expect the integral to be of the form $\alpha \log \mu + \beta$ up to terms that vanish as $\mu \rightarrow 0$. We are interested in calculating $\alpha$ and $\beta$. To do that, we introduce an auxiliary parameter $\mathcal{ N }$ and divide the integration region into two parts and write the integral as $J = J_1(\mu, \mathcal{ N}) + J_2(\mu, \mathcal{N})$. $J_1(\mu, \mathcal{ N})$ is the integral over spheres of radius $\mu \mathcal{N}$ around the two points $(a,b,c) = (0,0, \pm 1)$ and $J_2(\mu, \mathcal{N})$ is the integral over rest of $\mathbb{R}^3$. The integrals $J_{1,2} (\mu, \mathcal{N})$ simplify in the limit $\mathcal{N} \rightarrow \infty, \mu \mathcal{N} \rightarrow 0 $ if we do not care about the terms that vanish as $\mu \rightarrow 0$. Let's first look at $J_1(\mu, \mathcal{ N})$ and focus on the sphere around $(0,0,1)$. Making a substitution $a = \mu x, b = \mu y, c = 1 + \mu z$ gives 
\begin{equation}
\begin{split}
&\tilde{d}_{+} = \tilde{e}_{+} = 2 + \mu z + O(\mu^2) \\
&\tilde{d}_{-} = \mu \sqrt{\left(x - \frac{1}{2} \right)^2 + y^2  + z^2} + O(\mu^2), \hspace{1 cm} \tilde{e}_{-} = \mu \sqrt{\left(x + \frac{1}{2} \right)^2 + y^2  + z^2} + O(\mu^2).
\end{split}
\end{equation}
Similar things can also be said about the sphere around $(0,0,-1)$. The integral, up to terms that vanish as $\mu \rightarrow 0$ then becomes 
\begin{equation}
\begin{split}
J_1(\mu, \mathcal{ N}) &= \frac{1}{2} \int_{x^2 + y^2 + z^2 \le \mathcal{N}^2} dx dy dz \frac{1}{f_{+} f_{-} (f_+ + f_-)}; \hspace{1 cm} f_{\pm} = \sqrt{\left(x \pm \frac{1}{2} \right)^2 + y^2  + z^2} \\
& = 2 \pi \int_0^{\mathcal{N}} dx \frac{x + \frac{1}{2} -|x - \frac{1}{2}| + \sqrt{\mathcal{N}^2 + \frac{1}{4} - x} - \sqrt{\mathcal{N}^2 + \frac{1}{4} + x} }{2 x} \\
& = \pi \log 2 \mathcal{N} + O(1/\mathcal{N}).
\end{split}
\end{equation}
Now, let's shift our attention to $J_2(\mu, \mathcal{ N})$ which runs over the domain $D = \{(a,b,c) \in \mathbb{R}^3| a^2 + b^2 + (c\pm 1)^2 \ge (\mu \mathcal{N})^2 \}$. In the limit $\mathcal{N} \rightarrow \infty$, we can take $\tilde{d}_{\pm} = \tilde{e}_{\pm} = \sqrt{a^2 + b^2  + (c \pm 1)^2}$ and hence $4 c = \tilde{d}^2_{+} - \tilde{d}^2_{-}   $. The integral then simplifies to

\begin{equation}
\begin{split} 
J_2(\mu, \mathcal{N}) &= 2^{3 - 4 v} \int_D d a d b d c \frac{ (c^2)^{\frac{1}{2} - v} (\tilde{d}_+ + \tilde{d}_-)^{4 v} }{(\tilde{d}_+ \tilde{d}_-)^2 \left( (\tilde{d}_+ + \tilde{d}_-)^{4}- (4 c)^2 \right)} + (v \rightarrow 1 - v) \\
& =  \frac{1}{2} \int_D d a d b d c \frac{1 }{(\tilde{d}_+ \tilde{d}_-)^3 } \left( \frac{\tilde{d}_+ + \tilde{d}_-}{|\tilde{d}_+ - \tilde{d}_-|} \right)^{2 v - 1} + (v \rightarrow 1 - v).
\end{split}
\end{equation}
We then scale the variables by two followed by an inversion to change the variables as
\begin{equation}
\begin{split} \label{InversionScaling}
a'' = \frac{2 a }{ a^2 + b^2 + (c - 1)^2 }, \ \ b'' = \frac{2 b }{ a^2 + b^2 + (c - 1)^2 }, \ \ c'' - \frac{1}{2} = \frac{2 (c - 1) }{ a^2 + b^2 + (c - 1)^2 } \\
\implies \tilde{d}_{+} = 2 \sqrt{\frac{a''^2 + b''^2 + \left( c'' + \frac{1}{2}\right)^2}{a''^2 + b''^2 + \left( c'' - \frac{1}{2}\right)^2}}, \ \ \tilde{d}_{-} = \frac{2}{ \sqrt{a''^2 + b''^2 + \left( c'' - \frac{1}{2}\right)^2}} 
\end{split}
\end{equation}
and the integral simplifies to 
\begin{equation}
\begin{split}
J_2(\mu,  \mathcal{N}) &= \frac{1}{16} \int_{D''} d a'' d b'' d c'' \frac{1}{r_+^3} \left( \left( \frac{r_+ + 1}{ |r_+ - 1|} \right)^{2 v - 1} +  \left( \frac{r_+ + 1}{ |r_+ - 1|} \right)^{1 - 2 v} \right), \\  r_+ &= \sqrt{a''^2 + b''^2 + \left( c'' + \frac{1}{2}\right)^2 }. 
\end{split}   
\end{equation}
When we scale by two, the integration domain gets mapped to everywhere outside the spheres centered at $(0,0,1/2)$ and $(0,0,-1/2)$ with radius $\mu \mathcal{N}/2$. Then, when we invert about $(0,0,1/2)$, to leading order in $\mu \mathcal{N}$, it gets mapped to everywhere between the sphere of radius $\mu \mathcal{N}/2$ centered at $(0,0,-1/2)$ and the sphere of radius $2 / \mu \mathcal{N}$ centered at $(0,0,1/2)$. So $D'' = \{(a'',b'',c'') \in \mathbb{R}^3| a''^2 + b''^2 + (c''\pm 1/2)^2 \gtrless (\mu \mathcal{N}/2)^{\pm 2} \}$. Again, to leading order in $\mu \mathcal{N}$, we can shift the outer sphere of radius $2 / \mu \mathcal{N}$ so that it is also centered at $(0,0,-1/2)$ so the integration domain becomes everywhere between concentric spheres with radii inverse of each other and the integral becomes
\begin{equation}
\begin{split}
J_2(\mu,  \mathcal{N}) &= \frac{\pi}{4} \int_{\frac{\mu \mathcal{N}}{2}}^{\frac{2}{\mu \mathcal{N}}} d r_+ \frac{1}{r_+} \left( \left( \frac{r_+ + 1}{ |r_+ - 1|} \right)^{2 v - 1} +  \left( \frac{r_+ + 1}{ |r_+ - 1|} \right)^{1 - 2 v} \right) \\
& = \frac{\pi}{2} \left(-2 \log 2 -  \left(H^{1 - v}+H^{v}\right)+ \frac{1}{v (1 - v)} - 4 \log \mu \mathcal{N}\right)
\end{split}
\end{equation}
So $\mathcal{N}$ cancels and the full integral is 
\begin{equation} \label{AppendixresultTwoPoint}
J =  \frac{\pi}{2} \left(- 2 \log \mu -  \left(H^{1 - v} + H^{v}\right)+ \frac{1}{v (1 - v)} \right).
\end{equation}

For the same calculation with $\Delta^-$ boundary condition, we need an additional integral 
\begin{equation}
J^- = \int \frac{d r d y d z}{r} \frac{(r')^2 }{d_+ d_- e_+ e_- }  \frac{ ((d_+ + d_-)(e_+ + e_-))^{2 v}}{(4 r' r)^{2 v}} -  (v \rightarrow - v) .
\end{equation}
Doing the same subsitutions as before, we get 
\begin{equation}
J^- = 2^{-1 - 4 v} \int_{\mathbb{R}^3} d a d b d c \frac{ (c^2)^{-\frac{1}{2} - v} (\tilde{d}_+ + \tilde{d}_-)^{2 v} (\tilde{e}_+ + \tilde{e}_-)^{2 v} }{\tilde{d}_+ \tilde{d}_- \tilde{e}_+ \tilde{e}_- }  -  (v \rightarrow - v) .
\end{equation}
This integral in fact does not have diverge as $\mu \rightarrow 0$, so we can just plug in $\mu = 0$, since we are only interested in $\mu$ independent piece for this calculation. For $\mu = 0$, we have  $\tilde{d}_{\pm} = \tilde{e}_{\pm} = \sqrt{a^2 + b^2  + (c \pm 1)^2}$ and $4 c = \tilde{d}^2_{+} - \tilde{d}^2_{-}   $ which gives a simpler integral 
\begin{equation}
J^- = 2\int_{\mathbb{R}^3} d a d b d c \frac{  (\tilde{d}_+ + \tilde{d}_-)^{2 v - 1} }{(\tilde{d}_+ \tilde{d}_-)^2 |\tilde{d}_+ - \tilde{d}_-|^{2 v + 1} }  -  (v \rightarrow - v). 
\end{equation} 
We then do a change of variables as in \eqref{InversionScaling} to get 
\begin{equation} \label{AppendixIntegralBulkTwoResM}
\begin{split}
J^- &= \frac{1}{4}\int_{\mathbb{R}^3} d a'' d b'' d c'' \frac{  (r_+ + 1)^{2 v - 1} }{r_+^2 |r_+ - 1|^{2 v + 1} }  -  (v \rightarrow - v), \hspace{1cm }  r_+ = \sqrt{a''^2 + b''^2 + \left( c'' + \frac{1}{2}\right)^2 } \\
&= \pi \int_{0}^{\infty} d r_+ \frac{  (r_+ + 1)^{2 v - 1} }{|r_+ - 1|^{2 v + 1} }  -  (v \rightarrow - v) \\
&= - \frac{\pi}{v} .
\end{split}
\end{equation}

\subsection{For coefficient of displacement} \label{AppendixDisplacement}
For extracting the coefficient of the displacement in subsection \ref{SectionDisplacementCoefficient}, we have to perform the following integral 
\begin{equation}
\mathcal{I} = \int dr dy dz \frac{  (r')^6 r^3  }{ d_+^2 d_-^2 e_+^2 e_-^2 (d_+ + d_-)^2(e_+ + e_-)^2}
\end{equation}
in the limit $\kappa = r'/ y' \rightarrow 0$. As before, we rescale the coordinates as $y = a y', z = b y', r = c y'$ to rewrite the integral as 
\begin{equation}
\begin{split}
& \mathcal{I}  = \kappa^6 I, \hspace{1cm}  I = \frac{1}{2} \int_{\mathbb{R}^3} da db dc  \frac{   (c^2)^{3/2}  }{ \tilde{d}_+^2 \tilde{d}_-^2 \tilde{e}_+^2 \tilde{e}_-^2 (\tilde{d}_+ + \tilde{d}_-)^2(\tilde{e}_+ + \tilde{e}_-)^2}\\
&\tilde{d}_{\pm} = \sqrt{\left(a - \frac{1}{2} \right)^2 + b^2  + (c \pm \kappa)^2}, \hspace{1 cm} \tilde{e}_{\pm} = \sqrt{\left(a + \frac{1}{2} \right)^2 + b^2  + (c \pm \kappa)^2}.
\end{split}
\end{equation} 
As before, this integral has a logarithmic divergence as $\kappa \rightarrow 0$ at $(a,b,c) = (\pm 1/2, 0,0)$  and we want to evaluate the integral $I$ up to terms that vanish as $\kappa \rightarrow 0$. As before, we divide the integration region into two parts $I = I_1 (\kappa, \mathcal{N}) + I_2 (\kappa, \mathcal{N})$ with $I_1 (\kappa, \mathcal{N})$ being the integral within two spheres of radius $\kappa \mathcal{N}$ around $(\pm 1/2, 0,0)$. We will look at the integral in the limit $\mathcal{N} \rightarrow \infty, \mathcal{N} \kappa \rightarrow 0$ ignoring the terms that vanish in this limit. Focusing on the sphere around $(1/2,0,0)$, we make a substitution $a = 1/2 + \kappa x, b = \kappa y, c = \kappa z$, and note that $\tilde{e}_{\pm} = 1 + O(\kappa)$
\begin{equation}
I_1 (\kappa, \mathcal{N}) = \frac{1}{4} \int_{{x^2 + y^2 + z^2 \le \mathcal{N}^2}} dx dy dz \frac{(z^2)^{3/2}}{f_+^2 f_-^2 (f_+ + f_-)^2}, \hspace{1cm}  f_{\pm} = \sqrt{x^2 + y^2  + (z \pm 1)^2}
\end{equation}
The integral over $x,y$ can be directly performed, and the rest can be performed after we expand in large $\mathcal{N}$. Up to terms that vanish as $\mathcal{N} \rightarrow 0$, we get 
\begin{equation}
\begin{split}
I_1 (\kappa, \mathcal{N}) &= -\frac{\pi }{16} \int_0^{\mathcal{N}} dz \left( z \log \left( \frac{|z -1|(z + 1)}{2 z^2} \right) + z \log 2 + \frac{z^3}{\mathcal{N}^4} \right) \\
& = \frac{\pi}{16} \log \mathcal{N} - \frac{\pi }{64}. 
\end{split}
\end{equation}
Let's now look at the other part of the integral 
\begin{equation}
I_2(\kappa, \mathcal{N}) = \frac{1}{2} \int_{D} da db dc  \frac{   (c^2)^{3/2}  }{ \tilde{d}_+^2 \tilde{d}_-^2 \tilde{e}_+^2 \tilde{e}_-^2 (\tilde{d}_+ + \tilde{d}_-)^2(\tilde{e}_+ + \tilde{e}_-)^2}
\end{equation}
where the domain of the integral is $D = \{(a,b,c) \in \mathbb{R}^3| (a \pm 1/2)^2 + b^2 + c^2 \ge (\kappa \mathcal{N})^2 \}$. At large $\mathcal{N}$, we can use $\tilde{d}_{+} = \tilde{d}_{-}, \tilde{e}_{+} = \tilde{e}_{-}$, so the integral becomes 
\begin{equation}
I_2(\kappa, \mathcal{N}) = \frac{1}{32} \int_{D} da db dc  \frac{   (c^2)^{3/2}  }{ r_+^6 r_-^6}, \hspace{1cm} r_{\pm} =  \sqrt{\left(a \pm \frac{1}{2} \right)^2 + b^2  + c^2}.
\end{equation} 
Next, we do an inversion centered at $(1/2, 0 , 0)$ which maps the integration domain to $D'$ which is the region between a sphere of radius $\kappa \mathcal{N}$ centered at $(-1/2, 0 , 0)$ and a sphere of radius $1/ \kappa \mathcal{N}$ centered at $(1/2, 0 , 0)$ up to terms of order $\kappa \mathcal{N}$. Under this inversion, the integral simplifies to
\begin{equation}
I_2(\kappa, \mathcal{N}) = \frac{1}{32} \int_{D'} da db dc  \frac{   (c^2)^{3/2}  }{ r_+^6}.
\end{equation}
We can then shift the outer sphere so that it is also centered at $(-1/2, 0 , 0)$ which only changes the result at subleading order in $\kappa \mathcal{N}$. Then the integral can be evaluated to be 
\begin{equation} \label{AppendixResultDisplacement}
I_2(\kappa, \mathcal{N}) = -\frac{\pi}{16} \log \kappa \mathcal{N} \implies \mathcal{I}  = -\frac{\pi \kappa^6}{16} \left(  \log \kappa + \frac{1}{4} \right).
\end{equation}

\subsection{For bulk OPE coefficient}
In this appendix, we show how to calculate the bulk OPE coefficient ${C_{\bar{\Phi};{\Phi}};}^{\bar{\Phi} \Phi}$. It can be extracted from the bulk four-point function of $\Phi$ in the absence of the defect in flat space. As we saw in section \ref{SectionDefectFourPoint}, such a four-point function can be decomposed into singlet and adjoint representations of $U(N)$
\begin{equation}
\langle \bar{\Phi}_I (x_1) \Phi^J (x_2) \bar{\Phi}_K (x_3) \Phi^L (x_4) \rangle = {\delta_I}^J {\delta_K}^L \mathcal{G}_{\textrm{sing}} + \left({\delta_I}^L {\delta_J}^K - \frac{{\delta_I}^J {\delta_K}^L}{N} \right) \mathcal{G}_{\textrm{adj}}.
\end{equation}
We expect the operator $\bar{\Phi} \Phi$ to be present in the singlet sector. For simplicity, we place the four operators on a line with $|x_{12}| = |x_{34}| = r $, and $|x_{13}| = |x_{24}| = r/ \mu$ and look at it in the limit $\mu << 1$ when we expect
\begin{equation}
\mathcal{G}_{\textrm{sing}} = \frac{1}{r^{4 \Delta_{\Phi}}} \left[ C^2_{\Phi} + ({C_{\bar{\Phi};{\Phi}};}^{\bar{\Phi} \Phi})^2 C_{\bar{\Phi} \Phi} \mu^{\Delta_{\bar{\Phi}\Phi}} (1 + O(\mu))   \right].
\end{equation}
In the following, we fix the normalization of the operator such that $C_{\bar{\Phi} \Phi} = N C^2_{\Phi}.$ In the free theory, we have $\Delta_{\Phi} = d/2 - 1, \Delta_{\bar{\Phi}\Phi} = d - 2$. The result for $\Delta_{\Phi}$ does not get corrected to first order in interaction. In the free theory, the correlator is just  
\begin{equation}
\langle \bar{\Phi}_I (x_1) \Phi^J (x_2) \bar{\Phi}_K (x_3) \Phi^L (x_4) \rangle_0 = \frac{{\delta_I}^J {\delta_K}^L C^2_{\Phi}}{r^{4 \Delta_{\Phi}}} + \frac{ \mu^{4 \Delta_{\Phi}} {\delta_I}^L {\delta_J}^K C^2_{\Phi}}{r^{4 \Delta_{\Phi}} (1 - \mu^2)^{2 \Delta_{\Phi}}}. 
\end{equation}
In the interacting theory, it gets corrected to 
\begin{equation}
\begin{split}
\langle \bar{\Phi}_I (x_1) \Phi^J (x_2) \bar{\Phi}_K (x_3) \Phi^L (x_4) \rangle_1 &= -\frac{\lambda }{2} ({\delta_I}^J {\delta_K}^L + {\delta_I}^L {\delta_J}^K ) \int d^4 x_0 \frac{C^4_{\Phi}}{x_{01}^2 x_{02}^2 x_{03}^2 x_{04}^2} \\
& = -\frac{\lambda \pi^2 }{2} ({\delta_I}^J {\delta_K}^L + {\delta_I}^L {\delta_J}^K ) \frac{\mu^4}{r^4} C^4_{\Phi} \bar{D}_{1,1,1,1} (u,v)
\end{split}
\end{equation}
where we used the fact that this conformal integral can be performed in terms of well known $\bar{D}$ functions (see for instance \cite{Giombi:2018vtc}). Conformal cross-ratios  $u,v$ can be defined by 
\begin{equation}
u = \frac{x_{12}^2 x_{34}^2}{x_{13}^2 x_{24}^2} = \mu^4, \hspace{0.5 cm} v = \frac{x_{14}^2 x_{23}^2}{x_{13}^2 x_{24}^2} = (1 - \mu^2)^2. 
\end{equation}
This $\bar{D}$ function can be expanded into a power series around $u = 0, v = 1$ as  \cite{Dolan:2000ut, Dolan:2000uw}
\begin{equation}
\bar{D}_{1,1,1,1} (u,v) = \sum_{m,n = 0}^{\infty} \frac{((m + n)!)^2}{n! (2 m + n + 1)!} u^m (1 - v)^n \left(2 \psi (2 + 2 m + n) - 2 \psi (1 +  m + n)  - \log u \right) 
\end{equation}
The singlet sector correlator, to leading order in interaction, and to leading order in $\mu$ is thus given by 
\begin{equation} 
\begin{split}
\mathcal{G}_{\textrm{sing}} &= \frac{C^2_{\Phi}}{r^{d-2}} \left( 1 + \frac{\mu^{d - 2}}{N} \right)  -\frac{\lambda \pi^2 (N + 1) }{2 N}  \frac{\mu^4}{r^4} C^4_{\Phi} \bar{D}_{1,1,1,1} (u,v) \\
& =  \frac{C^2_{\Phi}}{r^{2 d-4}} \left( 1 + \frac{\mu^{2 d - 4}}{N} \right)  -\frac{2 \pi^4 (N + 1) \epsilon }{ N (N + 4)}  \frac{\mu^4}{r^4} C^4_{\Phi} \left( - 4 \log \mu + 2 \right). 
\end{split}
\end{equation}
Using our convention that 
\begin{equation}
C_{\Phi} = \frac{\Gamma \left( \frac{d}{2} - 1 \right)}{2 \pi^{d/2}}
\end{equation}
we get 
\begin{equation} \label{BulkOPECoefficient}
\begin{split}
\Delta_{\bar{\Phi}\Phi} &= 2 -  \frac{3}{N + 4} \epsilon \\
{C_{\bar{\Phi};{\Phi}};}^{\bar{\Phi} \Phi} &= \frac{1}{N} \left( 1 - \frac{N + 1}{2 (N + 4)} \epsilon \right).
\end{split}
\end{equation}

\bibliographystyle{ssg}
\bibliography{TwistDefect-bib}

\begingroup\raggedright\begin{thebibliography}{10}

\bibitem{Billo:2016cpy}
M.~Billò, V.~Gonçalves, E.~Lauria, and M.~Meineri, ``{Defects in conformal
  field theory},'' {\em JHEP} {\bf 04} (2016) 091,
  \href{https://arxiv.org/abs/1601.02883}{{\tt 1601.02883}}.

\bibitem{Gadde:2016fbj}
A.~Gadde, ``{Conformal constraints on defects},'' {\em JHEP} {\bf 01} (2020)
  038, \href{https://arxiv.org/abs/1602.06354}{{\tt 1602.06354}}.

\bibitem{Lauria:2017wav}
E.~Lauria, M.~Meineri, and E.~Trevisani, ``{Radial coordinates for defect
  CFTs},'' {\em JHEP} {\bf 11} (2018) 148,
  \href{https://arxiv.org/abs/1712.07668}{{\tt 1712.07668}}.

\bibitem{Lemos:2017vnx}
M.~Lemos, P.~Liendo, M.~Meineri, and S.~Sarkar, ``{Universality at large
  transverse spin in defect CFT},'' {\em JHEP} {\bf 09} (2018) 091,
  \href{https://arxiv.org/abs/1712.08185}{{\tt 1712.08185}}.

\bibitem{Guha:2018snh}
S.~Guha and B.~Nagaraj, ``{Correlators of Mixed Symmetry Operators in Defect
  CFTs},'' {\em JHEP} {\bf 10} (2018) 198,
  \href{https://arxiv.org/abs/1805.12341}{{\tt 1805.12341}}.

\bibitem{Isachenkov:2018pef}
M.~Isachenkov, P.~Liendo, Y.~Linke, and V.~Schomerus, ``{Calogero-Sutherland
  Approach to Defect Blocks},'' {\em JHEP} {\bf 10} (2018) 204,
  \href{https://arxiv.org/abs/1806.09703}{{\tt 1806.09703}}.

\bibitem{Lauria:2018klo}
E.~Lauria, M.~Meineri, and E.~Trevisani, ``{Spinning operators and defects in
  conformal field theory},'' {\em JHEP} {\bf 08} (2019) 066,
  \href{https://arxiv.org/abs/1807.02522}{{\tt 1807.02522}}.

\bibitem{Liendo:2019jpu}
P.~Liendo, Y.~Linke, and V.~Schomerus, ``{A Lorentzian inversion formula for
  defect CFT},'' \href{https://arxiv.org/abs/1903.05222}{{\tt 1903.05222}}.

\bibitem{Herzog:2020bqw}
C.~P. Herzog and A.~Shrestha, ``{Two Point Functions in Defect CFTs},''
  \href{https://arxiv.org/abs/2010.04995}{{\tt 2010.04995}}.

\bibitem{Lauria:2020emq}
E.~Lauria, P.~Liendo, B.~C. Van~Rees, and X.~Zhao, ``{Line and surface defects
  for the free scalar field},'' \href{https://arxiv.org/abs/2005.02413}{{\tt
  2005.02413}}.

\bibitem{Nishioka:2021uef}
T.~Nishioka and Y.~Sato, ``{Free energy and defect $C$-theorem in free scalar
  theory},'' \href{https://arxiv.org/abs/2101.02399}{{\tt 2101.02399}}.

\bibitem{Billo:2013jda}
M.~Billò, M.~Caselle, D.~Gaiotto, F.~Gliozzi, M.~Meineri, and R.~Pellegrini,
  ``{Line defects in the 3d Ising model},'' {\em JHEP} {\bf 07} (2013) 055,
  \href{https://arxiv.org/abs/1304.4110}{{\tt 1304.4110}}.

\bibitem{Gaiotto:2013nva}
D.~Gaiotto, D.~Mazac, and M.~F. Paulos, ``{Bootstrapping the 3d Ising twist
  defect},'' {\em JHEP} {\bf 03} (2014) 100,
  \href{https://arxiv.org/abs/1310.5078}{{\tt 1310.5078}}.

\bibitem{Yamaguchi:2016pbj}
S.~Yamaguchi, ``{The $\epsilon$-expansion of the codimension two twist defect
  from conformal field theory},'' {\em PTEP} {\bf 2016} (2016), no.~9 091B01,
  \href{https://arxiv.org/abs/1607.05551}{{\tt 1607.05551}}.

\bibitem{Soderberg:2017oaa}
A.~S\"oderberg, ``{Anomalous Dimensions in the WF O($N$) Model with a Monodromy
  Line Defect},'' {\em JHEP} {\bf 03} (2018) 058,
  \href{https://arxiv.org/abs/1706.02414}{{\tt 1706.02414}}.

\bibitem{Kapustin:2005py}
A.~Kapustin, ``{Wilson-'t Hooft operators in four-dimensional gauge theories
  and S-duality},'' {\em Phys. Rev. D} {\bf 74} (2006) 025005,
  \href{https://arxiv.org/abs/hep-th/0501015}{{\tt hep-th/0501015}}.

\bibitem{Hung:2014npa}
L.-Y. Hung, R.~C. Myers, and M.~Smolkin, ``{Twist operators in higher
  dimensions},'' {\em JHEP} {\bf 10} (2014) 178,
  \href{https://arxiv.org/abs/1407.6429}{{\tt 1407.6429}}.

\bibitem{Klebanov:2011uf}
I.~R. Klebanov, S.~S. Pufu, S.~Sachdev, and B.~R. Safdi, ``{Renyi Entropies for
  Free Field Theories},'' {\em JHEP} {\bf 04} (2012) 074,
  \href{https://arxiv.org/abs/1111.6290}{{\tt 1111.6290}}.

\bibitem{Casini:2010kt}
H.~Casini and M.~Huerta, ``{Entanglement entropy for the n-sphere},'' {\em
  Phys. Lett. B} {\bf 694} (2011) 167--171,
  \href{https://arxiv.org/abs/1007.1813}{{\tt 1007.1813}}.

\bibitem{Casini:2011kv}
H.~Casini, M.~Huerta, and R.~C. Myers, ``{Towards a derivation of holographic
  entanglement entropy},'' {\em JHEP} {\bf 05} (2011) 036,
  \href{https://arxiv.org/abs/1102.0440}{{\tt 1102.0440}}.

\bibitem{Hung:2011nu}
L.-Y. Hung, R.~C. Myers, M.~Smolkin, and A.~Yale, ``{Holographic Calculations
  of Renyi Entropy},'' {\em JHEP} {\bf 12} (2011) 047,
  \href{https://arxiv.org/abs/1110.1084}{{\tt 1110.1084}}.

\bibitem{Balakrishnan:2016ttg}
S.~Balakrishnan, S.~Dutta, and T.~Faulkner, ``{Gravitational dual of the
  R\'enyi twist displacement operator},'' {\em Phys. Rev. D} {\bf 96} (2017),
  no.~4 046019, \href{https://arxiv.org/abs/1607.06155}{{\tt 1607.06155}}.

\bibitem{Bianchi:2016xvf}
L.~Bianchi, S.~Chapman, X.~Dong, D.~A. Galante, M.~Meineri, and R.~C. Myers,
  ``{Shape dependence of holographic R\'enyi entropy in general dimensions},''
  {\em JHEP} {\bf 11} (2016) 180, \href{https://arxiv.org/abs/1607.07418}{{\tt
  1607.07418}}.

\bibitem{Belin:2013dva}
A.~Belin, A.~Maloney, and S.~Matsuura, ``{Holographic Phases of Renyi
  Entropies},'' {\em JHEP} {\bf 12} (2013) 050,
  \href{https://arxiv.org/abs/1306.2640}{{\tt 1306.2640}}.

\bibitem{Belin:2013uta}
A.~Belin, L.-Y. Hung, A.~Maloney, S.~Matsuura, R.~C. Myers, and T.~Sierens,
  ``{Holographic Charged Renyi Entropies},'' {\em JHEP} {\bf 12} (2013) 059,
  \href{https://arxiv.org/abs/1310.4180}{{\tt 1310.4180}}.

\bibitem{Belin:2014mva}
A.~Belin, L.-Y. Hung, A.~Maloney, and S.~Matsuura, ``{Charged Renyi entropies
  and holographic superconductors},'' {\em JHEP} {\bf 01} (2015) 059,
  \href{https://arxiv.org/abs/1407.5630}{{\tt 1407.5630}}.

\bibitem{Bianchi:2015liz}
L.~Bianchi, M.~Meineri, R.~C. Myers, and M.~Smolkin, ``{R\'enyi entropy and
  conformal defects},'' {\em JHEP} {\bf 07} (2016) 076,
  \href{https://arxiv.org/abs/1511.06713}{{\tt 1511.06713}}.

\bibitem{Paulos:2016fap}
M.~F. Paulos, J.~Penedones, J.~Toledo, B.~C. van Rees, and P.~Vieira, ``{The
  S-matrix bootstrap. Part I: QFT in AdS},'' {\em JHEP} {\bf 11} (2017) 133,
  \href{https://arxiv.org/abs/1607.06109}{{\tt 1607.06109}}.

\bibitem{Carmi:2018qzm}
D.~Carmi, L.~Di~Pietro, and S.~Komatsu, ``{A Study of Quantum Field Theories in
  AdS at Finite Coupling},'' {\em JHEP} {\bf 01} (2019) 200,
  \href{https://arxiv.org/abs/1810.04185}{{\tt 1810.04185}}.

\bibitem{Herzog:2019bom}
C.~P. Herzog and I.~Shamir, ``{On Marginal Operators in Boundary Conformal
  Field Theory},'' {\em JHEP} {\bf 10} (2019) 088,
  \href{https://arxiv.org/abs/1906.11281}{{\tt 1906.11281}}.

\bibitem{Herzog:2020lel}
C.~P. Herzog and N.~Kobayashi, ``{The $O(N)$ model with $\phi^6$ potential in
  ${\mathbb R}^2 \times {\mathbb R}^+$},''
  \href{https://arxiv.org/abs/2005.07863}{{\tt 2005.07863}}.

\bibitem{Giombi:2020rmc}
S.~Giombi and H.~Khanchandani, ``{CFT in AdS and boundary RG flows},'' {\em
  JHEP} {\bf 11} (2020) 118, \href{https://arxiv.org/abs/2007.04955}{{\tt
  2007.04955}}.

\bibitem{Aharony:2015hix}
O.~Aharony, M.~Berkooz, A.~Karasik, and T.~Vaknin, ``{Supersymmetric field
  theories on AdS$_{p} \times$ S$^{q}$},'' {\em JHEP} {\bf 04} (2016) 066,
  \href{https://arxiv.org/abs/1512.04698}{{\tt 1512.04698}}.

\bibitem{Rodriguez-Gomez:2017kxf}
D.~Rodriguez-Gomez and J.~G. Russo, ``{Free energy and boundary anomalies on
  $\mathbb{S}^a\times \mathbb{H}^b$ spaces},'' {\em JHEP} {\bf 10} (2017) 084,
  \href{https://arxiv.org/abs/1708.00305}{{\tt 1708.00305}}.

\bibitem{Pittelli:2018rpl}
A.~Pittelli, ``{Supersymmetric localization of refined chiral multiplets on
  topologically twisted $H^2$ \texttimes{} $S^1$},'' {\em Phys. Lett. B} {\bf
  801} (2020) 135154, \href{https://arxiv.org/abs/1812.11151}{{\tt
  1812.11151}}.

\bibitem{Klebanov:1999tb}
I.~R. Klebanov and E.~Witten, ``{AdS / CFT correspondence and symmetry
  breaking},'' {\em Nucl. Phys. B} {\bf 556} (1999) 89--114,
  \href{https://arxiv.org/abs/hep-th/9905104}{{\tt hep-th/9905104}}.

\bibitem{Kobayashi:2018lil}
N.~Kobayashi, T.~Nishioka, Y.~Sato, and K.~Watanabe, ``{Towards a $C$-theorem
  in defect CFT},'' {\em JHEP} {\bf 01} (2019) 039,
  \href{https://arxiv.org/abs/1810.06995}{{\tt 1810.06995}}.

\bibitem{Giombi:2014xxa}
S.~Giombi and I.~R. Klebanov, ``{Interpolating between $a$ and $F$},'' {\em
  JHEP} {\bf 03} (2015) 117, \href{https://arxiv.org/abs/1409.1937}{{\tt
  1409.1937}}.

\bibitem{McAvity:1995zd}
D.~M. McAvity and H.~Osborn, ``{Conformal field theories near a boundary in
  general dimensions},'' {\em Nucl. Phys.} {\bf B455} (1995) 522--576,
  \href{https://arxiv.org/abs/cond-mat/9505127}{{\tt cond-mat/9505127}}.

\bibitem{Beccaria:2017nco}
M.~Beccaria and A.~Tseytlin, ``{On induced action for conformal higher spins in
  curved background},'' {\em Nucl. Phys. B} {\bf 919} (2017) 359--383,
  \href{https://arxiv.org/abs/1702.00222}{{\tt 1702.00222}}.

\bibitem{doi:10.1063/1.530850}
R.~Camporesi and A.~Higuchi, ``Spectral functions and zeta functions in
  hyperbolic spaces,'' {\em Journal of Mathematical Physics} {\bf 35} (1994),
  no.~8 4217--4246,
  \href{https://arxiv.org/abs/https://doi.org/10.1063/1.530850}{{\tt
  https://doi.org/10.1063/1.530850}}.

\bibitem{Bytsenko:1994bc}
A.~A. Bytsenko, G.~Cognola, L.~Vanzo, and S.~Zerbini, ``{Quantum fields and
  extended objects in space-times with constant curvature spatial section},''
  {\em Phys. Rept.} {\bf 266} (1996) 1--126,
  \href{https://arxiv.org/abs/hep-th/9505061}{{\tt hep-th/9505061}}.

\bibitem{Diaz:2007an}
D.~E. Diaz and H.~Dorn, ``{Partition functions and double-trace deformations in
  AdS/CFT},'' {\em JHEP} {\bf 05} (2007) 046,
  \href{https://arxiv.org/abs/hep-th/0702163}{{\tt hep-th/0702163}}.

\bibitem{Lewkowycz:2013laa}
A.~Lewkowycz and J.~Maldacena, ``{Exact results for the entanglement entropy
  and the energy radiated by a quark},'' {\em JHEP} {\bf 05} (2014) 025,
  \href{https://arxiv.org/abs/1312.5682}{{\tt 1312.5682}}.

\bibitem{Witten:2001ua}
E.~Witten, ``{Multitrace operators, boundary conditions, and AdS / CFT
  correspondence},'' \href{https://arxiv.org/abs/hep-th/0112258}{{\tt
  hep-th/0112258}}.

\bibitem{Giombi:2016ejx}
S.~Giombi, ``{Higher Spin --- CFT Duality},'' in {\em {Theoretical Advanced
  Study Institute in Elementary Particle Physics}: {New Frontiers in Fields and
  Strings}}, pp.~137--214, 2017.
\newblock \href{https://arxiv.org/abs/1607.02967}{{\tt 1607.02967}}.

\bibitem{Petkou:1994ad}
A.~Petkou, ``{Conserved currents, consistency relations and operator product
  expansions in the conformally invariant O(N) vector model},'' {\em Annals
  Phys.} {\bf 249} (1996) 180--221,
  \href{https://arxiv.org/abs/hep-th/9410093}{{\tt hep-th/9410093}}.

\bibitem{Metlitski:2009iyg}
M.~A. Metlitski, C.~A. Fuertes, and S.~Sachdev, ``{Entanglement Entropy in the
  O(N) model},'' {\em Phys. Rev. B} {\bf 80} (2009), no.~11 115122,
  \href{https://arxiv.org/abs/0904.4477}{{\tt 0904.4477}}.

\bibitem{PhysRevD.7.2911}
K.~G. Wilson, ``Quantum Field - Theory Models in Less Than 4 Dimensions,'' {\em
  Phys. Rev. D} {\bf 7} (May, 1973) 2911--2926.

\bibitem{Fitzpatrick:2011dm}
A.~L. Fitzpatrick and J.~Kaplan, ``{Unitarity and the Holographic S-Matrix},''
  {\em JHEP} {\bf 10} (2012) 032, \href{https://arxiv.org/abs/1112.4845}{{\tt
  1112.4845}}.

\bibitem{Liu:1998ty}
H.~Liu and A.~A. Tseytlin, ``{On four point functions in the CFT / AdS
  correspondence},'' {\em Phys. Rev. D} {\bf 59} (1999) 086002,
  \href{https://arxiv.org/abs/hep-th/9807097}{{\tt hep-th/9807097}}.

\bibitem{DHoker:1999kzh}
E.~D'Hoker, D.~Z. Freedman, S.~D. Mathur, A.~Matusis, and L.~Rastelli,
  ``{Graviton exchange and complete four point functions in the AdS / CFT
  correspondence},'' {\em Nucl. Phys. B} {\bf 562} (1999) 353--394,
  \href{https://arxiv.org/abs/hep-th/9903196}{{\tt hep-th/9903196}}.

\bibitem{Dolan:2000ut}
F.~A. Dolan and H.~Osborn, ``{Conformal four point functions and the operator
  product expansion},'' {\em Nucl. Phys.} {\bf B599} (2001) 459--496,
  \href{https://arxiv.org/abs/hep-th/0011040}{{\tt hep-th/0011040}}.

\bibitem{Hijano:2015zsa}
E.~Hijano, P.~Kraus, E.~Perlmutter, and R.~Snively, ``{Witten Diagrams
  Revisited: The AdS Geometry of Conformal Blocks},'' {\em JHEP} {\bf 01}
  (2016) 146, \href{https://arxiv.org/abs/1508.00501}{{\tt 1508.00501}}.

\bibitem{Jepsen:2019svc}
C.~B. Jepsen and S.~Parikh, ``{Propagator identities, holographic conformal
  blocks, and higher-point AdS diagrams},'' {\em JHEP} {\bf 10} (2019) 268,
  \href{https://arxiv.org/abs/1906.08405}{{\tt 1906.08405}}.

\bibitem{Horowitz:2014gva}
G.~T. Horowitz, N.~Iqbal, J.~E. Santos, and B.~Way, ``{Hovering Black Holes
  from Charged Defects},'' {\em Class. Quant. Grav.} {\bf 32} (2015) 105001,
  \href{https://arxiv.org/abs/1412.1830}{{\tt 1412.1830}}.

\bibitem{Giombi:2018vtc}
S.~Giombi, V.~Kirilin, and E.~Perlmutter, ``{Double-Trace Deformations of
  Conformal Correlations},'' {\em JHEP} {\bf 02} (2018) 175,
  \href{https://arxiv.org/abs/1801.01477}{{\tt 1801.01477}}.

\bibitem{Dolan:2000uw}
F.~A. Dolan and H.~Osborn, ``{Implications of N=1 superconformal symmetry for
  chiral fields},'' {\em Nucl. Phys.} {\bf B593} (2001) 599--633,
  \href{https://arxiv.org/abs/hep-th/0006098}{{\tt hep-th/0006098}}.

\end{thebibliography}\endgroup

\end{document}